%% file: lfv-muc-prd-v1.tex
\documentclass[letterpaper, 11pt]{article}
\pdfoutput=1
\usepackage[margin=1.0in]{geometry}
\usepackage[T1]{fontenc}
\usepackage[utf8]{inputenc}

\usepackage{amsmath,amssymb,graphicx}
\usepackage{dsfont}
\usepackage{xcolor}
\definecolor{Mahogany}{rgb}{0.62,0.24,0.15}
\definecolor{DarkRed}{rgb}{0.6,0,0}
\definecolor{DarkGreen}{rgb}{0,0.6,0}
\definecolor{DarkBlue}{rgb}{0,0,0.6}
\definecolor{DarkOrange}{rgb}{0.9,0.5,0.0}
\definecolor{gray}{RGB}{128,128,128}
\usepackage[colorlinks=true,linktocpage=true,linkcolor=DarkBlue,citecolor=DarkGreen,urlcolor=DarkGreen]{hyperref}
\usepackage{slashed}
\usepackage{verbatim}

\newlength{\Oldarrayrulewidth}
\newcommand{\Cline}[2]{%
  \noalign{\global\setlength{\Oldarrayrulewidth}{\arrayrulewidth}}%
  \noalign{\global\setlength{\arrayrulewidth}{#1}}\cline{#2}%
  \noalign{\global\setlength{\arrayrulewidth}{\Oldarrayrulewidth}}}
  
\usepackage{setspace}
\usepackage{upgreek}
\usepackage{bbm} % for identity matrix

\usepackage[font=small,labelfont=bf]{caption}
\usepackage{cite}
\let\OLDthebibliography\thebibliography
\renewcommand\thebibliography[1]{
    \small
    \OLDthebibliography{#1}
    \setlength{\parskip}{0.0pt plus 0.5pt}
    \setlength{\itemsep}{3.5pt plus 2.0pt minus 1.0pt}
}

\usepackage{multirow}

\newcommand{\lab}[1]{{\mathrm{#1}}}
\DeclareMathOperator{\BR}{BR}
\DeclareMathOperator{\CR}{CR}
\renewcommand{\Re}{\mathrm{Re}}

\def\dd{\text{d}}
\newcommand{\hc}{{\textrm{h.c.}}}

\newcommand{\minus}{{\scalebox {0.75}[1.0]{$-$}}}

\newcommand{\lrDmu}{\overset{\scalebox{1.1}[0.75]{$\leftrightarrow$}}{D}_{\!\mu}}
\newcommand{\lrDmuI}{\overset{\scalebox{1.1}[0.75]{$\leftrightarrow$}}{D}_{\!\mu}{}^{\!\!\!I}}
\newcommand{\lrD}[1]{\overset{\scalebox{1.1}[0.75]{$\leftrightarrow$}}{D}_{\!#1}}
\newcommand{\lrDsuper}[2]{\overset{\scalebox{1.1}[0.75]{$\leftrightarrow$}}{D}_{\!#1}{}^{\!\!\!{#2}}}

\definecolor{customred}{rgb}{0.9, 0.36, 0.36}
\definecolor{customyellow}{rgb}{.92, 0.79, 0.0}
\definecolor{customorange}{rgb}{0.95, 0.55, 0.00}
\definecolor{customgreen}{rgb}{0.17, 0.67, 0.11}
\definecolor{custompurple}{rgb}{0.41, 0.16, 0.68}
\definecolor{custombrown}{rgb}{0.45, 0.25, 0.1}

\begin{document}

\begin{flushright}
\small{PITT-PACC-2510}
\end{flushright}

\begin{center}
{\LARGE \bf
Lepton Flavor Violation:\\[0.25em]From Muon Decays to Muon Colliders}
\end{center}

\begin{center}
Pouya Asadi$^{1,2}$, 
Hengameh Bagherian$^{3,4}$, 
Katherine Fraser$^{5,6}$,\\[0.25em] 
Samuel Homiller$^{7,8}$ and 
Qianshu Lu$^{9,10}$\\[0.5em]
{\small \color{custompurple} 
\texttt{pasadi@ucsc.edu},~ 
\texttt{hengameh@uchicago.edu},~ 
\texttt{kfraser@berkeley.edu}},\\[0.0em]
{\small \color{custompurple} 
\texttt{shomiller@pitt.edu},~ 
\texttt{qianshu.lu@ias.edu}}\\[0.5em]
{\small${}^1$Department of Physics and Santa Cruz Institute for Particle Physics,}\\[-0.25em]
{\small University of California Santa Cruz, Santa Cruz, CA 95064}\\[0.25em]
{\small${}^2$Institute for Fundamental Science and Department of Physics,}\\[-0.25em] 
{\small University of Oregon, Eugene, OR 97403}\\[0.25em]
{\small${}^3$Enrico Fermi Institute and Leinweber Institute for Theoretical Physics,}\\[-0.25em]
{\small University of Chicago, Chicago, IL 60637}\\[0.25em]
{\small${}^4$Department of Physics, Harvard University, Cambridge, MA 02138}\\[0.25em]
{\small ${}^5$Leinweber Institute for Theoretical Physics and Miller Institute}\\[-0.25em]
{\small University of California, Berkeley, CA 94720}\\[0.25em]
{\small ${}^6$Theoretical Physics Group, Lawrence Berkeley National Laboratory, Berkeley, CA 94720}\\[0.25em]
{\small ${}^7$Pitt PACC, Department of Physics and Astronomy,}\\
[-0.25em]
{\small University of Pittsburgh, Pittsburgh, PA 15260}\\[0.25em]
{\small ${}^8$Laboratory for Elementary Particle Physics, Cornell University, Ithaca, NY 14853}\\[0.25em]
{\small${}^9$School of Natural Sciences, Institute for Advanced Study, Princeton, NJ 08540}\\[0.25em]
{\small${}^{10}$Center for Cosmology and Particle Physics, Department of Physics,}\\[-0.25em]
{\small New York University, New York, NY 10003}
\end{center}

\begin{abstract}
We investigate the unique potential of a high-energy muon collider to probe lepton-flavor-violating signals arising from physics beyond the Standard Model (SM). 
Low-energy, precision searches for charged lepton-flavor violation (LFV) are projected to dramatically improve their sensitivity  
in the coming years, and could provide the first evidence of new physics. 
We interpret the sensitivity of these searches in terms of a set of LFV operators in the SM Effective Field Theory. 
The same operators are then probed at the TeV scale via new, high-energy processes only available at a high-energy muon collider, such as $\mu\mu\to\mu \tau$ or the scattering of a muon of an electroweak gauge boson into LFV final states.
We find that for most operators, a muon collider could confirm signals if they are seen at future low energy experiments, whereas for certain flavor combinations it extends the reach to scales well beyond those accessible at lower energies. 
We also project the sensitivity of a muon collider to lepton-flavor-violating decays of the SM Higgs boson and demonstrate improved sensitivity to $h \to e\tau$ and $h \to \mu\tau$ by an order of magnitude compared to the High-Luminosity LHC. 
The importance of having multiple, complementary probes is illustrated by considering both various combinations of operators and relative sizes of flavor-violating transitions between generations under various assumptions for the flavor structure of new physics. 
\end{abstract}

\newpage 

\tableofcontents
\vskip 0.5cm

\iffalse
{
\sh{Consistency Checks}
\begin{itemize}

    \item Standardize reference to Sections, Subsection, Table, Fig., Eq., etc.

    \item include Ref(s). when appropriate in citing papers.

    \item Standardize hyphenation vs. en-dash for $\mu-e$, $\tau$--$\mu$, etc.\\
    These should be done as $\tau$\,--\,$\mu$ (en-dash with small spaces around it). 
    
    \item Decide on hyphenation when "lepton-flavor-violating" is used as an adjective vs. noun

    \item Check $\ell$ vs. $l$ for light leptons.\ql{$\ell$ doesn't exist in the document right now}

    \item muon collider not Muon Collider.

    \item $\slashed{E}_{T}$ not $E_{\slashed{T}}$ or $E_{T,\textrm{miss}}$

    \item flavor indices are $e$, $\mu$, $\tau$, not 1, 2, 3.

    \item boldface for figure panels

    \item Check figure placement

    \item delete extra commented out text
    
\end{itemize}
}
\fi

%%%%%%%%%%%%%%%%%%%%%%%%%%%%%%%%%%%%%%%%%%%%%%%%%%%%%%%%%%%%%%%%%%%%%%%%%%
%%%%%%%%%%%%%%%%%%%%%%%%%%%%%%%%%%%%%%%%%%%%%%%%%%%%%%%%%%%%%%%%%%%%%%%%%%
\section{Introduction}

The Standard Model (SM) of particle physics has been remarkably successful in explaining disparate phenomena across a vast range of energy scales. 
The Large Hadron Collider (LHC) in particular has successfully corroborated many predictions of the SM, producing both astonishingly precise measurements of SM parameters and the discovery of its final piece: the Higgs boson.

Despite this progress, there are many features of the SM which we do not  understand. 
These include the vanishingly small charge parity (CP) odd gluon coupling (the Strong CP problem), the inexplicably small Higgs mass (the Hierarchy Problem), and the observed patterns of masses and mixings between different generations of SM fermions (the SM flavor puzzle). 
Resolutions of these puzzles require additional, unknown dynamics in the ultraviolet (UV). 
For the flavor puzzle in particular, these new dynamics generically generate flavor-changing processes that can be searched for experimentally. 
These signatures are tightly constrained by high-intensity and precision experiments at lower energies which can indirectly probe flavor physics at much higher energy scales.

Examples of such precision measurements include electric and magnetic dipole moments as well as charged lepton-flavor-violating (LFV) decays (such as $\mu \to e \gamma$). The Standard Model rates for these processes are suppressed by tiny neutrino masses and are negligibly small. 
At the same time, the existence of neutrino masses indicates that lepton flavor is not an exact symmetry of the SM, making it natural that any UV completion of the SM responsible for the observed mass pattern will also include LFV signals.
This expectation has strongly motivated the continual improvement of searches for LFV across multiple channels. Low-energy experiments have been remarkably successful in constraining LFV, with especially strong constraints on $\mu$\,--\,$e$ transitions. Their extraordinary reach in probing high scales makes them promising candidates to provide the first hints of new physics, potentially even before direct discoveries at colliders. 

Despite their great reach, these small-scale experiments have their own limitations: unlike the LHC or other high-energy facilities, they typically probe only a narrow range of flavor observables and BSM models. 
Additionally, if a small-scale experiment does discover new LFV physics,  comprehensively characterizing the underlying physics behind the signal would require a direct search at a high-energy collider. 
In this sense, the absence of direct BSM signals at the LHC elevates the role of precision flavor measurements: they may serve as harbingers of new physics and, in doing so, help guide the design and priorities of future colliders, shaping the long-term trajectory of particle physics.

A prominent proposal for future energy-frontier colliders is a high-energy muon collider~\cite{Ankenbrandt:1999cta, Wang:2015xoa, Boscolo:2018ytm, Neuffer:2018yof, Delahaye:2019omf, Shiltsev:2019rfl} (see also Refs.~\cite{MuonCollider:2022xlm, MuonCollider:2022glg, MuonCollider:2022nsa, MuonCollider:2022ded, Black:2022cth, Accettura:2023ked, InternationalMuonCollider:2024jyv, InternationalMuonCollider:2025sys} for recent reviews). 
Muon colliders outperform proton colliders in energy reach for a fixed size due to the elementary nature of the muon, while also avoiding the significant synchrotron energy loss of electron colliders due to the muon's larger mass. 
The report of the National Academies of Science, Engineering and Medicine lists the development of a path towards a high-energy muon collider as their highest priority recommendation for the future of particle physics~\cite{NationalAcademiesofSciencesEngineeringandMedicine:2025bix}, which motivates work on such machines from a variety of angles. 
Major technological developments towards building such a machine have been pointed out in Refs.~\cite{Ankenbrandt:1999cta, Wang:2015xoa, Boscolo:2018ytm, Neuffer:2018yof, Delahaye:2019omf, Shiltsev:2019rfl}, especially on the cooling front \cite{MICE:2019jkl}, while also outlining additional steps that are necessary before a final design can be developed. 
These studies have identified a 10 TeV muon collider with a final integrated luminosity of $10$~ab$^{-1}$ as a benchmark target. 

To complement and motivate accelerator-physics studies, parallel efforts have been devoted to assessing the potential of a future muon collider to probe well-motivated scenarios of new physics. 
In particular, it has been emphasized that such a machine effectively functions as an electroweak gauge boson collider \cite{Han:2020uid, AlAli:2021let, Han:2021kes, Ruiz:2021tdt, Garosi:2023bvq}, thereby enabling exploration of a wide range of colorless new physics models that remain comparatively less constrained at the LHC. 
In this spirit, numerous studies have investigated the reach of a muon collider in Higgs and precision electroweak measurements \cite{Eichten:2013ckl, Chakrabarty:2014pja, Buttazzo:2018qqp, Chiesa:2020awd, Han:2020pif, Bandyopadhyay:2020otm, Buttazzo:2020uzc, Liu:2021jyc, Han:2021udl, Franceschini:2021aqd, Chiesa:2021qpr, Chen:2022msz, Spor:2022hhn, Forslund:2022xjq, Forslund:2023reu, Ruhdorfer:2023uea, Amarkhail:2023xsc, Homiller:2024uxg}, 
flavorful new physics \cite{Capdevilla:2020qel, Buttazzo:2020ibd, Yin:2020afe, Huang:2021nkl, Capdevilla:2021rwo, Chen:2021rnl, Huang:2021biu, Asadi:2021gah, Bandyopadhyay:2021pld, Qian:2021ihf, Homiller:2022iax, Azatov:2022itm, Yang:2023ojm, Altmannshofer:2023uci, Jana:2023ogd, Ghosh:2023xbj, Han:2023njx, Bhattacharya:2023beo, Glioti:2025zpn}, 
dark-matter candidates \cite{Han:2020uak, Bottaro:2021srh, Liu:2022byu, Jueid:2023zxx, Vignaroli:2023rxr, Belfkir:2023vpo, Dasgupta:2023zrh, Asadi:2023csb, Cesarotti:2024rbh, Asadi:2024jiy}, 
and a broad spectrum of other BSM scenarios \cite{DiLuzio:2018jwd, Costantini:2020stv, Liu:2021akf, AlAli:2021let, Casarsa:2021rud, Bao:2022onq, Inan:2022rcr, Lv:2022pts, Chen:2022yiu, Kwok:2023dck, Inan:2023pva, Chowdhury:2023imd, Belfkir:2023lot, Chigusa:2023rrz, Capdevilla:2024bwt, Ma:2024ayr, Han:2025wdy}. 
Proposals have also been put forward to exploit muon-beam dumps as powerful tools to probe hidden-sector and other BSM models \cite{Cesarotti:2022ttv, Cesarotti:2023sje}.

In this work, we analyze the capabilities of a muon collider in observing signals associated with physics that violates lepton flavor. 
We will identify a number of processes initiated by muons or electroweak gauge bosons that probe lepton flavor violation, and highlight how a high-energy muon collider is uniquely positioned to study them.
LFV signals at a muon collider have previously been studied in the context of specific models \cite{Homiller:2022iax, AlAli:2021let}, but here we take a more model-agnostic approach. 
Specifically, we investigate the reach of a muon collider to detect LFV SM effective field theory (SMEFT) operators.
The SMEFT captures the most general set of higher dimension operators that can be written with the SM field content while respecting its gauge symmetries, and is a useful model-independent framework for BSM setups; see Refs.~\cite{Brivio:2017vri, Isidori:2023pyp} for recent reviews. Lepton-flavor-violating SMEFT operators have also been studied in the context of electron-positron colliders~\cite{Calibbi:2021pyh, Altmannshofer:2023tsa, Jahedi:2024kvi}, but as we will see, both the higher-energy and the fact that muons are participating in the collisions will lead to signatures at $\mu^+\mu^-$ machines that are unique.

We consider all SMEFT dimension six operators with no quark fields that can give rise to $\tau$\,--\,$\mu$ LFV signals.\footnote{We omit, for instance, four-fermion operators involving two quark fields that violate lepton flavor. For a comprehensive study of their low-energy effects, see Ref.~\cite{Plakias:2023esq}; for an overview of the corresponding signals at a high-energy muon collider, see the recent work in Ref.~\cite{Glioti:2025zpn}.} 
Not counting the generation indices, there are nine independent operators that violate lepton flavor and that do not involve quark fields. These can be written as:\footnote{We have suppressed the subscripts $L$ and $R$ on the fermions for clarity, and defined 
\begin{equation*}
(H^{\dagger}i \lrDmu H) \equiv i H^{\dagger}(D_{\mu}H) - i (D_{\mu}H)^{\dagger}H, \qquad
(H^{\dagger}i \lrDmuI H) \equiv i H^{\dagger} \tau^I (D_{\mu} H) - i (D_{\mu}H)^{\dagger} \tau^I H .
\end{equation*}
}
\begin{equation}
\label{eq:smeft_operators}
\begin{gathered}
\mathcal{O}_{e W} = (\bar{L}_i \sigma^{\mu\nu} e_j) \tau^I H W_{\mu\nu}^I \,, \quad 
\mathcal{O}_{e B} = (\bar{L}_i \sigma^{\mu\nu} e_j) H B_{\mu\nu} \,, \quad
\mathcal{O}_{e H} = (H^{\dagger}H) (\bar{L}_i e_j H) \,,
\\[0.5em]
\mathcal{O}_{Hl}^{(1)} = (H^{\dagger} i \lrD{\mu} H)(\bar{L}_i \gamma^{\mu} L_j) \,, \quad  
\mathcal{O}_{Hl}^{(3)} = (H^{\dagger} i \lrDsuper{\mu}{I} H) (\bar{L}_i \tau^I \gamma^{\mu} L_j) \,, \quad  
\mathcal{O}_{He} = (H^{\dagger}i \lrDmu H) (\bar{e}_i \gamma^{\mu} e_j) \,,
\\[0.75em]
\mathcal{O}_{ll} = (\bar{L}_i \gamma_{\mu} L_j)(\bar{L}_k \gamma^{\mu} L_m) \,, \quad
\mathcal{O}_{ee} = (\bar{e}_i \gamma_{\mu} e_j)(\bar{e}_k \gamma^{\mu} e_m) \,, \quad
\mathcal{O}_{le} = (\bar{L}_i \gamma_{\mu} L_j)(\bar{e}_k \gamma^{\mu} e_m) \, .
\end{gathered}
\end{equation}
Casting the effect of a UV model in the context of these operators allows us to properly compare the constraints from the high-energy collider searches and the IR precision observables; this is possible thanks to extensive works on the renormalization group equations (RGEs) of operators in SMEFT and their matching to the low energy field theory (LEFT) \cite{Jenkins:2013zja, Jenkins:2013wua, Alonso:2013hga, Jenkins:2017jig, Jenkins:2017dyc}.\footnote{Although we only consider turning on SMEFT operators with no quarks in the UV, we do consistently include quark operators in the low energy which are generated by running and matching.}
We use these results to properly compare the reach of a 10~TeV muon collider with an integrated luminosity of 10~ab$^{-1}$ to existing and near future experiments for probing LFV effects in the IR. 
We find that, in some cases, a future muon collider could provide better sensitivity to $\tau$\,--\,$\mu$ LFV signals than IR observables, while in other cases it could confirm LFV signals if they are detected at future low-energy experiments.

We examine the sensitivity of many different scattering processes to $\tau$\,--\,$\mu$ flavor violation, and conduct detailed background studies for those with the greatest reach. These include $\mu V \to \tau h$ and $\mu W \rightarrow \tau W$ for probing all operators involving the Higgs except $\mathcal{O}_{eH}$, and $\mu\mu \to \mu\tau$, $\mu V \to \mu \tau \tau$, and $\mu V \to \tau \tau \tau$ for probing the four-fermion operators. We find that after implementing cuts to suppress backgrounds, these processes probe scales of new LFV physics between 10\,--\,100s of TeV. For simplicity, we restrict to simple cut-and-count searches for the signal of these operators. While we expect a more complicated analysis using tools such as machine learning (ML) to provide additional improvement, this simple, conservative strategy will suffice to understand the approximate sensitivity for a future collider. We also constrain all three flavor combinations ($\tau$\,--\,$\mu$, $\tau$\,--\,$e$ and $\mu$\,--\,$e$) of $\mathcal{O}_{eH}$ from LFV Higgs decays, finding that a muon collider would improve on LFV branching ratios of Higgs by around one order of magnitude compared to current LHC bounds. 

The rest of this paper is organized as follows: first in Section~\ref{sec:low_energy},  we calculate the reach in the scale of SMEFT operators of flavor violating $\mu$, $\tau$, and $Z$ decays. In Section~\ref{sec:lfv_higgs}, we discuss constraints from flavor violating Higgs decays, including the reach of current experiments and the potential reach of a future muon collider. Next in Section~\ref{sec:collider}, we calculate additional constraints from scattering processes at a future muon collider, including $\mu V \rightarrow \tau h$, $\mu W \rightarrow \tau W$, $\mu \mu \rightarrow \mu \tau$, $\mu V \rightarrow \tau \tau \tau$, and $\mu^\mp V \rightarrow \mu^\pm \tau^\mp \tau^\mp$ with same sign $\tau$ leptons. Finally, we compare these constraints in Section~\ref{sec:summary}, including studying combinations of operators and different flavor ansatz, before concluding in Section~\ref{sec:conclusion}.

%%%%%%%%%%%%%%%%%%%%%%%%%%%%%%%%%%%%%%%%%%%%%%%%%%%%%%%%%%%%%%%%%%%%%%%%%%
%%%%%%%%%%%%%%%%%%%%%%%%%%%%%%%%%%%%%%%%%%%%%%%%%%%%%%%%%%%%%%%%%%%%%%%%%%
\section{Low-Energy Probes of Lepton Flavor Violation \label{sec:low_energy}}

In this section we review the existing and projected bounds from different precision experiments on LFV signals beyond the SM. 
These experiments include various $l \to l' \gamma$, lepton conversions or three-body decays, as well as flavor-violating decays of the $Z$ boson. We summarize the current and projected bounds on these observables in Table~\ref{tab:precision_bounds}. The following discussion, extending into the next section, reviews constraints from various experiments, and recast the sensitivity of current and proposed experiments in the context of the SMEFT operators in Eq.~\eqref{eq:smeft_operators}.

%%%%%
\input{precision-summary-table}

%%%%%

%%%%%%%%%%%%%%%%%%%%%%%%%%%%%%%%%%%%%%%%%%%%%%%%%%%%%%%%%%%%%%%%%%%%%%%%%%
\subsection{Decays or Conversions of Muons and Taus}

Some of the most stringent constraints on lepton flavor violation come from searches for flavor-violating decays of muons and tau leptons, or from searches for muons converting into electrons when scattering off of atomic nuclei.

Processes involving a muon and electron are particularly well-constrained. The MEG experiment, for instance, searches for the $\mu \to e\gamma$ decay via a $\mu^+$ beam incident on a thin plastic target giving rise to a back-to-back signal of a positron and a photon~\cite{MEGdesign:2013vqa}. 
The best limit, $\BR(\mu \to e\gamma) < 1.5 \times 10^{-13}$ was just recently set by the MEG-II experiment~\cite{MEGII:2025gzr}, and is expected to improve substantially with more data collection that continues to 2026~\cite{Cattaneo:2025bnk}. A complementary suite of experiments search for the decay $\mu^+\rightarrow e^+ e^-e^+$\cite{SINDRUMII:1993gxf,  Mu3e:2020gyw}.
Over the coming decade, the most significant improvements are projected for the muon-electron conversion processes, where a muon rapidly stops and binds with the atomic nucleus in a target, and subsequently converts to an electron with energy slightly below the muon mass\cite{COMET:2018auw}. The conversion rate (CR) is the flavor-changing capture rate normalized with respect to the total nucleus capture rate which is dominated by the Standard Model process $\mu^- N(A,Z)\rightarrow \nu_{\mu}N(A,Z-1)$. The Mu2e and COMET experiments \cite{Bartoszek:2014mya,COMET:2018auw} are expected to improve upon the limits set by \textsc{sindrum-ii}~\cite{SINDRUMII:1993gxf, Bertl:2006up} on Titanium and Gold nuclei by over four orders of magnitude using an aluminum nucleus.

There are also constraints on lepton-flavor-violating processes involving the $\tau$. Due to its larger mass and shorter lifetime, $\tau$ leptons are more challenging to produce and manipulate at small-scale experiments. However, they are produced abundantly in collider experiments operating at the $\Upsilon$ threshold: the so-called ``$B$-factory'' experiments such as BaBar, Belle and Belle II. Most of the strongest limits on processes such as $\tau \to \mu\gamma$ or $\tau \to 3e$ were set by the previous generation of $B$-factory experiments. 
SuperKEKB is targeting an increase in the total luminosity available to Belle II by a factor of 40 compared to Belle, so all of these limits are expected to significantly improve in the next decade. See Table~\ref{tab:precision_bounds} for a comprehensive summary.\footnote{These collider experiments can also constrain the semi-hadronic LFV $\tau$ decays in the form of $\tau\rightarrow l h$, $l = e,\mu$ and $h$ is a hadronic system\cite{BelleTau:2024wrw}. As we are omitting SMEFT operators that involve quarks from our study, we will not study the bounds from these decays: for purely leptonic operators, the purely leptonic decays are always more constraining.}

\bigskip

The aforementioned observables are all measured at energy scales fixed by the masses of the corresponding leptons. As such, a more apt description of potential new physics effects is not in terms of the SMEFT, but the LEFT.
As we will discuss more below, using the LEFT allows us to systematically incorporate higher order perturbative effects arising from flavor-violating SMEFT operators into the low-energy calculations, from both one- and two-loop matching of the SMEFT onto LEFT at the weak scale, and from renormalization group evolution of the LEFT coefficients down to the relevant lepton mass scale.
Systematic studies of LFV signals in the language of LEFT can be found in Refs.~\cite{Pruna:2014asa, Crivellin:2017rmk, Fernandez-Martinez:2024bxg}.

In the notation of Refs.~\cite{Jenkins:2017jig, Jenkins:2017dyc}, the LEFT operators relevant for flavor-violating muon and tau decays to leptonic final states are:
\begin{equation}
\label{eq:left_operators}
\begin{gathered}
\mathcal{O}_{e\gamma, ij} = \bar{e}_{L,i} \sigma^{\mu\nu} e_{R,j} F_{\mu\nu}\,, ~~
\mathcal{O}_{ee,prst}^{V,LL} = (\bar{e}_{L,p} \gamma^{\mu} e_{L,r})(\bar{e}_{L,s}\gamma_{\mu} e_{L,t}) \,,~~
\mathcal{O}_{ee,prst}^{V,RR} = (\bar{e}_{R,p} \gamma^{\mu} e_{R,r})(\bar{e}_{R,s}\gamma_{\mu} e_{R,t}) \,, \\[0.5em]
\mathcal{O}_{ee,prst}^{V,LR} = (\bar{e}_{L,p} \gamma^{\mu} e_{L,r})(\bar{e}_{R,s}\gamma_{\mu} e_{R,t})\,,\quad 
\mathcal{O}_{ee,prst}^{S,RR} = (\bar{e}_{L,p} e_{R,r})(\bar{e}_{L,s} e_{R,t}) \,.
\end{gathered}
\end{equation}
where $ij$ and $prst$ are flavor indices. In addition, we must also consider LEFT operators involving quarks, which are generated by renormalization group evolution below the weak scale even if we start with only leptonic operators in the UV, and are relevant for $\mu \to e$ conversion in atomic nuclei:
\begin{equation}
\begin{gathered}
\mathcal{O}_{eq,prst}^{V,LL} = (\bar{e}_{L,p} \gamma^{\mu} e_{L,r})(\bar{q}_{L,s} \gamma_{\mu} q_{L,t}) \,,\quad
\mathcal{O}_{eq,prst}^{V,RR} = (\bar{e}_{R,p} \gamma^{\mu} e_{R,r})(\bar{q}_{R,s} \gamma_{\mu} q_{R,t})\,,\\[0.5em]
\mathcal{O}_{eq,prst}^{V,LR} = (\bar{e}_{L,p} \gamma^{\mu} e_{L,r})(\bar{q}_{R,s} \gamma_{\mu} q_{R,t})\,, \\[0.5em]
\mathcal{O}_{eq,prst}^{S,RR} = (\bar{e}_{L,p} e_{R,r})(\bar{q}_{L,s} q_{R,t})\,,\quad
\mathcal{O}_{eq,prst}^{S,RL} = (\bar{e}_{L,p} e_{R,r})(\bar{q}_{R,s} q_{L,t})\,.\quad
\end{gathered}
\end{equation}
Here, $q = u$ or $d$.

For example, $\mu \to e \gamma$ decay can be written as a function of the Wilson coefficients of the LEFT operators as
\begin{equation}
\label{eq:mu_to_e_gamma_left}
\Gamma(\mu \to e\gamma) = \frac{m_{\mu}^3}{4\pi} \Big( |L_{e\gamma,12}(m_{\mu})|^2 + |L_{e\gamma,21}(m_{\mu})|^2 \Big) \, .
\end{equation} 
In this expression, the $L_{e\gamma,ij}(m_{\mu})$ are the Wilson coefficients of the operators $\mathcal{O}_{e\gamma,ij}$ evaluated at the muon mass scale. In general, operator coefficients at the muon mass scale (or the tau mass scale depending on the relevant experiment) can be written as a function of the Wilson coefficients of the SMEFT operators, defined at the weak scale, using
\begin{equation}
\label{eq:L_evolved}
L_{\mathcal{O}}(m_{\mu}) = L^{[0]}_{\mathcal{O}}(M_Z) + \frac{1}{16\pi^2}\dot{L}^{[0]}_{\mathcal{O}}(M_Z) \log\frac{m_{\mu}}{M_Z}
+ L_{\mathcal{O}}^{[1]}(m_{\mu}) + L_{\mathcal{O}}^{[2]}(m_{\mu}) + \dots 
\end{equation}
In this expression, $L^{[0]}_{\mathcal{O}}$ indicates the value of the Wilson coefficient obtained from a tree-level matching of SMEFT onto LEFT at the scale $\mu_R = M_Z$. The full catagloue of such matching can be found in Ref.~\cite{Jenkins:2017jig}. The $\dot{L}$ indicate the (one-loop) anomalous dimensions,
\begin{equation}
\dot{L} \equiv 16\pi^2 \mu \frac{\dd}{\dd\mu} L \ ,
\end{equation}
which were computed in Ref.~\cite{Jenkins:2017dyc} and can be written in terms of LEFT coefficients obtained from the tree-level matching at $M_Z$. For all LEFT operators other than $\mathcal{O}_{e\gamma, ij}$, we only include the $L^{[0]}_{\mathcal{O}}$ and $\dot{L}^{[0]}_{\mathcal{O}}$ contributions, i.e., the LEFT operator coefficient at the muon mass scale is computed to the leading log order.

For $\mathcal{O}_{e\gamma, ij}$ only, as indicated in Eq.~\eqref{eq:L_evolved}, we include two further contributions, $L^{[1]}$ and $L^{[2]}$, coming from a $1$- and $2$-loop matching from SMEFT to LEFT. These are evaluated directly at $\mu_R = m_{\mu}$, without re-summing the potentially large log from $M_Z$ to $m_{\mu}$ which would require unknown 2- and 3-loop anomalous dimensions.
The one-loop matching onto the dipole operators is given in Refs.~\cite{Crivellin:2013hpa, Pruna:2014asa}. 
For the dipole operators, the two-loop matching contributions from $\mathcal{O}_{eH}$ are also numerically important: because the dipole transition requires a chirality flip, the one-loop contributions are suppressed by a factor of the lepton mass, while the two-loop contributions can exchange this suppression for a factor of the top mass or a gauge coupling (and an additional loop factor). We therefore include these contributions, adapting from the results in Refs.~\cite{Chang:1993kw, Harnik:2012pb}.

For completeness, we also collect here the expressions for the decays $l_i \to l_j \gamma$ and $l_i \to 3l_j$, which can be written in terms of LEFT coefficients at the appropriate lepton mass scale: 
\begin{equation}
\textrm{BR}(l_i \to l_j \gamma) = \frac{48\pi^2}{G_F^2 m_i^2} \Big( |L_{e\gamma,ji}|^2 + |L_{e\gamma,ij}^*|^2 \Big) \ ,
\end{equation}
and 
\begin{equation}
\begin{aligned}
\textrm{BR}(l_i \to 3 l_j) & = 
\frac{16\pi \alpha_{\textsc{em}}}{G_F^2 m_i^2} \Big( \log\frac{m_i^2}{m_j^2} - \frac{11}{4} \Big) \big( |L_{e\gamma,ji}|^2 + |L_{e\gamma,ij}^*|^2 \big) 
\\[0.25em]
& \kern-4.5em
- \frac{e}{G_F^2 m_i} \Re\Big[ L_{e\gamma,ij}^* (L_{ee,jjji}^{V,LR} + 2 L_{ee,jijj}^{V,RR})^* + L_{e\gamma,ji} (L_{ee,jijj}^{V,LR} + 2 L_{ee,jijj}^{V,LL}) \Big] 
\\[0.25em]
& \kern-4.5em
+ \frac{1}{64 G_F^2} \Big( 
|L_{ee,jjij}^{S,RR\,*}|^2 + |L_{ee,jijj}^{S,RR}|^2 
+ 16 |L_{ee,jijj}^{V,LL}|^2 + 16 |L_{ee,jijj}^{V,RR}|^2  
+ 8 |L_{ee,jijj}^{V,LR}|^2 + 8 | L_{ee,jjji}^{V,LR}|^2 \Big) \, .
\end{aligned}
\end{equation}
Expressions for the mixed final state decays can be found in Ref.~\cite{Altmannshofer:2023tsa}. These are unique among the low-energy probes in their importance for testing mixed-flavor four-fermion operators such as $\mathcal{O}_{ee,ee\mu\tau}^{V,RR}$.

For many four-fermion operators that violate lepton flavor by two units (e.g., $\mathcal{O}_{le,\tau\mu\tau\mu}$), there are no kinematically allowed decays of a $\tau$ or $\mu$ to charged lepton final states. For $\mathcal{O}_{le}$ and $\mathcal{O}_{ll}$, these operators contribute to decays such as $\tau \to \mu \nu \nu$, where they compete with the SM leptonic decay modes. In terms of the LEFT coefficients, we have for instance,
\begin{equation}
\label{eq:ell_to_ellnunu_decay}
\frac{\Gamma_{l_i \to l_j\nu\nu}}{\Gamma^{\textrm{SM}}_{l_i\to l_j\nu\nu}} = 
1 + \frac{4 |L_{ee,ijij}^{V,LL}|^2 + |L_{ee,ijij}^{V,LR}|^2}{8 G_F^2} \, .
\end{equation}
For operators involving multiple $\tau$ indices, these can be constrained for example, via measurements of the lepton flavor universality ratio,
\begin{equation}
R_{\mu} \equiv \frac{\textrm{BR}(\tau^- \to \mu^- \bar{\nu}_{\mu}\nu_{\tau})}{\textrm{BR}(\tau^- \to e^- \bar{\nu}_{e}\nu_{\tau})}
= R_{\mu}^{\textrm{SM}} \Big( 1 + 4 |L_{ee,3232}^{V,LL}|^2 + |L_{ee,3232}^{V,LR}|^2 - 4|L_{ee,3131}^{V,LL}|^2 - |L_{ee,3131}^{V,LR}|^2 \Big) \, .
\end{equation}
The most recent Belle~II result finds $R_{\mu} = 0.9675 \pm 0.0037$~\cite{Belle-II:2024vvr}, which is slightly lower than the SM value of $0.9726$. The combination with earlier measurements at BaBar and CLEO yields $R_{\mu} =  0.9735 \pm 0.0026$~\cite{CLEO:1996oro, BaBar:2009lyd, HFLAV:2022esi}, which is in good agreement with the SM. This uncertainty is already systematics dominated at Belle~II, so we do not make any projections for an improved future value. Note that this probe would be insensitive to $\tau$\,--\,$\mu$ and $\tau$\,--\,$e$ flavor-violating operators if they are both the same size.

To constrain similar operators involving $\mu$\,--\,$e$ flavor violation, we follow Ref.~\cite{Bigaran:2022giz} and interpret the effect in Eq.~\eqref{eq:ell_to_ellnunu_decay} for $\mu \to e\nu\nu$ as a shift in the Fermi constant, $G_{F,0} = G_{F,\mu}(1 + \delta G_F)$, with $G_{F,0}$ the value inferred from the electroweak theory and $G_{F,\mu}$ the value inferred from muon decays. A comparison to the determination using the weak-mixing angle yields the constraint $\delta G_F = (1.2 \pm 17.0) \times 10^{-5}$. From Eq.~\eqref{eq:ell_to_ellnunu_decay}, we find
\begin{equation}
\delta G_F = \frac{4 |L_{ee,2121}^{V,LL}|^2 + |L_{ee,2121}^{V,LR}|^2}{16 G_{F,0}^2} \, ,
\end{equation}
and the constraint on $\delta G_F$ above thus translates directly into a bound on the LEFT operators.

A separate bound on the $\mu$--$e$ flavor-violating operators comes from contributions to muonium--antimuonium conversion. Accounting for the magnetic field in the experimental setup at PSI that provides the strongest limit to date, the oscillation probability for muonium--antimuonium conversion can be parameterized as~\cite{Conlin:2020veq, Fukuyama:2021iyw, Fukuyama:2022fwi, Petrov:2022wau, Fukuyama:2023drl}
\begin{multline}
P_{M_{\mu}\to \bar{M}_{\mu}} \simeq \frac{7.58 \times 10^{-7}\,\textrm{GeV}^{-4}}{G_F^2} \Big( \big|L_{ee,2121}^{V,LL} + L_{ee,2121}^{V,RR} - 1.68 L_{ee,2121}^{V,LR}\big|^2 \\ 
+ 0.563 \big|L_{ee,2121}^{V,LL} + L_{ee,2121}^{V,RR} + 0.68 L_{ee,2121}^{V,LR}\big|^2 \Big) \, .
\end{multline}
Note that this rate is insensitive to the axial-vector combination, $L_{ee}^{V,LL} - L_{ee}^{V,RR}$. 
The PSI experiment set a limit on the conversion probability $P_{M_{\mu} \to \bar{M}_{\mu}} < 8.3 \times 10^{-11}$ at $90\%\,\textrm{C.L.}$~\cite{Willmann:1998gd}. This will be improved by the MACE experiment at the J-PARC muon facility, which aims to achieve a sensitivity $P_{M_{\mu} \to \bar{M}_{\mu}} \lesssim \times 10^{-13}$~\cite{Bai:2024skk}.

\medskip

Finally, for muon-to-electron conversion, we follow the procedure of Refs.~\cite{Kitano:2002mt, Cirigliano:2009bz, Crivellin:2017rmk}. 
Recently, the ``effective field theory tower'' for muon-to-electron conversion was extended down to the non-relativistic effective theory, allowing for a more general parameterization of the response of nuclear targets~\cite{Haxton:2024lyc}. Here, we will consider only the coherent conversion process, in which case the transition rate can be written 
\begin{equation}
\begin{aligned}
\omega_{\textrm{capt.}} \textrm{CR}(\mu \to e)_N & = 
\frac{m_{\mu}^3}{4} \bigg|
L_{e\gamma,21}^* D_N
+ 4 m_{\mu} \Big(\tilde{C}_{(p)}^{SL} S^{(p,N)} + \tilde{C}_{(p)}^{VR} V^{(p,N)} + (p \to n) \Big) \bigg|^2 
+ \big( L \leftrightarrow R \big)
\end{aligned}
\end{equation}
where $D_N$, $S^{(p/n,N)}$ and $V^{(p/n,N)}$ are numerical quantities related to the overlap integrals between the lepton wave-functions and the nucleon densities.
Ref.~\cite{Kitano:2002mt} evaluates these overlap integrals using several different methods appropriate for various nuclei. We simply take the average of the different results; for Aluminum nuclei, the differences are at the percent level, while for Gold the different methods yield results that differ from a few to a few tens of percent:
\begin{equation}
\begin{gathered}
D_{\textrm{Al}} = 0.03595\,, \quad
S^{(p,\textrm{Al})} = 0.0154 \,, \quad
S^{(n,\textrm{Al})} = 0.0165\,, \quad
V^{(p,\textrm{Al})} = 0.0160\,, \quad
V^{(n,\textrm{Al})} = 0.0171\,, \\[0.25em]
D_{\textrm{Au}} = 0.178\,, \quad
S^{(p,\textrm{Au})} = 0.05685\,, \quad
S^{(n,\textrm{Au})} = 0.0764\,, \quad
V^{(p,\textrm{Au})} = 0.09165\,, \quad
V^{(n,\textrm{Au})} = 0.127\, . \\[0.25em]
\end{gathered}
\end{equation}
The effective couplings, $\tilde{C}^{SL}$, $\tilde{C}^{VR}$, etc., are given by weighted sums over the nucleon form factors for the individual quarks,
\begin{equation}
\begin{gathered}
\tilde{C}_{(p/n)}^{VR} = 
\sum_{q=u,d,s} \!\!\Big( (L_{eq,21qq}^{V,LR})^* + L_{eq,12qq}^{V,RR} \Big) f_{V\,p/n}^{(q)} , 
\quad 
\tilde{C}_{(p/n)}^{VL} = 
\sum_{q=u,d,s}\!\! \Big( L_{eq,12qq}^{V,LR} + L_{eq,12qq}^{V,LL} \Big) f_{V\,p/n}^{(q)} ,
\\[0.25em]
\tilde{C}_{(p/n)}^{SL} = 
\sum_{q=u,d,s}\!\! \Big( (L_{eq,21qq}^{S,RR})^* + (L_{eq,21qq}^{S,RL})^* \Big) f_{S\,p/n}^{(q)} , %\\[0.25em]
\quad
\tilde{C}_{(p/n)}^{SR} = 
\sum_{q=u,d,s}\!\! \Big( L_{eq,12qq}^{S,RR} + L_{eq,12qq}^{S,RL} \Big) f_{S\,p/n}^{(q)} .
\end{gathered}
\end{equation}
%This has simplified somewhat from the expression in \cite{Crivellin:2017rmk}, as we can neglect the gluonic operators (what they call $C^L_{gg}$).
The nucleon form factors are given by
\begin{equation}
\begin{gathered}
f^{(u)}_{Sp} = (20.8 \pm 1.5) \times 10^{-3}\,, \quad
f^{(d)}_{Sp} = (41.1 \pm 2.8) \times 10^{-3}\,, \quad
f^{(s)}_{Sp} = (43 \pm 11) \times 10^{-3}\,, \\[0.25em]
f^{(u)}_{Sn} = (18.9 \pm 1.4) \times 10^{-3}\,, \quad 
f^{(d)}_{Sn} = (45.1 \pm 2.7) \times 10^{-3}\,, \quad
f^{(s)}_{Sn} = (43 \pm 11) \times 10^{-3}\,, \\[0.25em]
f^{(u)}_{Vp} = 2\,, \quad
f^{(d)}_{Vp} = 1\,, \quad
f^{(s)}_{Vp} = 0\,, \quad %\\[0.25em]
f^{(u)}_{Vn} = 1\,, \quad 
f^{(d)}_{Vn} = 2\,, \quad
f^{(s)}_{Vn} = 0\,.
\end{gathered}
\end{equation}
Here, the form factors for vector operators are determined by vector-current conservation. For the scalar operators, we follow Refs.~\cite{Crivellin:2014cta, Crivellin:2017rmk} and take the values of the up- and down-quark factors from the two-flavor chiral perturbation theory calculation described in Refs.~\cite{Crivellin:2013ipa, Hoferichter:2015dsa}. 
The form factors for the strange quarks are taken from the average of lattice calculations, presented in Ref.~\cite{Junnarkar:2013ac}, which is in agreement with effective field theory calculations in Refs.~\cite{Alarcon:2011zs, Alarcon:2012nr}.
%\footnote{Note that Ref.~\cite{Crivellin:2017rmk} uses only the result of \cite{Junnarkar:2013ac}, as opposed to the lattice average.}
Note that we neglect the uncertainties on these form factors in our calculations, for simplicity.

%%%%%%%%%%%%%%%%%%%%%%%%%%%%%%%%%%%%%%%%%%%%%%%%%%%%%%%%%%%%%%%%%%%%%%%%%%
\subsection{Flavor-Violating Decays of the \texorpdfstring{$Z$}{Z} Boson}

Additional constraints on lepton-flavor violation can be derived from constraints on the decays of the $Z$ and Higgs bosons.
As flavor-violating decays of the Higgs will also be well-tested at a future muon collider, %the LHC
we postpone a discussion of these constraints until Section~\ref{sec:lfv_higgs}, and here focus on the decays $Z \to l_i^+l_j^-$.
As these processes involve decays of particles near the weak scale, a tree-level, leading order description of the decays directly in terms of the SMEFT operators evaluated at the weak scale will be satisfactory for our purposes.

The width for lepton-flavor violating $Z$ decays is given by
\begin{equation}
\Gamma(Z \to l_i^+ l_j^-) = \frac{M_Z^3}{6\pi v^2} 
\Big(\big|\delta g_{ij}^{l_L}\big|^2 + \big|\delta g_{ij}^{l_R}\big|^2 \Big)\,,
\end{equation}
where the anomalous $Z$ couplings are related to the SMEFT coefficients by
\begin{equation}
\delta g_{ij}^{l_L} = 
\minus \frac{1}{2} v^2 \big(C_{Hl,ij}^{(1)} + C_{Hl,ij}^{(3)}\big)\,, \qquad
\delta g_{ij}^{l_R} = \minus \frac{1}{2} v^2 C_{He,ij} \, .
\end{equation}
This competes with the total width in the SM, $\Gamma_{Z,\textsc{SM}} = 2.4955\,\textrm{GeV}$~\cite{ParticleDataGroup:2022pth}. 

The best constraints on LFV $Z$ decays are set by the ATLAS and CMS collaborations~\cite{ATLAS:2021bdj, ATLAS:2022uhq, CMS:2025wqy}.
These searches are statistically limited, and we obtain projections for the HL-LHC limit by scaling the expected results from Refs.~\cite{ATLAS:2021bdj, ATLAS:2022uhq, CMS:2025wqy} to $3\,\textrm{ab}^{-1}$, assuming that the searches remain statistics-limited. These constraints are unlikely to improve at future muon colliders, as the final states are relatively distinctive at hadron machines and the production cross section for on-shell $Z$ bosons is not substantially larger with high-energy muon collisions. FCC-ee is expected to improve further from the HL-LHC limit due to its superior momentum resolution and $Z$ mass resolution, by one to three orders of magnitude depending on potential improvement on the  $\mu\rightarrow e$ mistagging rate~\cite{Dam:2018rfz} (c.f.~\cite{Altmannshofer:2023tsa}). 

%%%%%%%%%%%%%%%%%%%%%%%%%%%%%%%%%%%%%%%%%%%%%%%%%%%%%%%%%%%%%%%%%%%%%%%%%%
%%%%%%%%%%%%%%%%%%%%%%%%%%%%%%%%%%%%%%%%%%%%%%%%%%%%%%%%%%%%%%%%%%%%%%%%%%
\section{Lepton-Flavor-Violating Higgs Decays}
\label{sec:lfv_higgs}

In this section, we consider flavor-violating decays of the Higgs boson, both at the LHC and future muon colliders. 
Such decays have previously been studied in other colliders \cite{Diaz-Cruz:1999sns, Harnik:2012pb, Qin:2017aju, Hou:2020tgl}.

In the context of flavor-violating Higgs decays, it is useful to temporarily step back from the effective field theory approach used in the rest of this paper, and work instead with a general (i.e., not necessarily diagonal) coupling matrix for the Higgs and the charged leptons,
\begin{equation}
\label{eq:higgs_yukawa_matrix}
\mathcal{L} \supset - Y^e_{ij}\, h \bar{e}_{L,i} e_{R,j} + \hc 
\end{equation}
In the SMEFT, these Yukawas couplings are of the form
\begin{equation}
\label{eq:higgs_yukawa_smeft}
Y^e_{ij} = \frac{m_i}{v} \delta_{ij} - \frac{v^2}{\sqrt{2} \Lambda^2} C_{eH,ij} \, .
\end{equation}
The parameterization in Eq.~\eqref{eq:higgs_yukawa_matrix} is more general for the purposes of understanding Higgs decays because it can also be applied to models with additional Higgs doublets with a more complex flavor structure. In this case, the new mass scales can be at the electroweak scale, so the EFT expansion is not a good description, even if the strongest constraints come from precision measurements of the $125\,\textrm{GeV}$ resonance.

The width of the decays $h \to l_i l_j$ is given by\footnote{In this manuscript, we always mean the combination of the modes $h \to l_i^+ l_j^-$ and $h \to l_i^- l_j^+$.} 
\begin{equation}
\Gamma(h \to l_i l_j) = \frac{m_h}{8\pi} \Big( |Y^e_{ij}|^2 + |Y^e_{ji}|^2 \Big) \, .
\label{eq:higgs_lfv_rate}
\end{equation}
This can be compared directly to the branching ratios measured by the experiments, using the SM value of the width $\Gamma_{h,\textsc{SM}} = 4.07\,\textrm{MeV}$~\cite{LHCHiggsCrossSectionWorkingGroup:2013rie}. 

The ATLAS and CMS collaborations have searched for lepton flavor violating decays of the Higgs $h \to e\mu$, $e\tau$ and $\mu\tau$~\cite{ATLAS:2019old, CMS:2021rsq, CMS:2023pte, ATLAS:2023mvd}, summarized in Table~\ref{tab:precision_bounds}.
The strongest of these constraints is on the $e\mu$ final state, whose branching ratio is constrained to be $< 4.4 \times 10^{-5}$~\cite{CMS:2023pte}. All of these searches are statistically limited, and the High-Luminosity phase of the LHC will improve each of these constraints by nearly an order of magnitude.\footnote{
The most recent search for $h \to \tau\mu$ and $h \to \tau e$ in ATLAS actually has a modest upward fluctuation, preferring $\textrm{BR}(h \to \tau e) \simeq \textrm{BR}(h \to \tau \mu) \simeq 0.1\%$~\cite{ATLAS:2023mvd}. These fluctuations are only at the $\sim 2\sigma$ level, but will be decisively tested with the next run of LHC data.} The projected constraints in Table~\ref{tab:precision_bounds} are obtained by simply rescaling the expected limits from Refs.~\cite{CMS:2023pte, ATLAS:2023mvd}, assuming the systematic uncertainties remain subdominant, improving by the same factor as the statistical component.

The prospects for LFV Higgs decays have also been studied in the context of $e^+e^-$ Higgs factories in Ref.~\cite{Qin:2017aju}. They were found to be similar to the HL-LHC projections mentioned above, with $\mathcal{O}(1)$ improvements on the $\mu\tau$ and $e\tau$ final states.

%%%%%%%%%%%%%%%%%%%%%%%%%%%%%%%%%%%%%%%%%%%%%%%%%%%%%%%%%%%%%%%%%%%%%%%%%%
\subsection{LFV Higgs Decays at a High-Energy Muon Collider}
\label{subsec:lfv_higgs_muc}

We now wish to understand how well these flavor-violating decays can be measured at a high-energy muon collider.
Muon collisions at TeV energies yield large rates for Higgs production via vector boson fusion in an environment that is relatively clean from hadronic backgrounds. This makes muon colliders powerful machines for Higgs precision studies. For instance, Refs.~\cite{Forslund:2022xjq, Forslund:2023reu} studied the attainable precision on all the SM Higgs couplings using a \texttt{Delphes}~\cite{deFavereau:2013fsa} fast simulation of a hypothetical muon collider detector. These projections were validated with a more detailed detector simulation, including beam-induced background, in Ref.~\cite{Andreetto:2024rra}.

In this section, we will extend these studies to the beyond the Standard Model decays $h \to l_i l_j$. As in later sections, we will focus on the capabilities of a $\sqrt{s} = 10\,\textrm{TeV}$ collider. 
We generate samples of Higgs boson production via vector-boson fusion using \texttt{MadGraph5}~\cite{Alwall:2014hca}, and decay the Higgs to the different lepton-flavor-violating final states using \texttt{MadSpin}~\cite{Artoisenet:2012st}. The lepton-flavor-violating interactions were implemented by including the dimension-6 operators $\mathcal{O}_{eH}$ using the \textsc{SMEFTsim} model file~\cite{Brivio:2017btx,Brivio:2020onw}, with the most general flavor scheme. For the $e\tau$ and $\mu\tau$ final states, we focus on the hadronic decays of the $\tau$ only, which are generated by interfacing to \texttt{Pythia8}~\cite{Bierlich:2022pfr}.

To obtain a realistic estimate of the attainable sensitivity on the LFV Higgs branching ratios, we perform a parameterized detector simulation using the muon collider detector card shipped with \texttt{Delphes}~\cite{deFavereau:2013fsa}.
This is the same card that was used in the studies in Refs.~\cite{Forslund:2022xjq, Forslund:2023reu}, and more detailed information on its performance can be gleaned from the analyses therein. 
Jets were clustered using \texttt{FastJet}~\cite{Cacciari:2011ma} with the $k_t$ algorithm~\cite{Catani:1993hr, Ellis:1993tq} with $R = 0.5$.
Consistent with Refs.~\cite{Forslund:2022xjq, Forslund:2023reu}, a flat tau-tagging efficiency of $80\%$ was assumed for jets originating from hadronic $\tau$ decays with a reconstructed transverse momentum $p_{T,j} > 10\,\textrm{GeV}$ within $|\eta_j| < 2.5$. These parameters are consistent with a $2\%$ mis-tag rate for QCD jets and a $0.1\%$ mis-tag rate for electron energy depositions, both of which are small enough that the reducible backgrounds in our study are insignificant compared to the irreducible ones discussed below, the only exception being the $\mu^- V \to \mu^+ \tau^- \tau^-$ process discussed in Section~\ref{subsec:muVmutautau}.

\medskip

\begin{figure}
\centering
\includegraphics[width=5cm]{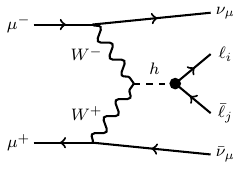}
\quad
\includegraphics[width=5cm]{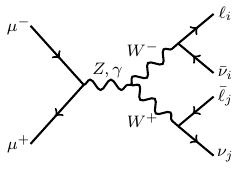}
\quad 
\includegraphics[width=4cm]{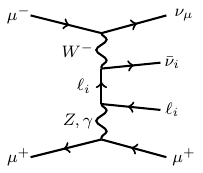}
\caption{Feynman diagrams for the VBF production of a Higgs, decaying to lepton-flavor-violating final states (\textbf{left}) and for SM backgrounds that produce different flavor final state leptons (\textbf{middle} and \textbf{right}).}
\label{fig:ww-bkg-diagrams}
\end{figure}

The irreducible backgrounds to the LFV Higgs decay processes are $l\nu l'\nu$ production through intermediate $W$ bosons and $\tau^+\tau^-$ pair production, with at least one of the $\tau$s decaying leptonically.
The former includes both processes which produce on-shell $W$ bosons that decay promptly, as well as non-resonant production of the final state leptons, e.g., through diagrams such as that shown in Fig.~\ref{fig:ww-bkg-diagrams}. 
Both $l\nu l'\nu$ and $\tau^+\tau^-$ backgrounds can be produced either through direct $\mu^+\mu^-$ annihilation or via vector-boson scattering processes, where the additional beam remnants are assumed to be undetected.
As these are the only direct ways of producing different flavor leptons in high-energy lepton collisions, these background processes will play a role in the other flavor-violating probes studied in Section~\ref{sec:collider} as well.
Additional backgrounds such as $t\bar{t}$ production, which are the dominant background at the LHC, are much easier to remove in the relatively clean hadronic environment of a muon collider, due to the additional objects in the event. 

For the annihilation and charged VBF production modes, the background processes can be simulated directly in \texttt{MadGraph}. 
For the neutral VBF backgrounds, there is a singularity associated with the muons scattering in the forward direction, so we instead employ the effective vector approximation (EVA)~\cite{Dawson:1984gx,Kane:1984bb} in \texttt{MadGraph}~\cite{Ruiz:2021tdt}. 
To validate this approach, we performed additional simulations including the forward muons explicitly: for $\tau^+\tau^-$ production, the forward singularity is integrable, and we verified explicitly that the bulk of the cross section arises from the forward limit, with only a tiny fraction of the events containing muons with appreciable transverse momentum and finite pseudorapidity, none of which passed the selection cuts described below.
For the $l\nu l'\nu$ background, avoiding the IR divergence in the forward direction requires explicit cuts on the transverse momentum and pseudorapidity of the muons. We verified that as these cuts were relaxed from $|\eta| < 4$ to $|\eta| < 8$ and $p_{T,\mu} > 20\,\textrm{GeV}$ to $p_{T,\mu} > 5\,\textrm{GeV}$, our estimate of the number of events passing all the selection criteria was unchanged and was always several orders of magnitude smaller than the background estimated with the EVA. Given these results, we restrict our attention to the EVA estimates of these backgrounds in what follows.

\begin{figure}[t]
\centering
\includegraphics[width=7.5cm]{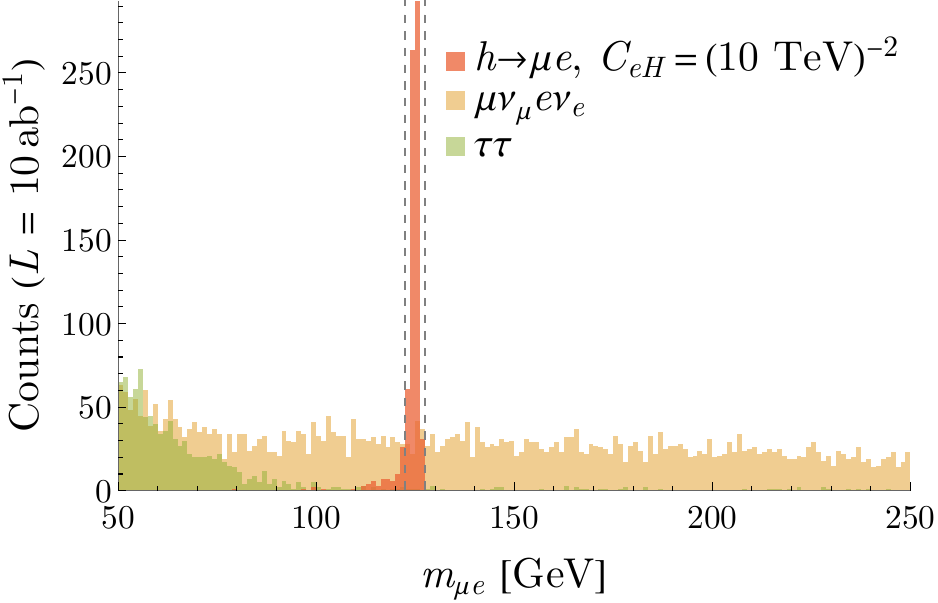}\quad
\includegraphics[width=7.5cm]{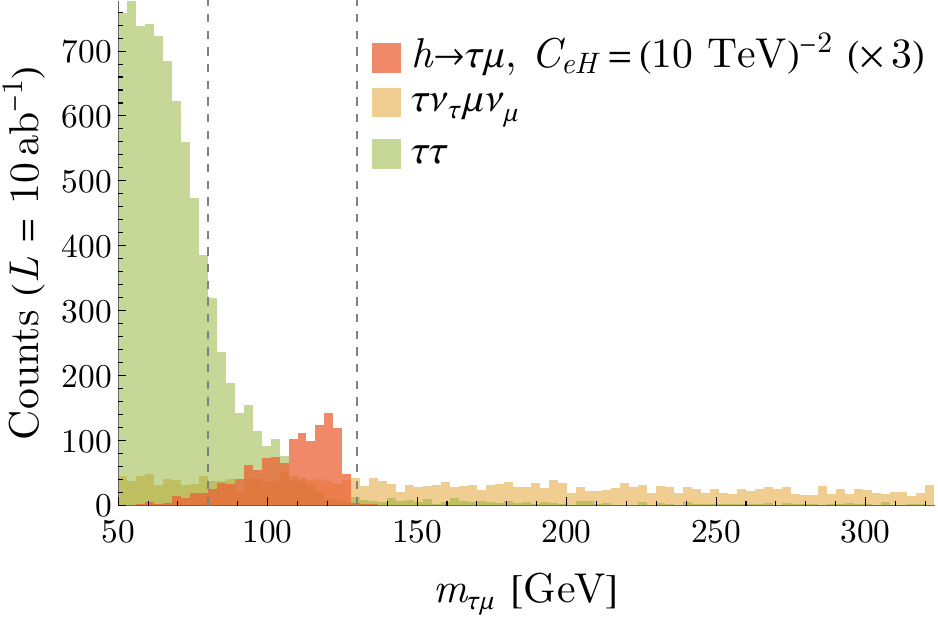}\\[0.5em]
\includegraphics[width=7.5cm]{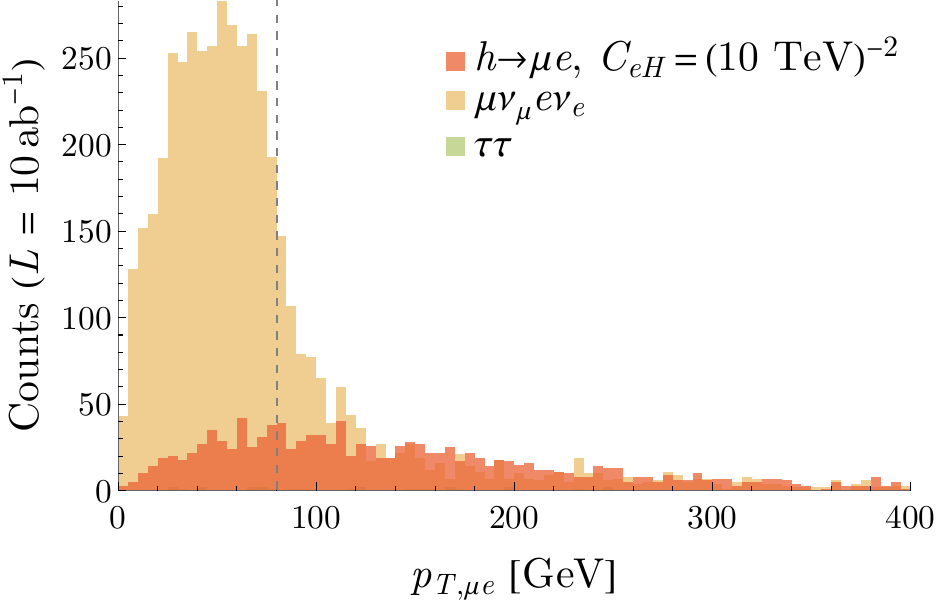}\quad
\includegraphics[width=7.5cm]{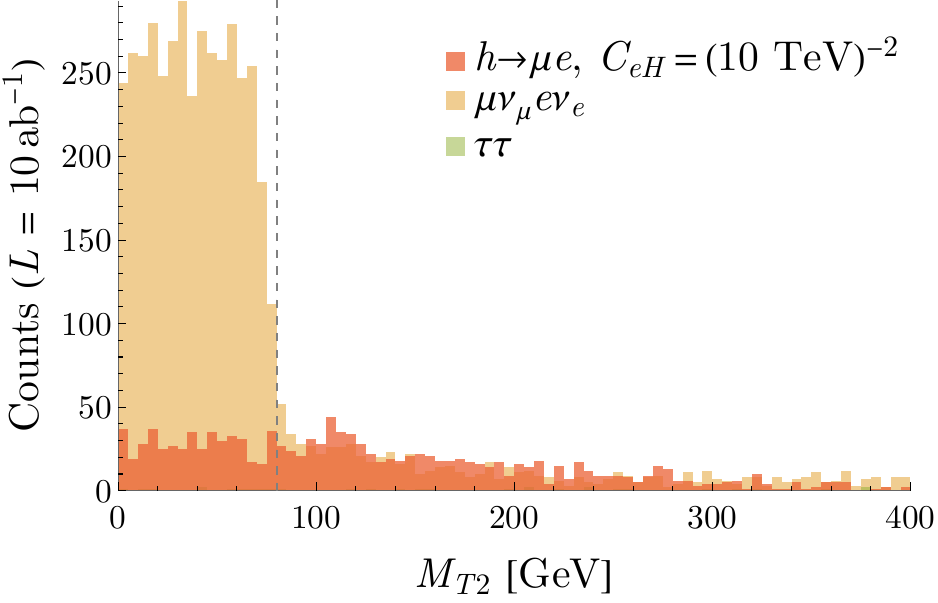}

\caption{ 
Histograms of kinematic observables from toy datasets of 
the $h \to l_i l_j$ signal and the different backgrounds, after applying basic pre-selection cuts demanding the correct number of reconstructed particle objects. The \textbf{top row} shows the invariant mass of the lepton pair for the $\mu e$ and $\tau\mu$ analyses (the plot for the $\tau e$ analysis is essentially identical to the latter). The \textbf{bottom row} shows the $p_T$ of the reconstructed Higgs and the $M_{T2}$ variable from the $\mu e$ analysis. The $\tau\tau$ background contribution is both small in magnitude and sufficiently diffuse to remain undetectable in the bottom-left and bottom-right histograms. The vertical dashed lines denote the cuts on the variable, see the text for details.}
\label{fig:lfv_higgs_histograms}
\end{figure}

To isolate the Higgs decay signal, we impose several selection cuts on the events.
These cuts are informed by various kinematic distributions, some of which are shown in Fig.~\ref{fig:lfv_higgs_histograms}.
For the $h \to e\mu$ signal we require exactly one reconstructed electron and muon in the central region with $|\eta|<2.5$, each with $p_T > 30\,\textrm{GeV}$. 
The primary discriminant is the invariant mass of the $e\mu$ pair, which we require to lie in the range $122.5 < m_{e\mu} < 127\,\textrm{GeV}$ (see Fig.~\ref{fig:lfv_higgs_histograms}, top left). 
We additionally require a modest boost of the reconstructed Higgs system, $p_{T,e\mu} > 80\,\textrm{GeV}$ (see Fig.~\ref{fig:lfv_higgs_histograms}, bottom left). 
To eliminate the significant component of the background arising from on-shell $W^+W^-$ or $\tau^+\tau^-$ pairs with decays to multiple neutrinos, we also compute the ``stransverse mass'' of the system, $M_{T2}$~\cite{Lester:1999tx, Barr:2003rg} using the library provided by Ref.~\cite{Lester:2014yga}. 
For the on-shell $W^+W^-$ and $\tau^+\tau^-$ components of the background with no additional missing transverse momentum component, $M_{T2}$ is bounded above by $M_W$ or $m_{\tau}$, so a cut $M_{T2} > 80\,\textrm{GeV}$ eliminates a significant component of the background events (see Fig.~\ref{fig:lfv_higgs_histograms}, bottom right). 
The remaining background events have a significant amount of missing transverse momentum that is not characteristic of the signal, so we impose a cut $E_{T,\textrm{miss}} < 350\,\textrm{GeV}$.

The $h \to \tau e$ and $h \to \tau\mu$ analyses are quite similar, except that we require one tau-tagged jet with $p_{T,j} > 30\,\textrm{GeV}$ in the final state along with the charged lepton. 
Due to the lower energy/momentum resolution for jets, along with the losses from the invisible part of the hadronic tau decays, keeping a large fraction of the signal requires a much larger, asymmetric window for the invariant mass of the reconstructed Higgs: we require $80 < m_{\tau e}, m_{\tau\mu} < 130\,\textrm{GeV}$ (see Fig.~\ref{fig:lfv_higgs_histograms}, top right). 
The cuts on $p_{T,h}$, $M_{T2}$ and $E_{T,\textrm{miss}}$ are the same as above.

%%%%%
\input{lfv-higgs-summary-table}
%%%%%

In Table~\ref{tab:lfv_higgs_summary} we summarize the number of events from the signal and the relevant backgrounds for each LFV Higgs decay channel. The signal events assume the relevant $C_{eH,ij} / \Lambda^2 = 1 / (10\,\textrm{TeV})^2$ for illustration, and we assume a total integrated luminosity of $10\,\textrm{ab}^{-1}$. 
The $\sigma \times \textrm{BR}$ column shows the total cross section for each signal or background process {\em after} requiring the correct number of final state particles are reconstructed (e.g., exactly one of each flavor lepton considered with sufficient transverse momentum and within the central region $|\eta|<2.5$). Note that this includes the efficiency for identifying the jets as hadronic $\tau$ decays. This allows for a more direct comparison between the different backgrounds where the cross section may be computed inclusively or exclusively but the forward particles are not observed. The efficiency column shows the fraction of events passing all other cuts (on the invariant mass, the Higgs $p_T$ and the $M_{T2}$ of the event), while the last just shows the expected number of events assuming $10\,\textrm{ab}^{-1}$ total integrated luminosity.

Given the expected number of background events, we derive an upper limit on the branching ratio of the Higgs to each final state at $95\%~\textrm{C.L.}$, using the Gaussian approximation for the likelihood and neglecting any systematic uncertainties. The projected branching ratio constraints are
\begin{equation}
\textrm{BR}(h \to \mu e) < 9.9 \times 10^{-6}\,, \qquad
\textrm{BR}(h \to \tau e) < 8.4 \times 10^{-5}\,, \qquad
\textrm{BR}(h \to \tau\mu) < 7.7 \times 10^{-5}\, .
\end{equation}
The projected constraint on $h \to \mu e$ is essentially the same as the one from HL-LHC---this is not unexpected, as the $e\mu$ final state does not suffer from significant hadronic backgrounds, so the large cross section for Higgs production in proton collisions can be leveraged for a strong constraint, and the advantages of muon collisions are less apparent. For the $\tau e$ and $\tau\mu$ final states on the other hand, the muon collider projections represent an order of magnitude improvement compared to HL-LHC. 

\begin{figure}[t!]
\centering
\includegraphics[width=8cm]{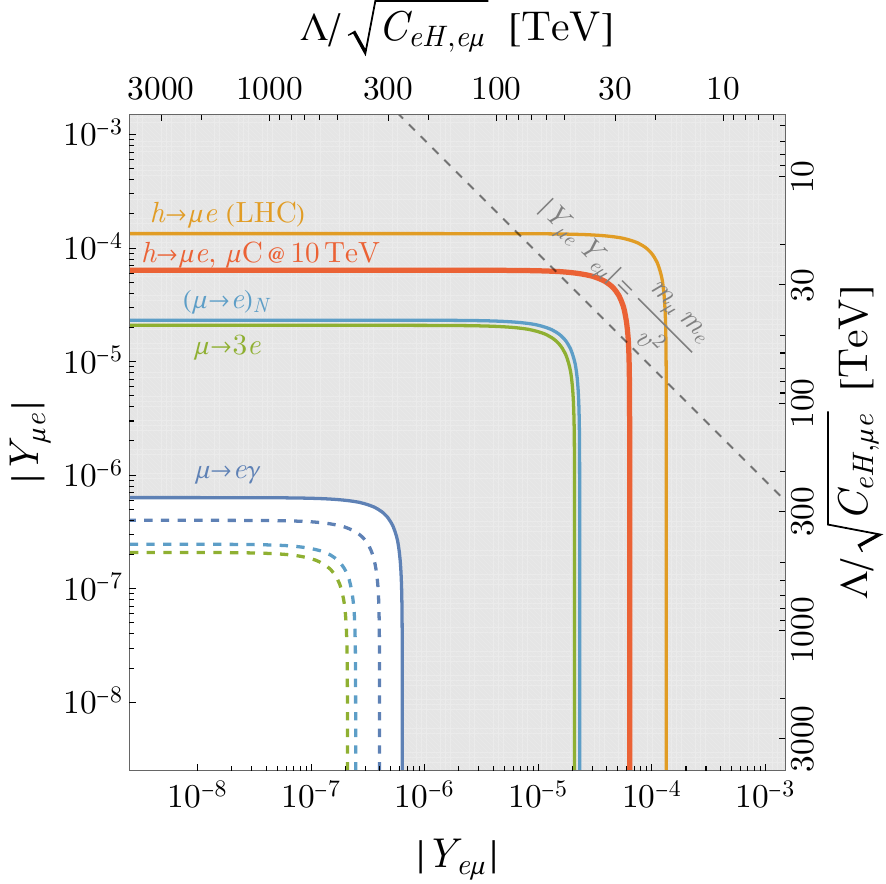} \quad
\includegraphics[width=8cm]{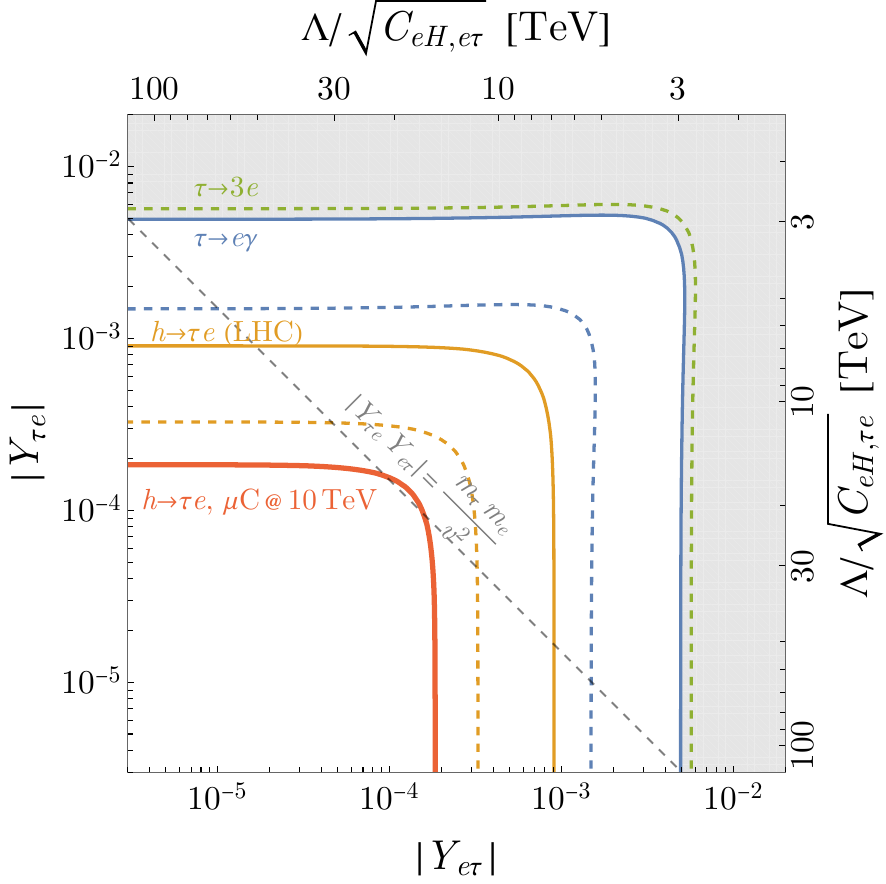} \\[0.5em]
\includegraphics[width=8cm]{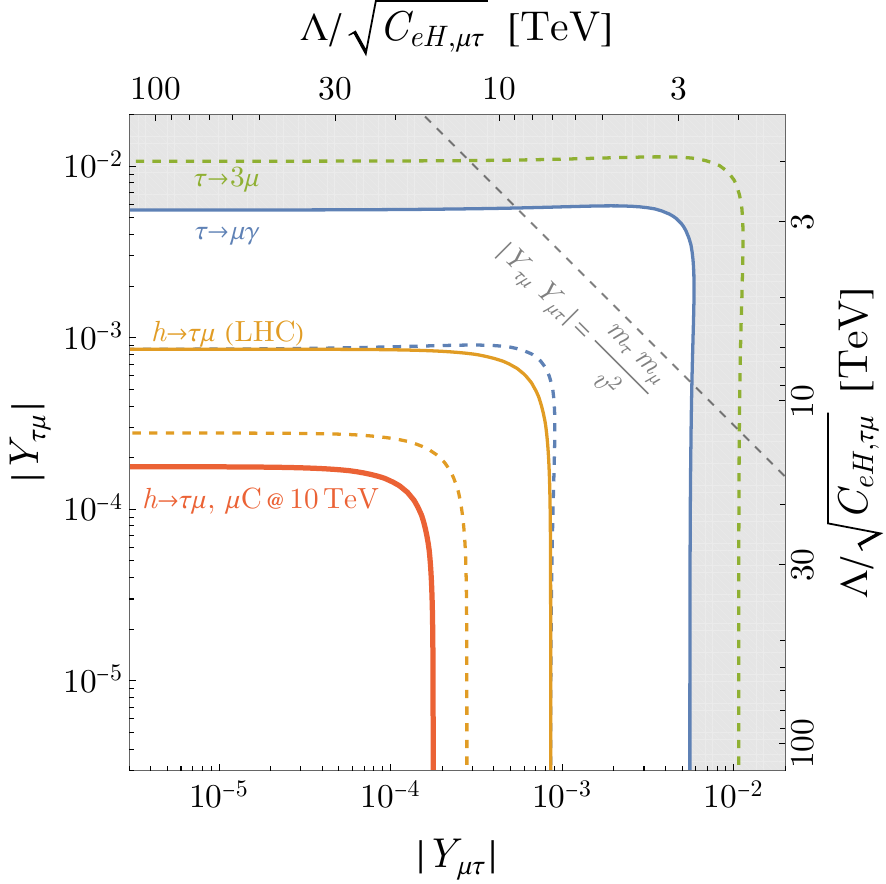}
\caption{Summary of the constraints on flavor-violating Higgs Yukawa couplings. We show the constraints for $\mu$\,--\,$e$, $\tau$\,--\,$e$, and $\tau$\,--\,$\mu$ flavor violation in the \textbf{top-left}, \textbf{top-right} and \textbf{bottom} panels, respectively. Each plot shows the constraints in the plane of the two off-diagonal couplings; the opposite axes show the constraints interpreted as constraints on the scale of the effective operator $\mathcal{O}_{eH}$. The shaded region in each panel indicates the region excluded by various low-energy constraints (with projected low-energy constraints indicated by the corresponding dashed lines), while the solid (dashed) orange line show the current LHC bounds (HL-LHC projections). The solid red curve shows the projected muon collider constraint. The dashed gray lines indicate a ``fine-tuning'' constraint on the size of the off-diagonal elements, as discussed in the text.} 
\label{fig:lfv_higgs_decay_summary}
\end{figure}

The constraints from LFV Higgs decays on the off-diagonal entries of the Yukawa matrices as in Eq.~\eqref{eq:higgs_yukawa_matrix} are summarized in Fig.~\ref{fig:lfv_higgs_decay_summary}. 
Here, the constraints from low-energy processes such as $l_i \to l_j\gamma$ are included by translating the bounds derived on $C_{eH,ij}$ in the last section into constraints on $Y^e_{ij}$ using Eq.~\eqref{eq:higgs_yukawa_smeft}. 
In each case, we show the two relevant off-diagonal couplings for $\mu$\,--\,$e$, $\tau$\,--\,$e$ and $\tau$\,--\,$\mu$ transitions, respectively. The dashed gray line indicates the theoretical bound $|Y_{ij} Y_{ji}| = m_i m_j / v^2$, above which fine-tuning is required to obtain the observed, hierarchical masses for the SM fermions from electroweak symmetry breaking. The opposite axes show the SMEFT operator interpretation of the constraints as bounds on $\Lambda / \sqrt{C_{eH,ij}}$.
Note that the plots are symmetric around the main diagonal as $C_{eH,ij}$ and $C_{eH,ji}$ have identical contributions to LFV observable in the figure, see Eq.~\eqref{eq:higgs_yukawa_smeft}--\eqref{eq:higgs_lfv_rate}.

The shaded regions in each plot indicate the present constraints from low-energy precision experiments, predominantly $\mu \to e\gamma$, $\tau \to e\gamma$ and $\tau \to \mu\gamma$, in each situation. For $\mu$\,--\,$e$ transitions, it is clear that the parameter space testable in direct Higgs decays at current or future colliders is already ruled out even by present precision constraints. For $\tau$\,--\,$e$ and $\tau$\,--\,$\mu$ transitions, on the other hand, existing LHC constraints set the most stringent limits, competitive even with projected future constraints from Belle~II. A high-energy muon collider represents the best possibility for testing this important aspect of the Higgs sector.

Future improvements in sensitivity to LFV Higgs decays may arise from sophisticated kinematic variables like the collinear mass, which is particularly effective for $\tau$-involving channels (see Ref.~\cite{CMS:2021rsq}). 
Moreover, deploying modern tools such as machine learning techniques can yield major gains in signal discrimination. 
Combined with the simple strategy outlined here, these approaches highlight the extraordinary promise of a future muon collider for exploring BSM physics via LFV Higgs decays.

%%%%%%%%%%%%%%%%%%%%%%%%%%%%%%%%%%%%%%%%%%%%%%%%%%%%%%%%%%%%%%%%%%%%%%%%%%
%%%%%%%%%%%%%%%%%%%%%%%%%%%%%%%%%%%%%%%%%%%%%%%%%%%%%%%%%%%%%%%%%%%%%%%%%%
\section{Flavor-Violating Scattering Processes at a Muon Collider}
\label{sec:collider}

We now turn to additional probes of lepton-flavor-violating new physics which are accessible only at a high-energy muon collider.
In contrast to the precision Higgs decay measurements discussed in the previous section, here we focus on high-energy {\em scattering} processes that involve the lepton-flavor-violating interactions between the muon, tau and electroweak gauge/Higgs bosons in the Standard Model. We focus on $\mu$-$\tau$ transitions in particular because muon colliders offer unique opportunities to explore these processes in regimes that are inaccessible to current precision experiments.
As we will see, the capability to scatter muons at high energy in a relatively clean environment leads to distinctive signatures that can test the physics of lepton flavor in new ways. These processes will be particularly important for characterizing new physics if a signal is detected at next-generation precision machines, such as Belle~II or Mu2e. 

In the rest of this section, we focus on five different scattering processes which violate lepton flavor:
$\mu V \to \tau h$, $\mu W \to \tau W$, $\mu \mu \to \mu \tau$, $\mu V \to \tau\tau\tau$, and $\mu^\mp V \rightarrow \mu^\pm \tau^\mp \tau^\mp$ with same sign $\tau$ leptons, where $V = \gamma, Z$.
Each of these processes is sensitive to several combinations of the effective operators in Eq.~\eqref{eq:smeft_operators}.
Unlike the on-shell Higgs and $Z$ decays discussed above, they receive contributions from four-point vertices---either flavor-changing operators with electroweak gauge/Higgs bosons or flavor-changing four-fermion operators. 
These lead to kinematically distinct, high-momentum processes that take full advantage of the high energies available at a 10~TeV machine. 
A subset of these interactions was studied in Refs.~\cite{AlAli:2021let, Glioti:2025zpn}. 
Here, we extend these works with more detailed investigations covering the suite of LFV interactions from dimension-6 operators involving leptons or electroweak gauge/Higgs bosons.

In each case, we simulate the signal process, along with the primary, irreducible backgrounds, in \texttt{MadGraph} with the same methodology used in Section~\ref{subsec:lfv_higgs_muc}. The events are again passed through \texttt{Pythia} for decays, parton showering, and hadronization, and then passed through \texttt{Delphes} with the same muon collider Detector card discussed in Section~\ref{subsec:lfv_higgs_muc}.

In all cases, we will focus on the hadronic decays of the final state $\tau$. This choice takes full advantage of the relatively clean hadronic environment available at a high-energy lepton collider and keeps the largest fraction of the events without suffering additional flavor-conserving backgrounds.
However, properly identifying these processes as violating lepton-flavor requires that the resulting jets can be identified as arising from a $\tau$.
This ``$\tau$-tagging'' has been well-studied in the context of the LHC, where $\tau$ jets are distinguishable from light quark or gluon jets due to the small number of charged tracks associated with them, the distinctive jet mass, and their collimated geometry~\cite{Heldmann:2005zda, Bagliesi:2007qx}.
This is an active area of research in the LHC experiments, where recent advances in machine learning promise improvements in the efficiency of tagging tau jets and the rejection of QCD-induced backgrounds~\cite{Lange:2023gbe, ATLAS:2024njy}. While the principles are the same at high-energy lepton colliders, detailed studies are needed that properly examine the expected capabilities at future muon collider detectors. In particular, while the relatively clean hadronic environment reduces contamination, there are challenges associated with beam-induced backgrounds, and the typical boost for the jets will be much larger, which makes distinguishing between the low-multiplicity $\tau$ decays and QCD radiation inherently more challenging. In the absence of a detailed study, we will simply assume an $80\%$ efficiency for tagging $\tau$ jets, with a rejection rate for QCD jets of 2\% and electrons of 0.1\%. For most of the processes we consider, these mistagging rates are small enough to ignore any resulting reducible backgrounds, with the exception of $\mu^\mp V \rightarrow \mu^\pm \tau^\mp \tau^\mp$ in Section~\ref{subsec:muVmutautau}.

With all these choices, a relatively complete study of these processes can be performed.
As we will see, in each case a simple set of cuts is sufficient to reduce the backgrounds to the $\mathcal{O}(1-10)$~events level. 
More sophisticated analyses involving combinations of observables could easily reduce these backgrounds even further, but we will leave these studies to future work when the design and expected performance of future detectors are better understood. 
We also defer a comprehensive treatment of systematic uncertainties to future work, once the detector design has matured.

In Table~\ref{tab:lfv_muc_summary} we summarize the production cross section and the event count for all the signals and the most relevant backgrounds for each process. 
As before, we assume  $\sqrt{s}=10$~TeV and 10~ab$^{-1}$ data. 
For each process, we report the signal event count with only one SMEFT operator included. 
The quoted cross section and total event yield include contributions from both the process and its conjugate.
The following subsections detail our analyses for each of these processes and the kinematic cuts. 
A summary of the projected constraints and a comparison to low-energy signatures is left to Section~\ref{sec:summary}.

%%%%%
\input{high-energy-summary-table}
%%%%%

%%%%%%%%%%%%%%%%%%%%%%%%%%%%%%%%%%%%%%%%%%%%%%%%%%%%%%%%%%%%%%%%%%%%%%%%%%
\subsection{\texorpdfstring{$\mu V \to \tau h$}{mu V to tau h}}

We first consider the $\mu V \to \tau h$ process. 
This process simultaneously tests two distinct sources of LFV: 
a $\mu \tau \gamma h$ four-point interaction that arises from the dipole operators $\mathcal{O}_{eB}$ and $\mathcal{O}_{eW}$, and a $\mu \tau Z h$ interaction that arises from $\mathcal{O}_{He}$, $\mathcal{O}_{Hl}^{(1)}$ and $\mathcal{O}_{Hl}^{(3)}$.  

\begin{figure}
\centering
\includegraphics[width=5cm]{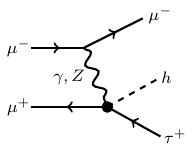}
~
\includegraphics[width=4.5cm]{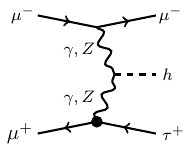}
~
\includegraphics[width=5cm]{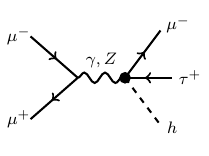}
\\
\includegraphics[width=4.5cm]{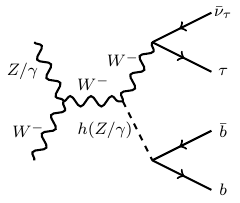}
~
\includegraphics[width=4cm]{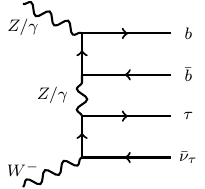}
\caption{
\textbf{Top row}: Example diagrams contributing to the $\mu^+ \mu^- \to \mu^\pm \tau^\mp h$ scattering process at a high-energy muon collider, arising from insertions of dimension-6 lepton-flavor-violating operators. The first diagram includes the $\mu V \to \tau h$ sub-process of interest, with the additional muon traveling in the forward direction. The other diagrams contribute at the same order, and are included in our simulations for consistency.
\textbf{Bottom row}: Example diagrams for processes that serve as irreducible backgrounds to the $\mu \mu \to \mu \tau h$ signal.
}
\label{fig:diagrams_muv_tah}
\end{figure}

In both cases, our primary interest is in the $\mu V \to \tau h$ sub-process contained in Fig.~\ref{fig:diagrams_muv_tah} (top left), in which the Higgs and $\tau$ are scattered with high momentum. 
The additional muon in the final state, which radiates the $\gamma$ or $Z$ is deflected at a small angle, and is thus generally not visible in the detector. As a result, this process can in principle be treated with the effective vector approximation. 
In practice, however, we simulate the full $2 \to 3$ process $\mu^+ \mu^- \to \mu^\pm \tau^\mp h$, keeping the initial- and final-state muon legs explicit to capture contributions from diagrams such as the top right-hand panel of Fig.~\ref{fig:diagrams_muv_tah}. 
These are subdominant in the region we are most interested in, but are included in our simulation for consistency. 

%%%%%
%%%%%

For the final state Higgs, we focus on the dominant branching fraction $h \to b\bar{b}$, again taking advantage of the relatively clean hadronic environment. 
As we will be primarily interested in the boosted regime to separate the signal from irreducible backgrounds, the two $b$-jets are nearly collinear. We thus consider jets clustered with a larger radius parameter $R = 1.0$, still with the $k_T$ algorithm, so that all the (hadronized) Higgs decay products are consistently clustered in a single jet object. 
Such jet objects containing $B$ hadrons are distinct from ordinary QCD jets due to the displaced secondary vertices and semi-leptonic decay products, and sophisticated algorithms have been developed by the ATLAS and CMS experiments to tag these ``$b$-jets''~\cite{CMS:2017wtu, ATLAS:2019bwq, CMS-DP-2022-041, ATLAS:2023lwk, ATLAS:2025dkv}. 
We assume that similar performance is attainable at a future high-energy muon collider, and adopt a flat, $80\%$ efficiency for tagging these ``double-$b$ jets'', with mis-tagging rates small enough that any reducible backgrounds can be neglected. 

Our signal is thus a single hadronic $\tau$, and a $h \to b\bar{b}$ resonance. The additional muon in the signal process is generally in the forward direction, so rather than selecting for it, we veto events with a central ($p_T > 30\,\textrm{GeV}$ and $|\eta| < 2.5$) muon. This helps mitigate other potential irreducible backgrounds with leptonically decaying $W$ bosons or $\tau$ leptons. 

%%%%%
\begin{figure}[t!]
\centering
\includegraphics[width=7.5cm]{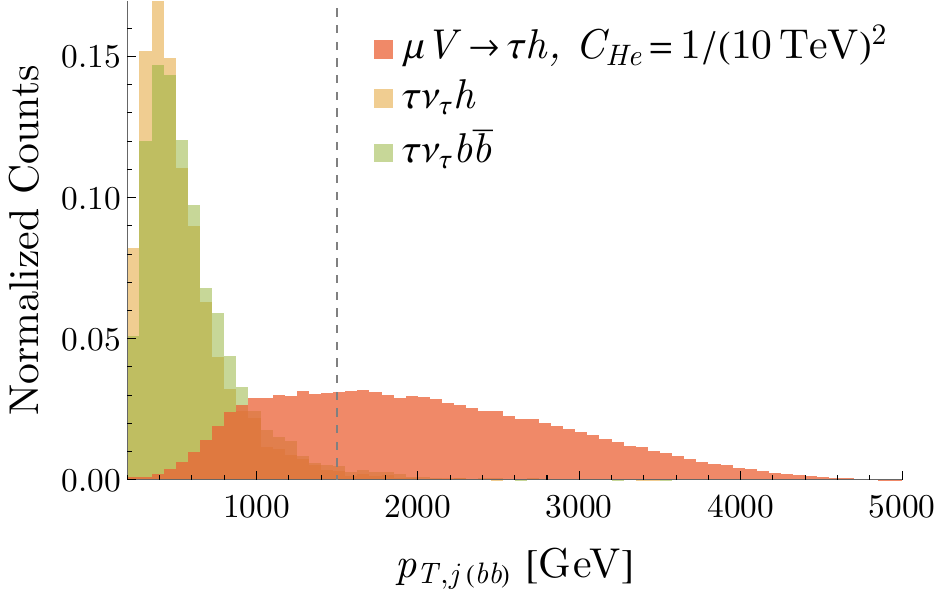}
~
\includegraphics[width=7.5cm]{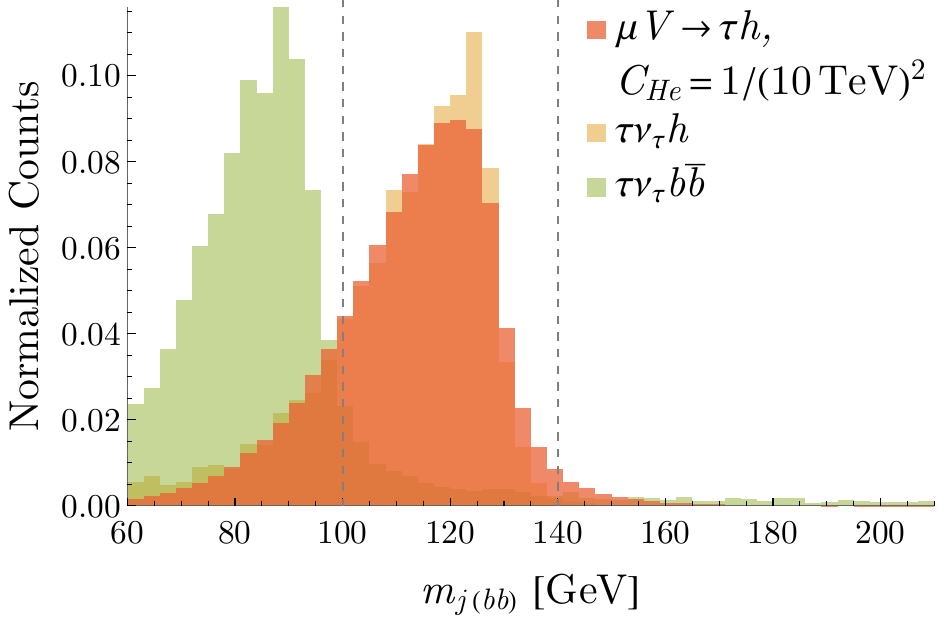}
\caption{Histograms of the $p_T$ and mass of reconstructed Higgs from a fat jet system with two $b$-tags for both the signal (red) and dominant $\tau\nu_{\tau}h$ (orange) and $\tau\nu_{\tau}b\bar{b}$ (green) backgrounds with each process independently normalized. The vertical dashed lines denote the cuts on the variable, see the text for details.}
\label{fig:histograms_muv_tah}
\end{figure}
%%%%%

The remaining, significant, irreducible backgrounds come from the production of $\tau \nu b\bar{b}$ final states, where the $\tau$ and $\nu$ arise from an intermediate $W$ leg, and the $b\bar{b}$ can come from either a resonant Higgs (e.g., in associated $Wh$ production) or from continuum QED processes.\footnote{We use the word ``continuum'', despite the fact that many of the events arise from resonant $Z \to b\bar{b}$ production.} 
Both of these backgrounds are produced via vector-boson fusion, as illustrated in the example diagrams in the bottom row of Fig.~\ref{fig:diagrams_muv_tah}. 

As with the $l\nu l'\nu$ backgrounds for the Higgs decays studied in Section~\ref{subsec:lfv_higgs_muc}, there are cases where the full background process including the final state muons/neutrinos suffers from a collinear divergence at fixed order. To simulate them, we thus again employ the effective vector approximation.
To ensure that we are not artificially excluding events where a small amount of transverse momentum for the radiated vector bosons would change the kinematics enough to avoid our selection cuts, we also simulate the full process with a small, explicit cut on the transverse momentum and pseudorapidity of the final state muon to avoid the collinear singularity. We find consistent results as these cuts are varied, and the selection criteria discussed below suppress this contribution to $< 0.1$~events, so we will focus only on the EVA-induced component in the remainder.

In contrast to the signal, which involves a high-energy scattering process from one of the colliding muons, the backgrounds are produced entirely from the scattering of radiated vector bosons, and the resulting final states are produced at correspondingly lower energy. This is illustrated in Fig.~\ref{fig:histograms_muv_tah} (left), where we show a histogram of the transverse momentum of the reconstructed $h\to b\bar{b}$ jet, with each process independently normalized. Requiring $p_{T,j(bb)} > 1.5\,\textrm{TeV}$ eliminates a substantial fraction of the VBF-induced backgrounds.
Aside from this, we require that the jet mass of the double-$b$-tagged fat jet be near the Higgs mass, $100\,\textrm{GeV} < m_{j(bb)} < 140\,\textrm{GeV}$, which, as illustrated in Fig.~\ref{fig:histograms_muv_tah} (right), significantly reduces the continuum $\tau\nu b\bar{b}$ background. Finally, we require $\slashed{E}_{T} > 60\,\textrm{GeV}$ on the missing transverse energy (MET), which helps eliminate potential collinear backgrounds, while the signal process generally has appreciable missing energy arising from the neutrino in the hadronic $\tau$ decay.

After these selection cuts, the mean number of total expected background events based on our simulations is $3.4$ events, 
coming in roughly equal parts from the $\tau\nu h$ and continuum $\tau\nu b\bar{b}$ components. See Table~\ref{tab:lfv_muc_summary} for a breakdown of signal and background counts. 
The signal processes, on the other hand, pass the $p_{T,j(bb)}$, $m_{j(bb)}$ and $\slashed{E}_{T}$ cuts with a $\sim 53\%$ efficiency. This efficiency is essentially unchanged irrespective of whether the signal arises from a dipole interaction or from a $\tau\mu Z h$ vertex.
A detailed breakdown of the efficiencies for the signal process and each background are given in the first section of Table~\ref{tab:lfv_muc_summary}. 

Assuming only one operator contributes at a time, we can compute the yield for the signal process, taking $C_i / \Lambda^2 = 1 / (10\,\textrm{TeV})^2$ as a benchmark. Including both a process and its conjugate, we find the resulting event yields $S_\mathcal{O}$ for operator $\mathcal{O}$ and for $10\,\textrm{ab}^{-1}$ data is given by 
\begin{equation}
\begin{gathered}
S_{\mathcal{O}_{eB}} \simeq 4500 \,\,\textrm{events}\, \times C_{eB}^2 \Big(\frac{10\,\textrm{TeV}}{\Lambda}\Big)^4 \ ,
\qquad 
S_{\mathcal{O}_{eW}} \simeq 1130\,\,\textrm{events}\, \times C_{eW}^2 \Big(\frac{10\,\textrm{TeV}}{\Lambda}\Big)^4 \ , \\[0.75em]
S_{\mathcal{O}_X} \simeq 75\,\,\textrm{events}\, \times C_X^2 \Big(\frac{10\,\textrm{TeV}}{\Lambda}\Big)^4 
\quad \text{for} \quad \mathcal{O}_X = \mathcal{O}_{Hl}^{(1)},~ \mathcal{O}_{Hl}^{(3)},~ \mathcal{O}_{He}\, , 
\end{gathered}
\label{eq:event_yield_muv_tah}
\end{equation}
where we have suppressed the flavor indices $ij = \mu\tau$ on the Wilson coefficients above for clarity. 
For more general scenarios where multiple Wilson coefficients are turned on, one must take into account the interference between the two dipole operators, $\mathcal{O}_{eB}$ and $\mathcal{O}_{eW}$, and between the two left-handed current operators, $\mathcal{O}_{Hl}^{(1)}$ and $\mathcal{O}_{Hl}^{(3)}$. The rest of the operators do not mutually interfere.

It is clear from Eq.~\eqref{eq:event_yield_muv_tah} that the dipole operators $\mathcal{O}_{eB}$, $\mathcal{O}_{eW}$ yield larger signal rates at a given scale; however, these compete with the significantly stronger precision constraints, as we'll discuss further in Section~\ref{sec:summary}. 

%%%%%%%%%%%%%%%%%%%%%%%%%%%%%%%%%%%%%%%%%%%%%%%%%%%%%%%%%%%%%%%%%%%%%%%%%%
\subsection{\texorpdfstring{$\mu W \to \tau W$}{mu W to tau W}}

We next consider the $\mu W \to \tau W$ process. It is sensitive to the same set of operators as $\mu V \to \tau h$, but tests different linear combinations of them, and---as we will see---provides stronger constraints on $\mathcal{O}_{He}$ and $\mathcal{O}_{eW}$. 
As before, we do not use the effective vector boson approximation for simulating the signal, and focus on diagrams like those in Fig.~\ref{fig:diagrams_muv_tah}, but with the Higgs or neutral vector boson replaced with a $W$ boson, together with the necessary $\mathrm{SU}(2)_L$ rotations on the final state leptons for the SM vertices. 

We again focus on fully hadronic final states. For the signal process, the hadronic decay products of the $W$ boson are highly collimated and are therefore clustered into a single jet. The event selection is thus similar to the $h\tau$ final state discussed above, consisting of a $\tau$ jet and a second fat jet. In this analysis, we do need to tag the $W$ jet, since there are no reducible backgrounds with a single fat jet that does not originate from a $W$.

There are two sources of significant backgrounds. One is from $\tau\nu_{\tau} jj$ with the two jets bundled into a single fat jet; this background dominantly arises from two intermediate $W$ bosons, with one decaying to $\tau \nu$ and one decaying hadronically. Another background is $\tau\tau$ production, with both $\tau$ leptons decaying hadronically but with only one correctly tagged as $\tau$ jet. 
Note that in both the signal and the $\tau\nu_{\tau} jj$ background processes we expect the fat jet to carry all of the intermediate $W$ energy, while the $\tau$-tagged jet only carries around half of the $\tau$ energy (with the other half carried by the undetected neutrino). Thus, for these processes we typically expect the $\tau$-tagged jet to be the softer of the two final state jets.

\begin{figure}
\centering
\includegraphics[width=5cm]{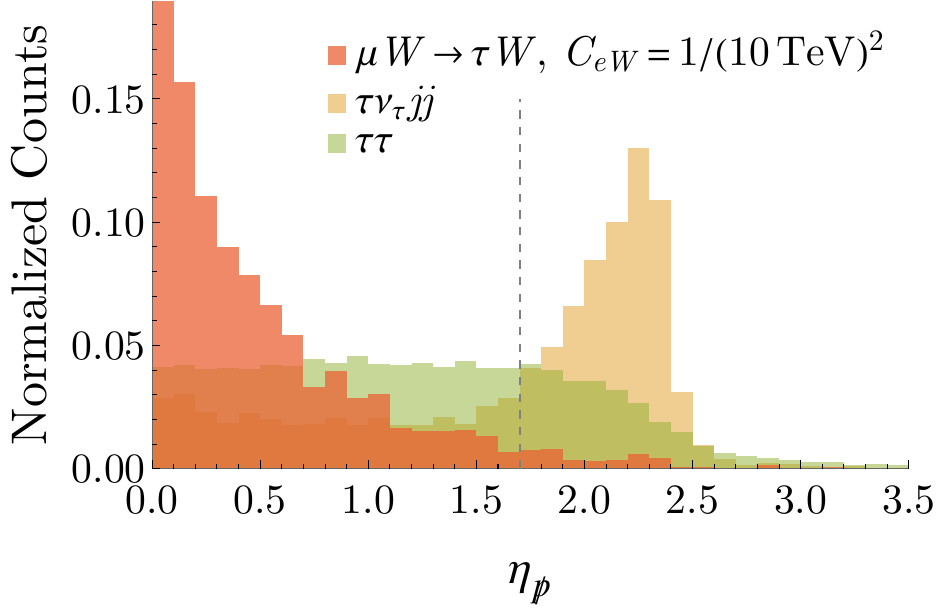}
~
\includegraphics[width=5cm]{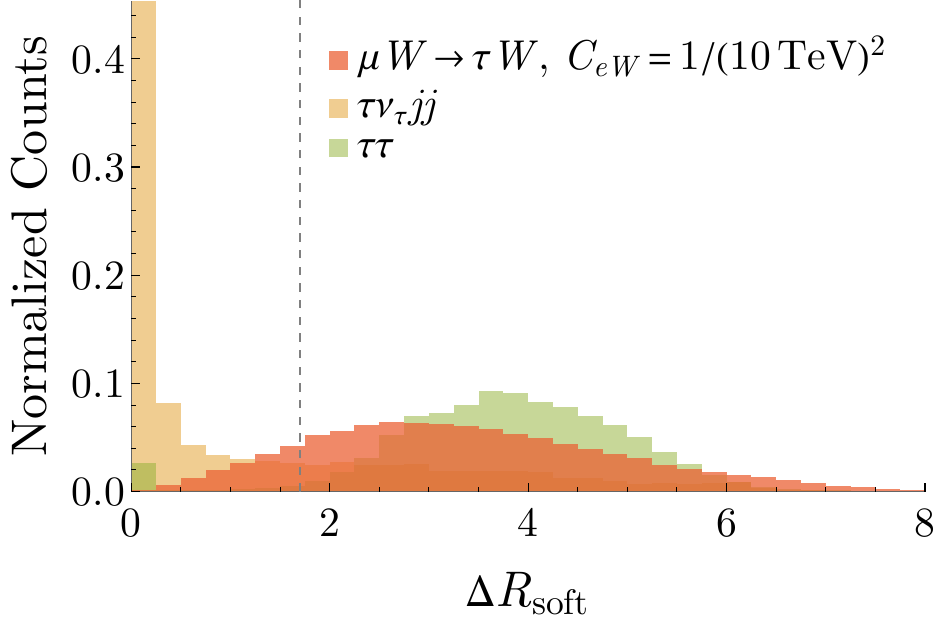}
~
\includegraphics[width=5cm]{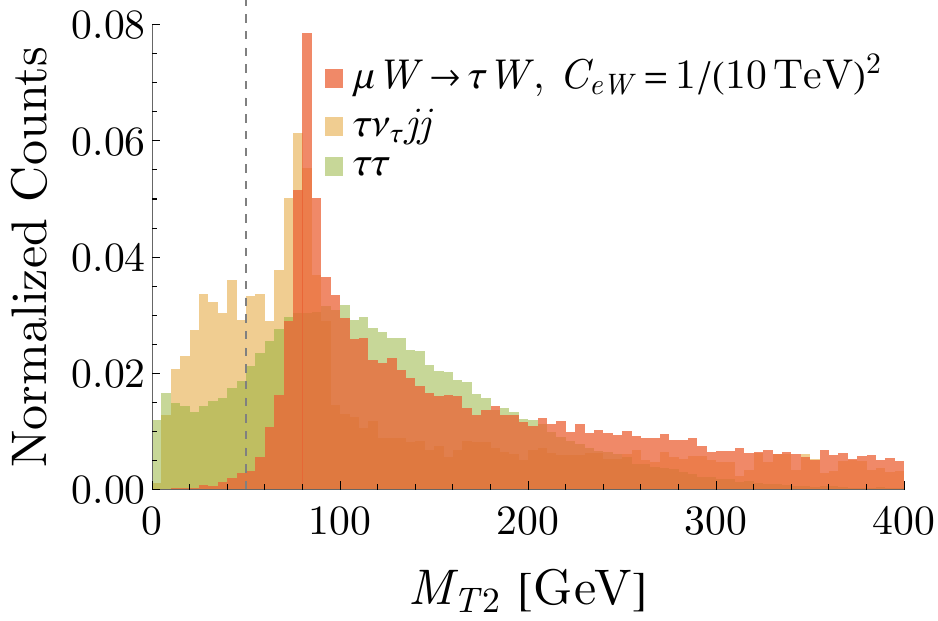}
\caption{Normalized histograms of the missing momentum pseudorapidity (\textbf{left}), the $\Delta R$ between the subleading jet and missing momentum vector $\slashed{p}$ (\textbf{middle}), and the stransverse mass $M_{T2}$ (\textbf{right}) for the signal in red and the leading $l\nu jj$ and $\tau^+\tau^-$ backgrounds in orange and green, respectively. The vertical dashed lines denote the cuts on the variable, see the text for details.}
\label{fig:histograms_muw_taw}
\end{figure}

As already noted, the signal arises from hard scattering of an incoming muon and a radiated gauge boson, whereas the background originates from the scattering of two radiated vector bosons, which carry only a fraction of the muon beam energy. This difference informs the cuts that can be used to subtract the background. 
Specifically, we impose lower bounds on the transverse momenta of the leading and sub-leading
final-state jets, requiring $p_{T,j1} > 1500\,\mathrm{GeV}$ and $p_{T,j2} > 500\,\mathrm{GeV}$.  
For the same reason, we expect the signal to be more forward than the background. Thus, we impose a cut on the pseudorapidity of the missing momentum vector, $\slashed{p}$,  $|\eta_{\slashed{p}}| > 2$. 

The distinct topologies of signal and background make the angular separation between the missing momentum and the softer jet, i.e.,
\begin{equation}
\Delta R_{\mathrm{soft}} \equiv \sqrt{(\eta_{j2} - \eta_{\slashed{p}})^2 + (\phi_{j2} - \phi_{\slashed{p}})^2} \, ,
\label{eq:DeltaR}
\end{equation}
a powerful discriminator. 
%As argued above, the $\tau$-jet is typically the softer jet.
In VBF backgrounds, forward leptons escaping down the beam pipe carry large energies whose momenta mostly cancel; the remaining missing momentum is then dominated by the neutrino from the $\tau$ decay and tends to be aligned with the $\tau$-tagged (softer) jet, yielding a small $\Delta R_{\mathrm{soft}}$.
In contrast, the signal is initiated by an incoming muon and a gauge boson, so a substantial fraction of the missing momentum is carried along the beam pipe and dominates the net missing momentum.
The visible system recoils against this, leading to a large separation between the softer jet and the missing momentum vector. 
Therefore imposing a lower bound on $\Delta R_{\mathrm{soft}}$ enhances the signal over the background. 

The combined selection cuts described above substantially suppress the $\tau\nu jj$ background. To demonstrate their effectiveness, Fig.~\ref{fig:histograms_muw_taw} shows representative histograms, with the chosen cut values indicated for each kinematic variable.
This figure illustrates that the $\tau \tau$ background can further be suppressed by a cut on $M_{T2}$ \cite{Lester:1999tx}.
Unlike the signal, for backgrounds with $\tau\tau$ production, $M_{T2}$ is bounded above by the $\tau$ mass, so imposing $M_{T2} > 50\,\mathrm{GeV}$ efficiently removes these events. 
Finally, and similar to the previous section, we require a lower bound on the missing transverse energy $\slashed{E}_T > 50\,\mathrm{GeV}$.

Putting all these cuts together, we can calculate the final signal event counts for different SMEFT operators:
\begin{equation}
\begin{gathered}
S_{\mathcal{O}_{eB}} \simeq 1.4 \,\,\textrm{events}\, \times C_{eB}^2 \Big(\frac{10\,\textrm{TeV}}{\Lambda}\Big)^4 \ ,
\qquad 
S_{\mathcal{O}_{eW}} \simeq 1290\,\,\textrm{events}\, \times C_{eW}^2 \Big(\frac{10\,\textrm{TeV}}{\Lambda}\Big)^4 \ , \\[0.75em]
S_{\mathcal{O}_{Hl}^{(1)}} \simeq 155\,\,\textrm{events}\, \times {C_{Hl}^{(1)}}^2 \Big(\frac{10\,\textrm{TeV}}{\Lambda}\Big)^4 \ ,
\qquad 
S_{\mathcal{O}_{Hl}^{(3)}} \simeq 154\,\,\textrm{events}\, \times {C_{Hl}^{(3)}}^2 \Big(\frac{10\,\textrm{TeV}}{\Lambda}\Big)^4 \ , \\[0.75em]
S_{\mathcal{O}_{He}} \simeq 216\,\,\textrm{events}\, \times C_{He}^2 \Big(\frac{10\,\textrm{TeV}}{\Lambda}\Big)^4\, .
\end{gathered}
\label{eq:event_yield_muw_taw}
\end{equation}
The proposed cuts also reduce each backgrounds to $\mathcal{O}$(1) events, see Table~\ref{tab:lfv_muc_summary}.

%%%%%

%%%%%%%%%%%%%%%%%%%%%%%%%%%%%%%%%%%%%%%%%%%%%%%%%%%%%%%%%%%%%%%%%%%%%%%%%%
\subsection{\texorpdfstring{$\mu \mu \to \mu \tau$}{mu mu to mu tau}}

\begin{figure}[t!]
\centering
\resizebox{0.8\columnwidth}{!}{
\includegraphics[width=4.5cm]{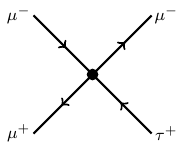}
~~~~~
\includegraphics[width=4.5cm]{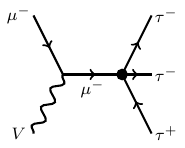}
~~~~~
\includegraphics[width=4.5cm]{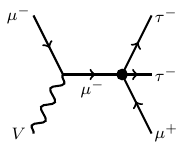}
}
\caption{
Example diagrams for four fermion signals $\mu\mu \to \mu\tau$ (\textbf{left}), $\mu V \rightarrow \tau \tau \tau$ (\textbf{center}), and $\mu^- V \rightarrow \mu^+ \tau^- \tau^-$ (\textbf{right}).
}
\label{Fig:FF_diags}
\end{figure}

Our next process is $\mu \mu \rightarrow \mu \tau$. This signal arises from a single vertex with an LFV four-fermion operator insertion (see the left panel of Fig.~\ref{Fig:FF_diags}). 
As a result, the cross section experiences a relative growth with the center of mass energy $\sqrt{s}$, making it a particularly attractive target for high-energy colliders. 
The operators that contribute include four different four-fermion operator helicity structures: $C_{ee}^{\mu\mu\mu\tau}$, $C_{ll}^{\mu\mu\mu\tau}$, $C_{le}^{\mu\mu\mu\tau}$, $C_{le}^{\mu\tau\mu\mu}$; see Eq.~\eqref{eq:smeft_operators} for their definitions. 
All of these have fairly large cross sections, as displayed in Table~\ref{tab:lfv_muc_summary}.
As before, we will focus on the hadronic tau decay, so that the final state in the detector will be a $\tau$-jet and a single visible muon. 

\begin{figure}[t!]
\centering
\includegraphics[width=5cm]{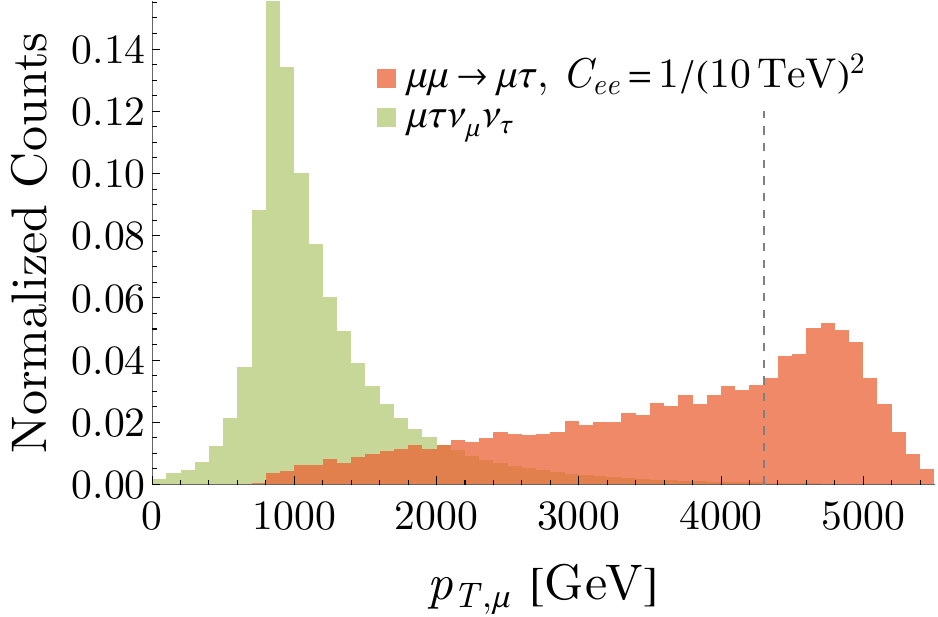}
~
\includegraphics[width=5cm]{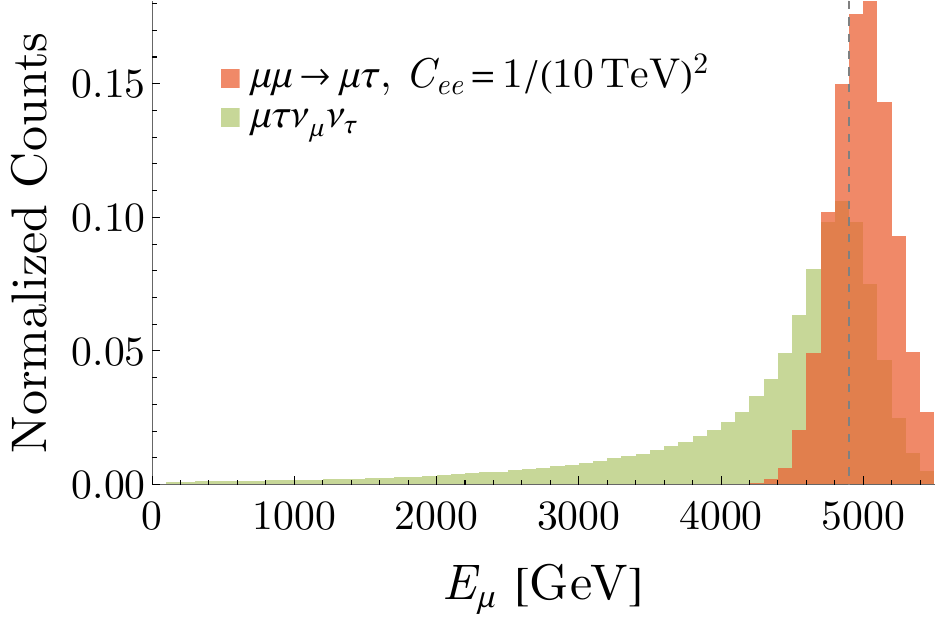}
~
\includegraphics[width=5cm]{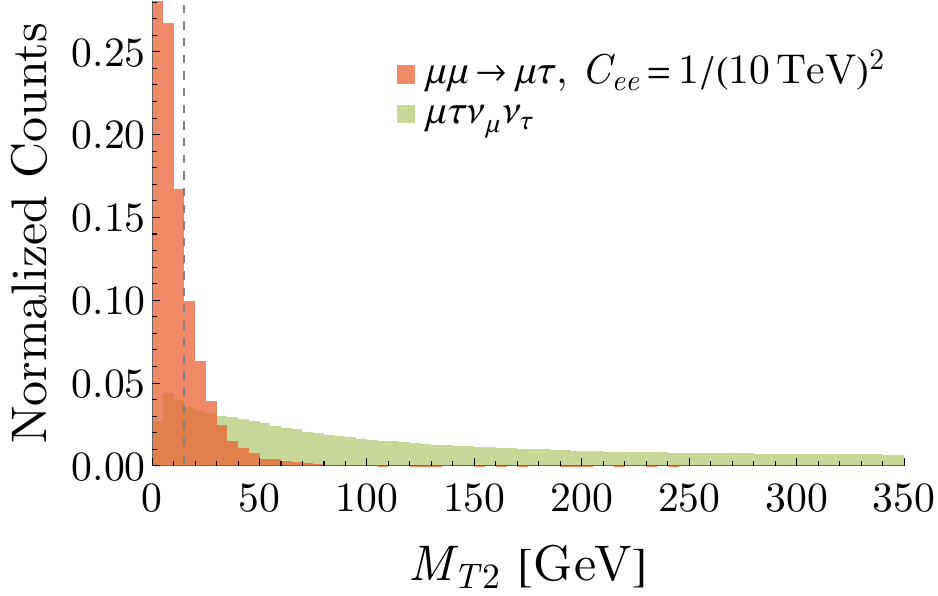}
\caption{Normalized histograms of the distributions of the muon $p_T$ (\textbf{left}), the muon energy (\textbf{middle}) and the $M_{T2}$ of the system (\textbf{right}) for the $\mu \mu \to \mu \tau$ signal (red) and the $\mu\tau\nu_{\mu} \nu_{\tau}$ background (green). The vertical dashed lines denote the cuts on the variable, see the text for details.}
\label{fig:histograms_mumu_mutau}
\end{figure}

The primary background for this signal is $\mu \mu \rightarrow \mu \nu_\mu \tau \nu_\tau$. This gets several contributions including the decay of on-shell tau pairs where one decays to a muon, and of on-shell $W$ pairs where one decays to a tau and the other decays to a muon. Each of these on-shell decays contributes roughly $\sim 2\%$ of the total background, though they are a larger fraction of the higher $p_T$ events. With only basic selection cuts (muon and tau $p_T \geq 100$ GeV and $|\eta| \leq 2.5$), this background appears large, with $4.0 \times 10^{4}$ events expected at a benchmark luminosity of 10 ab$^{-1}$.

To suppress this background, we impose more stringent cuts on the minimum muon transverse momentum and energy, requiring $p_T > 4300~\text{GeV}$ and $E > 4900~\text{GeV}$, respectively, along with an upper bound of $M_{T2} < 15~\text{GeV}$. These cuts are illustrated in Fig.~\ref{fig:histograms_mumu_mutau}. The muon $p_T$ requirement effectively separates signal from background, as relatively forward muons are far more common in background processes than in the signal. The additional cut on muon energy provides further discrimination, since in the signal the muon typically shares energy with fewer final-state particles and is therefore somewhat more energetic. The $M_{T2}$ variable also aids in distinguishing signal from background, as it is sensitive to missing energy and thereby to the number of neutrinos in the event. Although $M_{T2}$ was not originally designed for this purpose, it serves adequately for our study; further optimization, e.g. potentially involving alternative missing-energy observables or machine-learning–based methods, could provide even stronger separation.

After these selection cuts, only 5.0 expected background events remain, but signal passes with between 19\% and 25\% efficiency (see Table~\ref{tab:lfv_muc_summary}). Assuming we only consider one operator at a time, this gives signal yields of 
\begin{equation}
\begin{split}
    S_{\mathcal{O}_{ee}} \simeq 8.00 \times 10^{5} \mathrm{ events} \times C_{ee}^{\mu\mu\mu\tau} \Big(\frac{10 \mathrm{ TeV}}{\Lambda}\Big)^4, \quad
S_{\mathcal{O}_{le}^{\mu\mu\mu\tau}} \simeq 2.38 \times 10^{5} \mathrm{ events} \times C_{le}^{\mu\mu\mu\tau} \Big(\frac{10 \mathrm{ TeV}}{\Lambda}\Big)^4 \ , \\[0.5em]
S_{\mathcal{O}_{ll}} \simeq 8.76 \times 10^{5} \mathrm{ events} \times C_{ll}^{\mu\mu\mu\tau} \Big(\frac{10 \mathrm{ TeV}}{\Lambda}\Big)^4, \quad
S_{\mathcal{O}_{le}^{\mu\tau\mu\mu}} \simeq 2.57 \times 10^{5} \mathrm{ events} \times C_{le}^{\mu\tau\mu\mu} \Big(\frac{10 \mathrm{ TeV}}{\Lambda}\Big)^4 \ .
\end{split}
\end{equation}

One might also be concerned about similar final states arising from charged or neutral VBF processes. However, these backgrounds feature muons with transverse momenta and energies that are significantly lower than those in the signal, and are therefore efficiently removed by our selection cuts. For completeness, we also checked potential reducible backgrounds for this operator. In particular, we examined backgrounds such as $\mu \mu \to \mu \nu_\mu j j$ to account for the possibility that jets could be mistagged as $\tau$ leptons at a rate of $0.02$, and confirmed they are likewise eliminated by our selection criteria.

%%%%%%%%%%%%%%%%%%%%%%%%%%%%%%%%%%%%%%%%%%%%%%%%%%%%%%%%%%%%%%%%%%%%%%%%%%
\subsection{\texorpdfstring{$\mu V \to \tau \tau \tau$}{mu V to tau tau tau}}

Next we consider $\mu V \rightarrow \tau \tau \tau$. An example contribution for this process is shown in the center of Fig.~\ref{Fig:FF_diags}. The four-fermion operators which contribute are $C_{ee}^{\tau\tau\mu\tau}$, $C_{ll}^{\tau\tau\mu\tau}$, $C_{le}^{\mu\tau\tau\tau}$, $C_{le}^{\tau\tau\mu\tau}$, and we again restrict to hadronic tau decays.

All of the backgrounds for this signal are negligible with a relatively loose cut of 500~GeV on the transverse momenta of the hardest two jets and of 50~GeV on the missing transverse energy. 
The irreducible background for this signal is $W^+ Z/\gamma \rightarrow \overline{\nu}_{\tau} \tau^+ \tau^+ \tau^-$. Since it is computationally intractable to run this with a forward muon and neutrino, we studied this background using the effective vector approximation. Additionally, we imposed generator level cuts of 50 GeV on the $p_T$ of the softest $\tau$ jet and on the $p_T$ of $\nu_{\tau}$, as well as a cut of 450 GeV on the hardest two tau jets for sufficient sampling. The combination of these cuts is sufficient to reduce the background to a total of $\ll 1$ events, making it effectively negligible. These cuts work because the signal process includes one of the hard partons,  while the background is at lower pT because it comes from the interaction of two gauge bosons. In addition to this irreducible background, there are also  backgrounds from four $\tau$'s (produced either through annihilation, neutral or charged VBF) where only three are reconstructed, and from a $jj \tau \tau$ final state (again produced via annihilation or VBF) where the light jets are mistagged as a single $\tau$. In both cases, a loose pt cut of 500 on the hardest two jets is also sufficient to eliminate these backgrounds.\footnote{A lower bound on the invariant mass of the $\tau$ pair is also imposed at generator level to avoid IR divergences. Conservatively, this cut is also included in the signal, which only modifies the cross section by a few percent.}

Given the negligible backgrounds for this process, we only impose the loose cuts previously mentioned. This corresponds to signal yields of 
\begin{equation}
\begin{split}
    S_{\mathcal{O}_{ee}} \simeq 650 \;  \mathrm{ events} \times C_{ee}^{\tau\tau\mu\tau} \Big(\frac{10 \mathrm{ TeV}}{\Lambda}\Big)^4, \quad
S_{\mathcal{O}_{le}^{\mu\tau\tau\tau}} \simeq 78\; \mathrm{ events} \times C_{le}^{\mu\tau\tau\tau} \Big(\frac{10 \mathrm{ TeV}}{\Lambda}\Big)^4 \\[0.75em]
S_{\mathcal{O}_{ll}} \simeq 624 \; \mathrm{ events} \times C_{ll}^{\tau\tau\mu\tau} \Big(\frac{10 \mathrm{ TeV}}{\Lambda}\Big)^4, \quad
S_{\mathcal{O}_{le}^{\tau\tau\mu\tau}} \simeq 81\; \mathrm{  events} \times C_{le}^{\tau\tau\mu\tau} \Big(\frac{10 \mathrm{ TeV}}{\Lambda}\Big)^4
\end{split}
\end{equation}

%%%%%%%%%%%%%%%%%%%%%%%%%%%%%%%%%%%%%%%%%%%%%%%%%%%%%%%%%%%%%%%%%%%%%%%%%%
\subsection{\texorpdfstring{$\mu^\mp V \rightarrow \mu^\pm \tau^\mp \tau^\mp$}{mu V to mu tau tau}}\label{subsec:muVmutautau}

Finally we consider $\mu^\mp V \rightarrow \mu^\pm \tau^\mp \tau^\mp$. This process is similar to the one studied in the last section except for the existence of a muon instead of one of the final $\tau$s (see the right panel of Fig.~\ref{Fig:FF_diags}). It is LFV only if the two final state $\tau$s have the same charge, and arises from the four fermion operators $C_{ee}^{\mu\tau\mu\tau}$, $C_{ll}^{\mu\tau\mu\tau}$, $C_{le}^{\mu\tau\mu\tau}$ with $\Delta L_\mu = - \Delta L_\tau = \pm 2$. Again we require both $\tau$s decay hadronically, and tag the sign of the muon to make sure that we end up with two $\tau$s of the same sign.

For our predictions, we will assume we do not have additional information about the sign of the $\tau$ from the $\tau$ jet. This is a conservative assumption since it leads us to consider additional background with $\tau$s that have opposite signs. Since these backgrounds are already small, $\tau$ charge tagging is not expected to notably change the scales to which this process is sensitive, but could slightly reduce the number of background events. Either the observable jet charge \cite{Field:1977fa,Krohn:2012fg} or machine learning techniques which capture the same information \cite{Fraser:2018ieu, Chen:2019uar} could be utilized for this purpose.

There are several different possible irreducible backgrounds.
Two possibilities come from different kinematic regions of $W^+ V \to \nu_{\tau} \tau^+ \tau^+ \tau^-$. One is a beam muon moving out of the forward and into the central region while simultaneously detecting only two $\tau$ jets, and the other comes from one of the tau's decaying to a muon and the others being tagged hadronically. A $p_T$ cut of 500 GeV on the leading $\tau$ jet, as well as cuts of 50 GeV on the $p_T$ of the other $\tau$ jet, the muon, and the MET leaves the first of these to contribute $3.2$ events at a benchmark of 10 ab$^{-1}$ of data. The second is completely eliminated by the same $p_T$ cut on the leading $\tau$ jet.\footnote{This background specifically is run with the EVA. $p_T$ cuts of 50 on each of the taus and the tau neutrino are also required at generator level for \texttt{MadGraph} to effectively sample this background.} Another irreducible background is $\mu^+ \mu^- \rightarrow \mu^+ \mu^- \tau^+ \tau^-$, where only one muon goes down the beam pipe, but this is negligible due to kinematics. 
The final irreducible background we study is $\mu^+ \mu^- \rightarrow \mu^+ W^- \nu_{\mu} \tau^+ \tau^-$; again the same set of $p_T$ cuts leaves $3.9$ events for this background.

We also study three reducible backgrounds. One is the case where four $\tau$s are produced, one decays to a muon, and the others decay hadronically but only two jets are reconstructed. A second is where two light jets are produced in addition to two taus (one of which decays hadronically and the other to a muon), and a total of two jets are reconstructed and $\tau$-tagged. Both can be produced either in a $\mu^+ \mu^-$ hard process, or by charged or neutral VBF. The four tau annihilation case contributes 1.8 events after the same selection cuts; the same cuts leave a negligible number of events for the rest of these backgrounds. The third reducible background is $\mu^+ \mu^- \rightarrow \mu^+ \overline{\nu}_{\mu} \tau^- \nu_{\tau} j j $ where the muon is visible and the two jets are mistagged as a single tau. After the same set of cuts, assuming a mistag rate of 0.02 for light jets mistagged as taus, this contributes 8.3 expected events at 10 ab$^{-1}$.

Adding these same cuts to the signal gives yields of 
\begin{equation}
\begin{split}
    S_{\mathcal{O}_{ee}} \simeq 414 \;  \mathrm{ events} \times C_{ee}^{\mu\tau\mu\tau}& \Big(\frac{10 \mathrm{ TeV}}{\Lambda}\Big)^4, \quad
S_{\mathcal{O}_{le}^{\mu\tau\mu\tau}} \simeq 96\; \mathrm{ events} \times C_{le}^{\mu\tau\mu\tau} \Big(\frac{10 \mathrm{ TeV}}{\Lambda}\Big)^4 \\[0.75em]
S_{\mathcal{O}_{ll}} &\simeq 399 \; \mathrm{ events} \times C_{ll}^{\mu\tau\mu\tau} \Big(\frac{10 \mathrm{ TeV}}{\Lambda}\Big)^4
\end{split}
\end{equation}

%%%%%%%%%%%%%%%%%%%%%%%%%%%%%%%%%%%%%%%%%%%%%%%%%%%%%%%%%%%%%%%%%%%%%%%%%%
%%%%%%%%%%%%%%%%%%%%%%%%%%%%%%%%%%%%%%%%%%%%%%%%%%%%%%%%%%%%%%%%%%%%%%%%%%
\section{Summary of Low- and High-Energy Probes}
\label{sec:summary}

With all of the constraints in hand, we can compare the various limits on the effective operators from Eq.~\eqref{eq:smeft_operators}. As we have seen in some examples already, the constraints vary widely depending on the flavor-violating transition involved: the indirect constraints on $\mu$\,--\,$e$ transitions are quite stringent, while for $\tau$\,--\,$\mu$ and $\tau$\,--\,$e$ transitions, different probes are sensitive to roughly similar scales, depending on the operator.

We begin in Section~\ref{subsec:summary_1d} by comparing all existing and projected constraints in the case where only a single operator is turned on at a time. 
For many operators we find that the proposed muon collider searches from Section~\ref{sec:collider} can probe scales beyond the reach of both current and even future small-scale experiments. 
In line with the previous section, we focus on $\tau$\,--\,$\mu$ transitions in a high-energy muon collider.
While this setup allows for a clear comparison of the sensitivities of different experimental probes, it is not a fully realistic scenario since any UV completion is expected to generate multiple operators simultaneously. 
In Section~\ref{subsec:summary_2d} we therefore extend the analysis to the case where more than one SMEFT operator can be nonzero at a time. Here the importance of complementarity between probes becomes evident: for many operator combinations, the muon collider and small-scale experiments cover different regions of parameter space. Finally, in Section~\ref{subsec:summary_flavor} we compare constraints across different generations, exploring various assumptions about the flavor structure of the operator coefficients.

\begin{figure}[ht!]
\centering
\includegraphics[width=\linewidth]{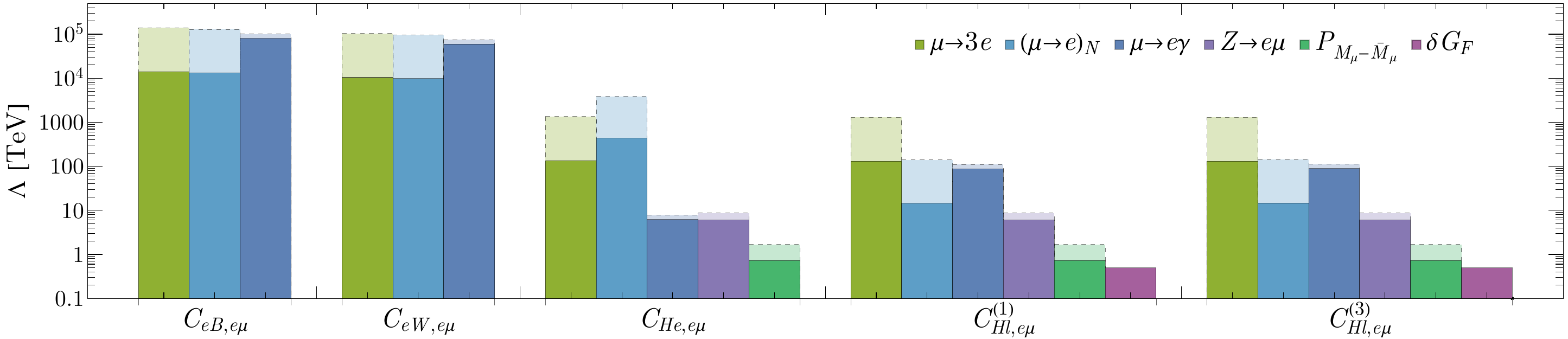}
\\[1em]
\includegraphics[width=\linewidth]{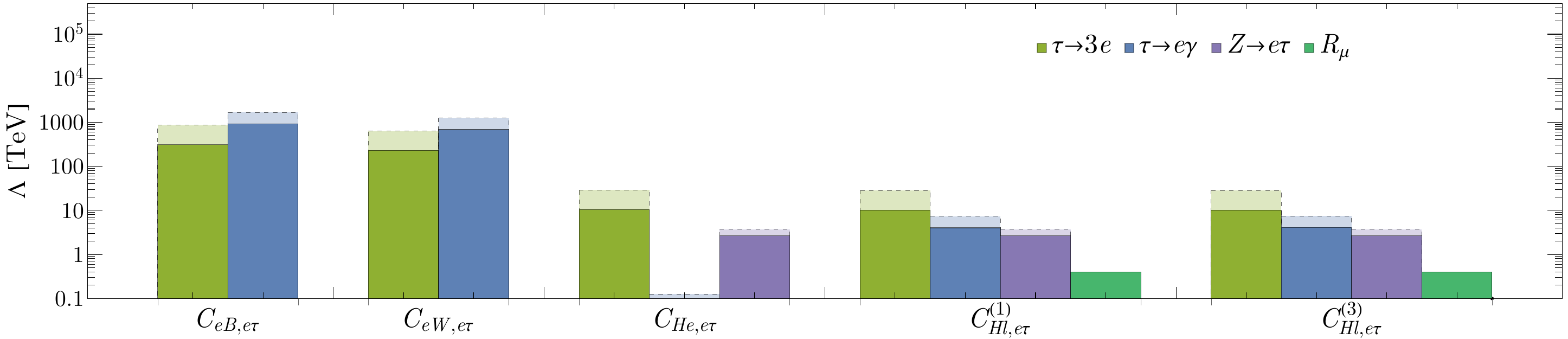}
\\[1em]
\includegraphics[width=\linewidth]{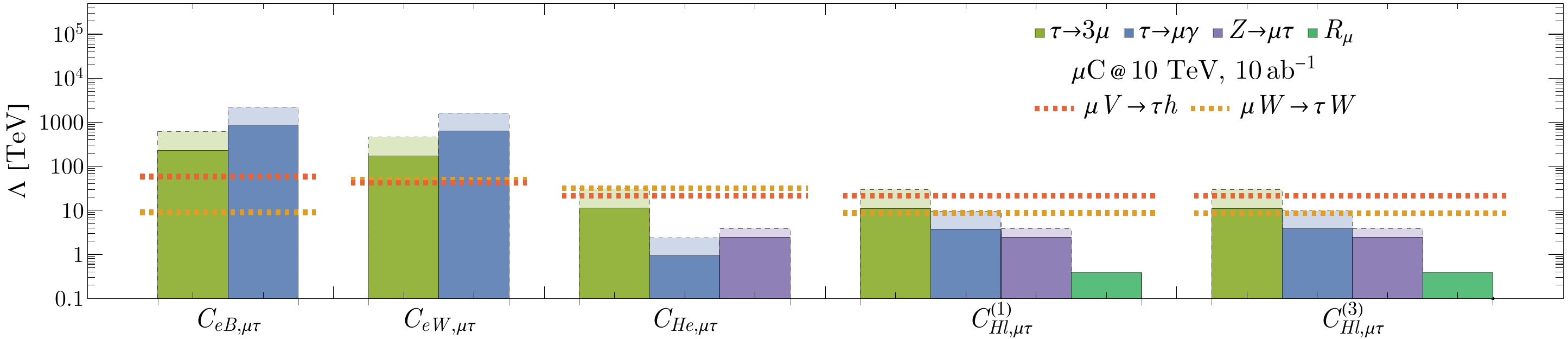}
\caption{
Constraints on lepton-flavor-violating dipole and Higgs-current effective operators from various low-energy experiments and from a high-energy muon collider. We show the constraints on $\mu$\,--\,$e$, $\tau$\,--\,$e$ and $\tau$\,--\,$\mu$ in the \textbf{top}, \textbf{middle} and \textbf{bottom} rows, respectively. The dark shaded regions indicate the current low-energy constraints while the lighter shaded regions with a dashed border indicate the future projections tabulated in Table~\ref{tab:precision_bounds}. For the $\tau$\,--\,$\mu$ transitions, we compare to the various muon collider probes discussed in Section~\ref{sec:collider}, assuming a $\sqrt{s} = 10\,\textrm{TeV}$ muon collider with $10\,\textrm{ab}^{-1}$ integrated luminosity.}
\label{fig:barchart}
\end{figure}

%%%%%%%%%%%%%%%%%%%%%%%%%%%%%%%%%%%%%%%%%%%%%%%%%%%%%%%%%%%%%%%%%%%%%%%%%%
\subsection{Comparison of One-Dimensional Constraints}
\label{subsec:summary_1d}

\begin{figure}[ht]
\centering
\includegraphics[width=\linewidth]{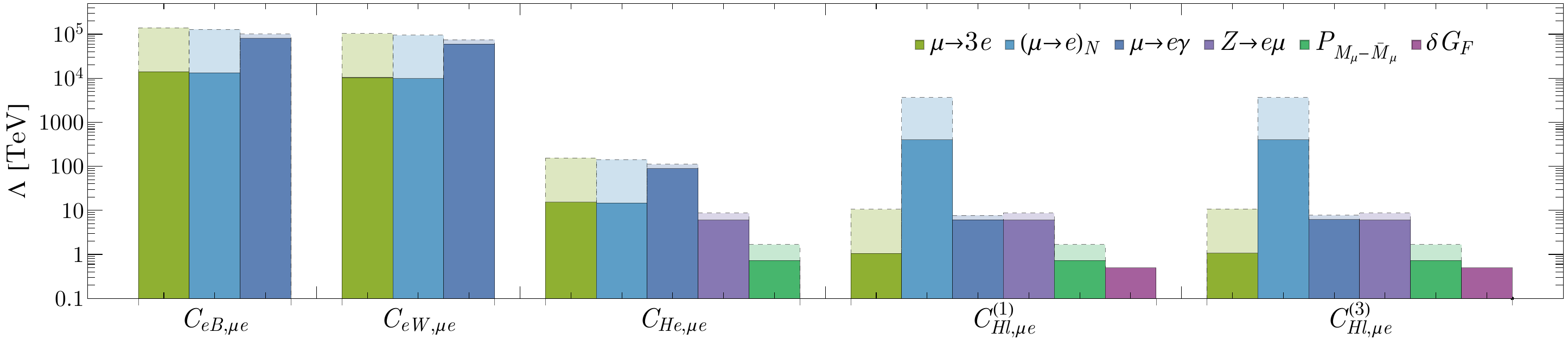}
\\[1em]
\includegraphics[width=\linewidth]{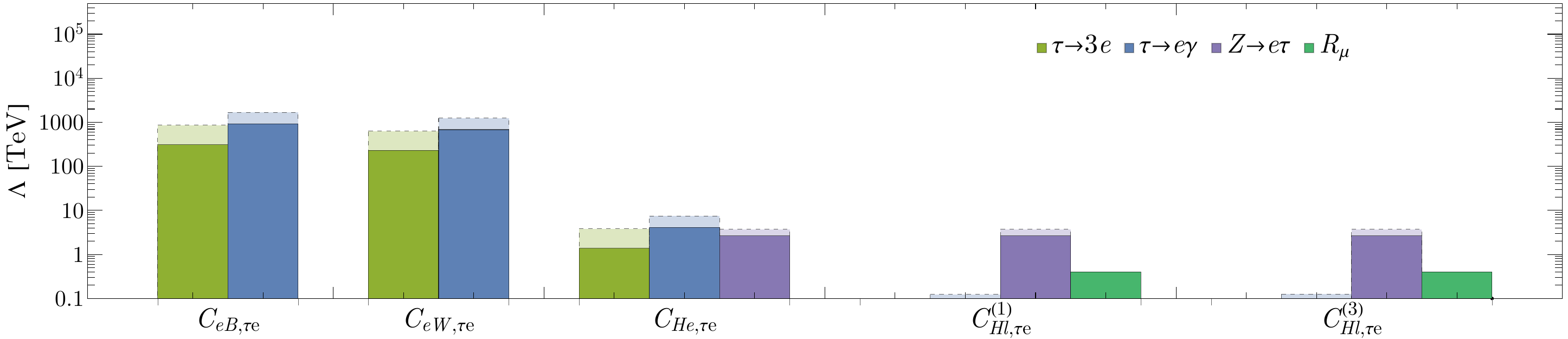}
\\[1em]
\includegraphics[width=\linewidth]{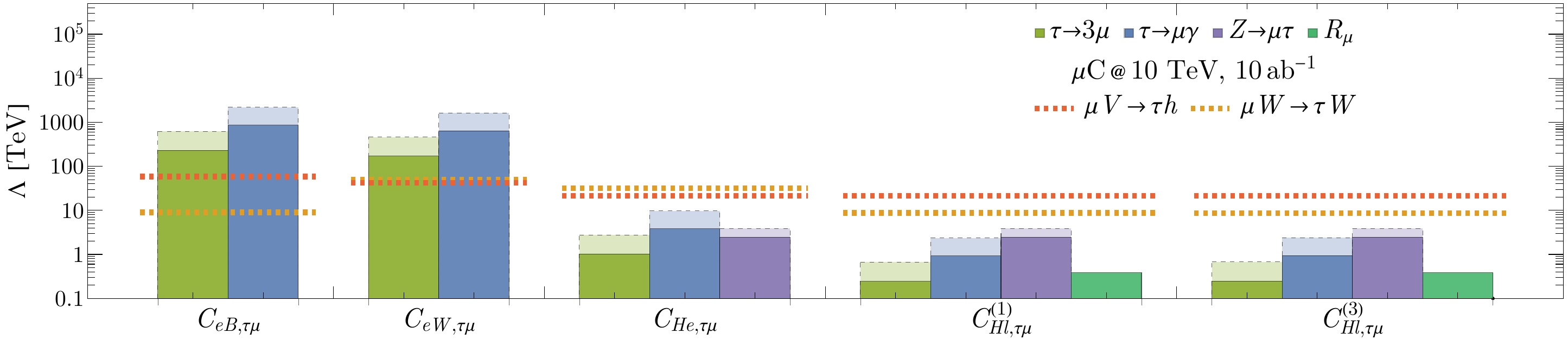}
\caption{The same as Fig.~\ref{fig:barchart}, but for the opposite combination of flavor-indices, for which some of the low-energy precision constraints are weaker due to factors of the lepton mass.}
\label{fig:barchart_extra}
\end{figure}

\begin{figure}[ht]
\centering
\includegraphics[width=\linewidth]{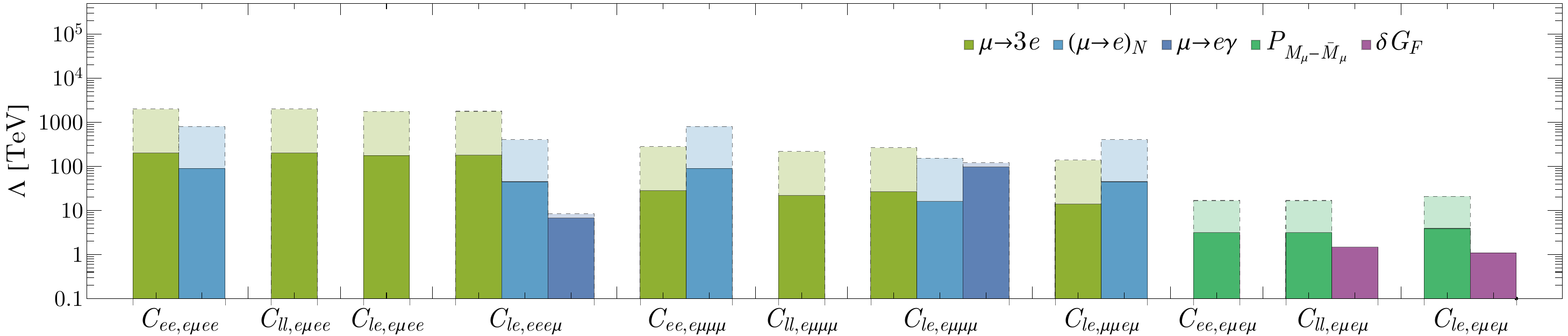}
\\[1em]
\includegraphics[width=\linewidth]{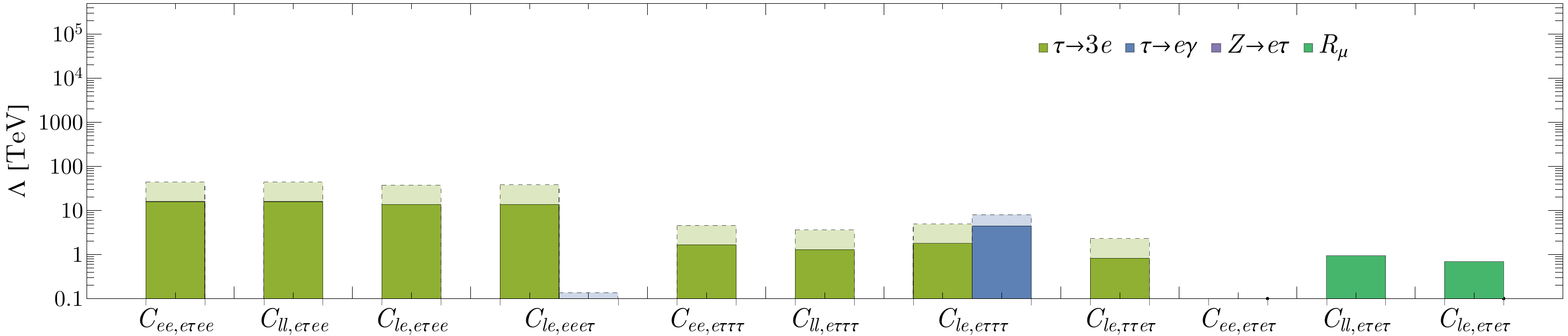}
\\[1em]
\includegraphics[width=\linewidth]{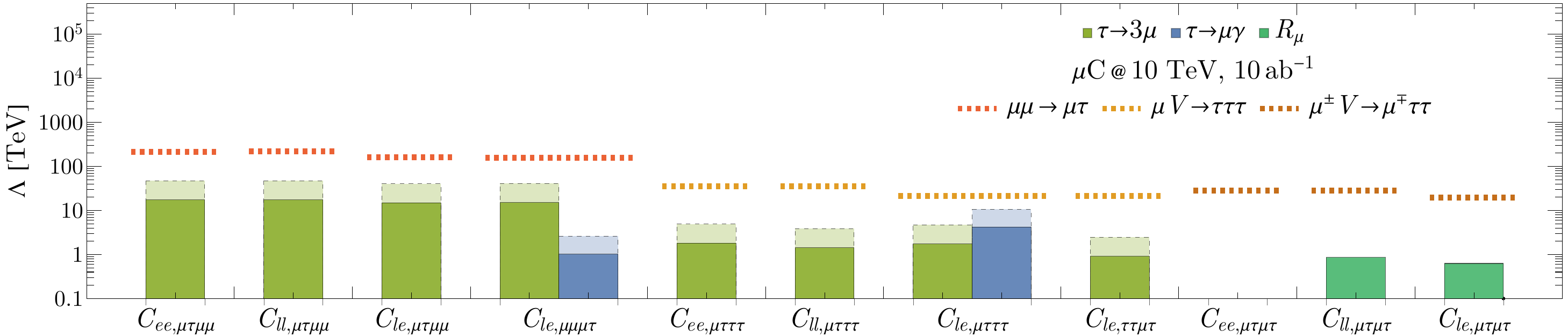}
\caption{Similar to Fig.~\ref{fig:barchart}, but for the considered four-fermion operators.}
\label{fig:barchart_4ferm}
\end{figure}

For each of the LFV operators in Eq.~\eqref{eq:smeft_operators}, and for each set of generation indices, we compute all the relevant low- and high-energy constraints as described in Sections~\ref{sec:low_energy}, \ref{sec:lfv_higgs} and \ref{sec:collider}. 
The results for each operator (excluding $\mathcal{O}_{eH}$, considered in Section~\ref{sec:lfv_higgs}) and for each generation are shown in Figs.~\ref{fig:barchart}--\ref{fig:barchart_4ferm}. The top, middle and bottom rows in each figure show the bounds on $\mu$\,--\,$e$, $\tau$\,--\,$e$ and $\tau$\,--\,$\mu$ transitions, respectively. 
The muon collider constraints appear as dashed lines only in the bottom row, as we considered exclusively $\tau$\,--\,$\mu$ flavor-violating effects.
The constraints on $C_{Hl}^{(1)}$ and $C_{Hl}^{(3)}$ are seen to be essentially identical, as they both result from the modified $Z$\,--\,$l_i$\,--\,$l_j$ coupling which is proportional to $(C_{Hl}^{(1)} - C_{Hl}^{(3)})$.

At leading order, collider constraints are insensitive to changing the order of flavor indices, e.g. the cross section and resulting sensitivity for processes involving $C_{He,\mu\tau}$ are identical to those for $C_{He,\tau\mu}$ at a muon collider. 
In contrast, many low-energy constraints do depend on this choice, since it determines which lepton mass flips the chirality in the loop diagrams that match onto the low-energy effective operators. 
For operators containing the Higgs, the two possible orders of lepton-flavor-violating indices are shown in Figs.~\ref{fig:barchart} and~\ref{fig:barchart_extra}. 
While the muon collider reach remains unchanged between these two figures, the sensitivity of some low-energy probes is significantly reduced once the indices are flipped. 
In particular, for the flavor combinations shown in Fig.~\ref{fig:barchart_extra}, low-energy probes become far less constraining, leaving the muon collider as the clearly dominant probe for these operators.

The four-fermion operators involving a muon-antimuon that can annihilate are easily the most constrained of all operators from collider searches, see Fig.~\ref{fig:barchart_4ferm}. 
This is anticipated, as the contact operator leads to a relative enhancement with the center of mass energy $\sqrt{s}$ of the collision. These searches thus benefit the most from the high energy available in $\mu^+\mu^-$ collisions at $\sqrt{s} = 10\,\textrm{TeV}$, probing scales up to $\Lambda \simeq 160 - 225~\textrm{TeV}$, depending on the chirality structure of the operator. This vastly exceeds the sensitivity available even to future $\tau \to 3\mu$ decay searches, which test only the $40 - 50\,\textrm{TeV}$ range.

All in all, we find that for many operators, the proposed processes at a muon collider (Section~\ref{sec:collider}) can probe scales well beyond the reach of even future low-energy experiments. This disparity is especially striking in the case of four-fermion operators, as illustrated in Fig.~\ref{fig:barchart_4ferm}. For certain operators and flavor structures, such as $\mathcal{O}_{ee,\mu\tau\mu\tau}$, no low-energy probes exist at all, making the muon collider the only available avenue to constrain these operators. Additionally, there are other operators for which a muon collider would be able to confirm and characterize a signal if one is seen in a future low-energy probe.

More generally, it also appears that the low-energy constraints on $\mu$\,--\,$e$ transitions are orders of magnitude more stringent than those on $\tau$\,--\,$e$ and $\tau$\,--\,$\mu$, with the latter two having quite comparable low-energy constraints. 
Of course, this assumes Wilson coefficients all of order unity, which is unlikely in realistic UV completions (given the rich flavor structure already present in the Standard Model). In reality, which processes are the most interesting for constraining new physics depend on the expected size of the Wilson coefficients, which is dictated by the underlying model of flavor. This motivates our discussion of how these constraints compare for various different flavor ans\"atze in Section~\ref{subsec:summary_flavor}.

%%%%%%%%%%%%%%%%%%%%%%%%%%%%%%%%%%%%%%%%%%%%%%%%%%%%%%%%%%%%%%%%%%%%%%%%%%
\subsection{Constraints on Multiple (Same-Flavor) Operators}
\label{subsec:summary_2d}

We expect multiple SMEFT operators to be generated below the scale of any realistic model of new physics in the UV. 
A unique capability of a high-energy muon collider, however, beyond the overall sensitivity demonstrated in the previous section, is that it allows for numerous independent probes of these operators. 
As we will now see, the combination of channels discussed in Section~\ref{sec:collider}---along with the low-energy probes---would allow for a nearly complete characterization of the operators responsible for any signal observed at planned future experiments.

\begin{figure}[ht!]
\centering
\includegraphics[width=0.65\linewidth]{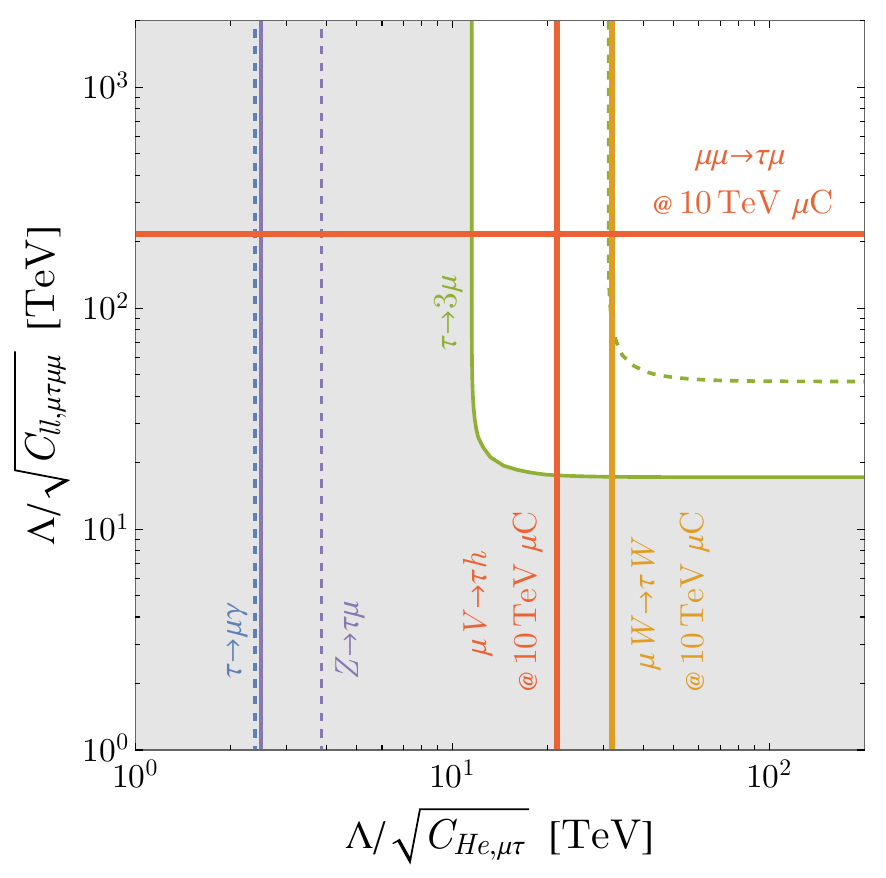}
\caption{Constraints on $\tau$\,--\,$\mu$ flavor-violating operators $C_{He}$ and $C_{ll}$, expressed as bounds on the scale. We show the muon collider constraints in red and orange, along with the low-energy constraints from $\tau \to \mu\gamma$, $\tau \to 3\mu$, and $Z \to \tau\mu$ in blue, green and purple, respectively (with projected low-energy constraints indicated by the corresponding dashed lines). The shaded region indicates the parameter space excluded by the best current constraint.}
\label{fig:2dplots_tamu_highlight}
\end{figure}

To illustrate this, we consider two-dimensional parameter planes spanned by pairs of SMEFT operators from Eq.~\eqref{eq:smeft_operators} with flavor indices $i,j = \mu, \tau$, and plot the $95\%$ C.L. constraints from all relevant low- and high-energy probes. An example is shown in Fig.~\ref{fig:2dplots_tamu_highlight}, which displays the reach of both current and future low-energy experiments alongside different processes at a high-energy muon collider on the plane of $\mathcal{O}_{He,\mu\tau}$ and $\mathcal{O}_{ll,\mu\tau\mu\mu}$. This figure highlights not only the ability of collider processes such as $\mu\mu \to \mu\tau$ to access scales beyond the reach of low-energy probes, but also their strong complementarity with rare decays such as $\tau \to 3\mu$, together covering a wide region of parameter space. Moreover, if a signal were to appear in future $\tau \to 3\mu$ searches and be attributable to these operators, a combination of $\mu W \to \tau W$ and $\mu\mu \to \mu\tau$ processes at a muon collider would provide a guaranteed path to both discover and characterize the underlying physics.
\footnote{A recent study found that a degree of discrimination between different operators is possible using differential distributions with only low-energy data, but several degenerate explanations would remain~\cite{Chang:2024xvc}.}

\begin{figure}
\centering
\includegraphics[width=0.42\linewidth]{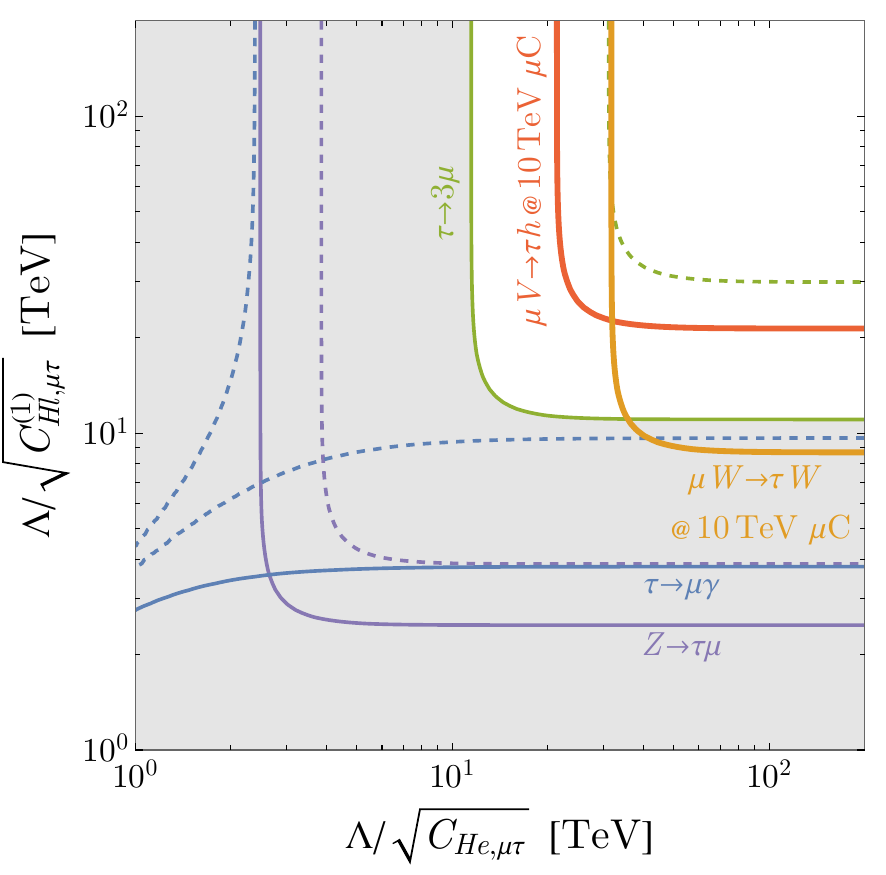}
\quad\quad
\includegraphics[width=0.42\linewidth]{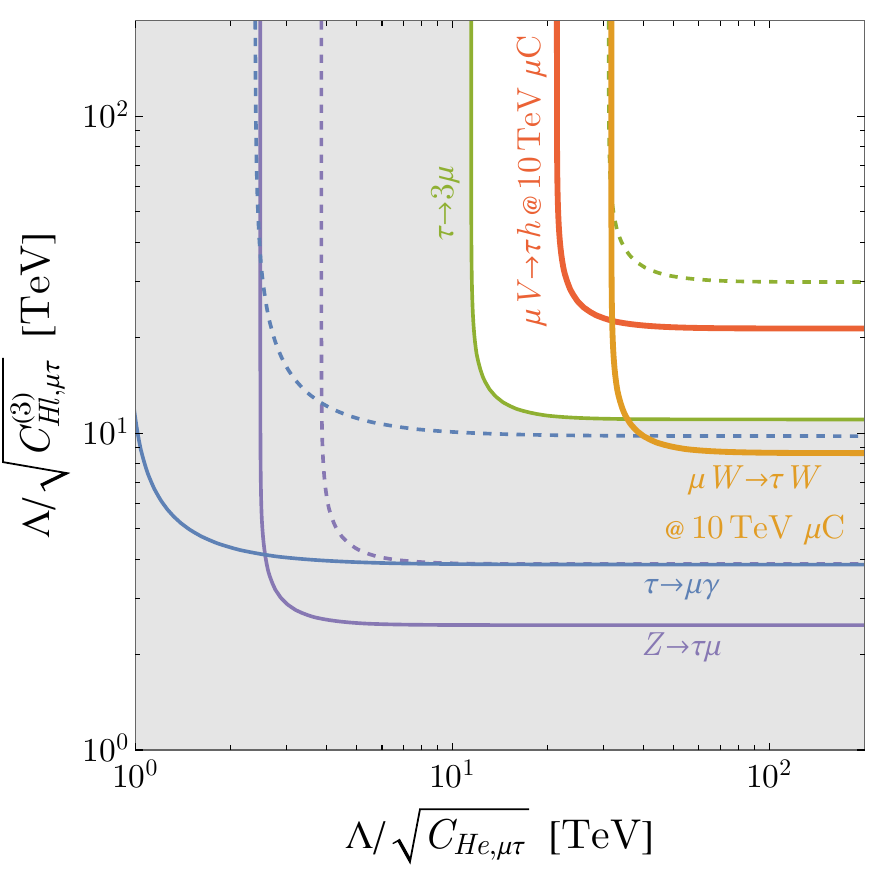} \\[0.5em]
\includegraphics[width=0.42\linewidth]{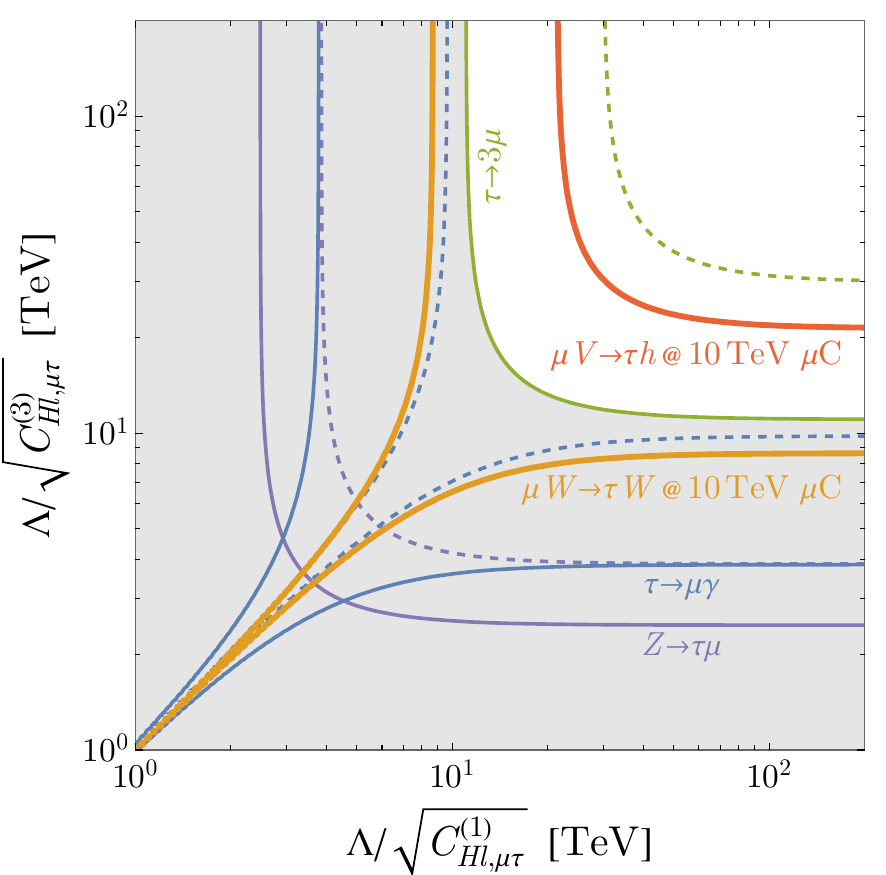}
\quad\quad
\includegraphics[width=0.42\linewidth]{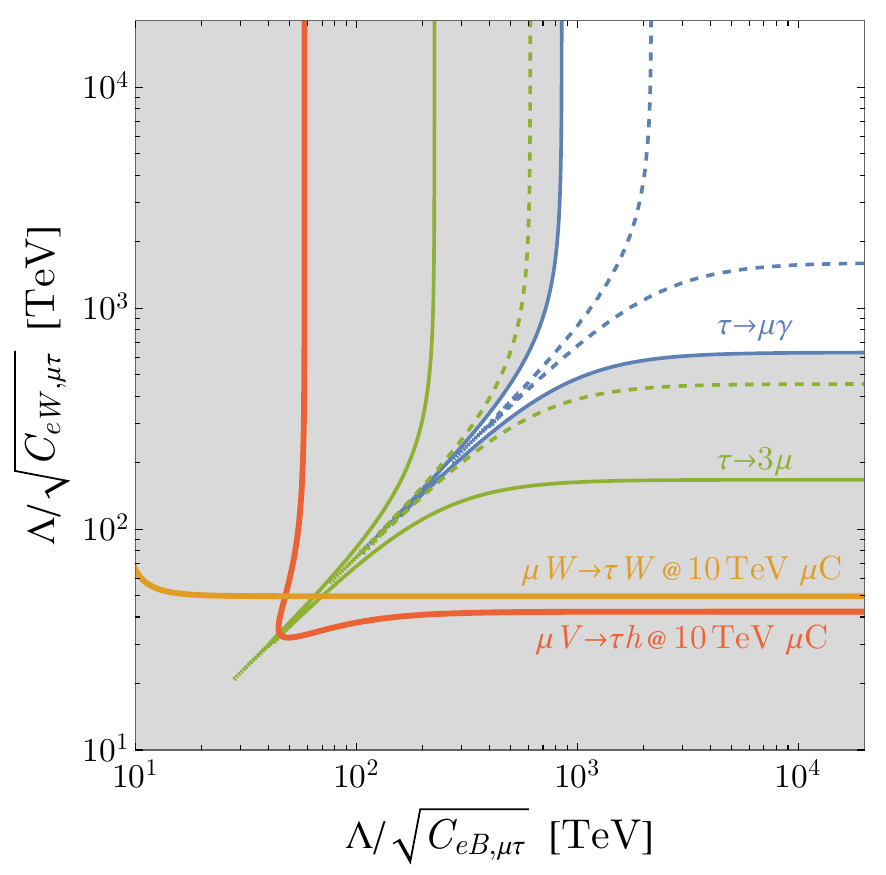} \\[0.5em]
\includegraphics[width=0.42\linewidth]{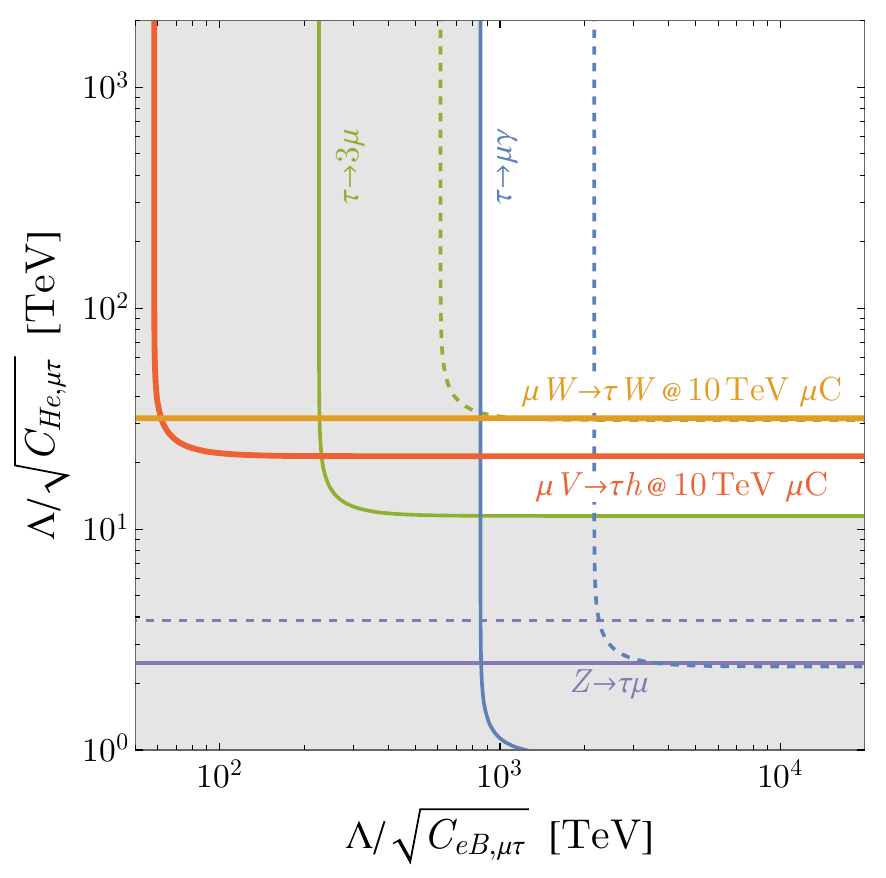}
\quad\quad
\includegraphics[width=0.42\linewidth]{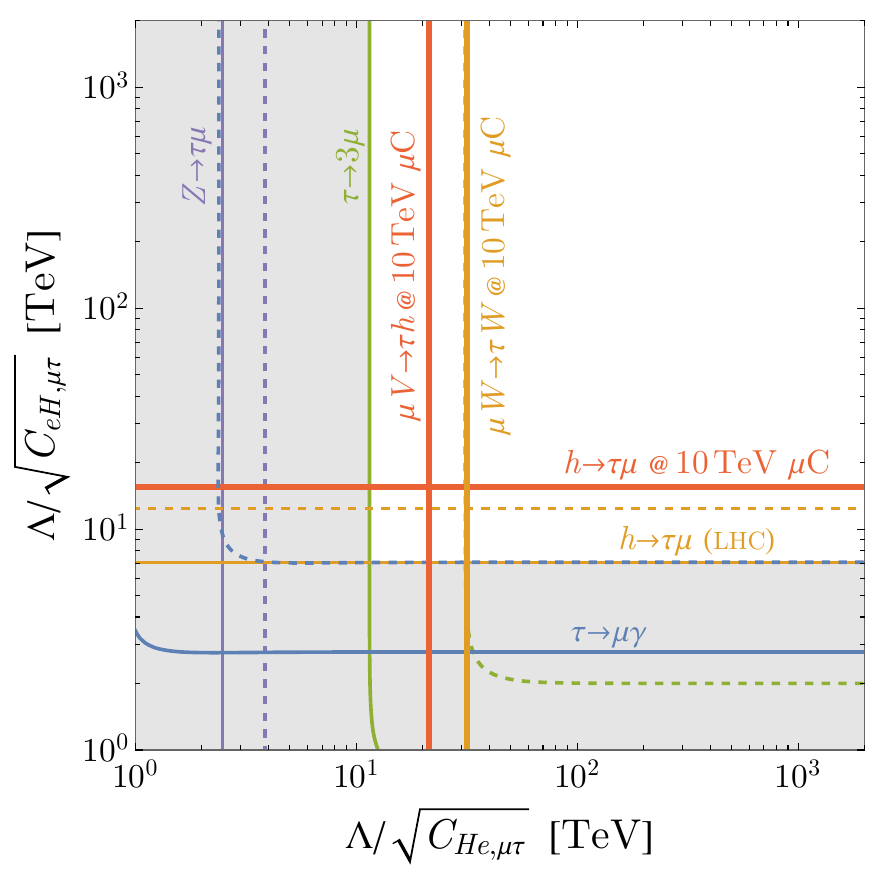} %\\[-1.5em]
\caption{
Similar to Fig.~\ref{fig:2dplots_tamu_highlight} for different benchmark pairs of operators involving leptons, the Higgs boson, and electroweak bosons.
%The muon collider constraints are shown in red and orange, along with the low-energy constraints from $\tau\to\mu\gamma$, $\tau\to 3\mu$ and $Z \to \tau\mu$ in blue, green and purple, respectively.
}
\label{fig:2dplots_tamu}
\end{figure}

\begin{figure}[ht]
\centering
\includegraphics[width=0.42\linewidth]{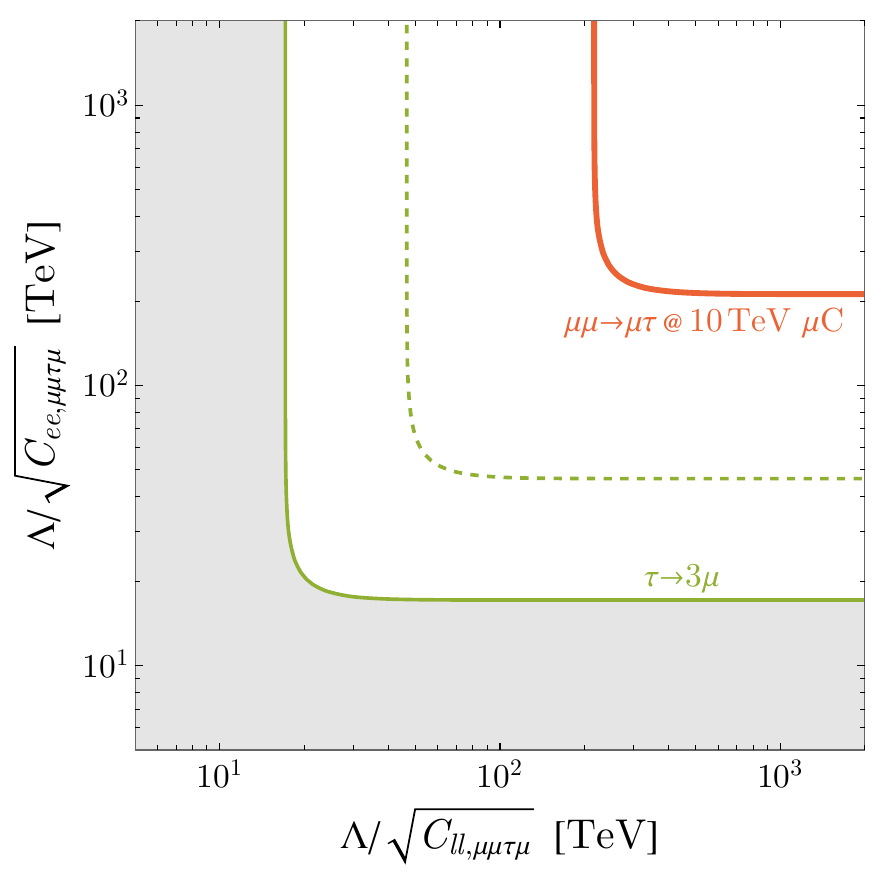}
\quad\quad
\includegraphics[width=0.42\linewidth]{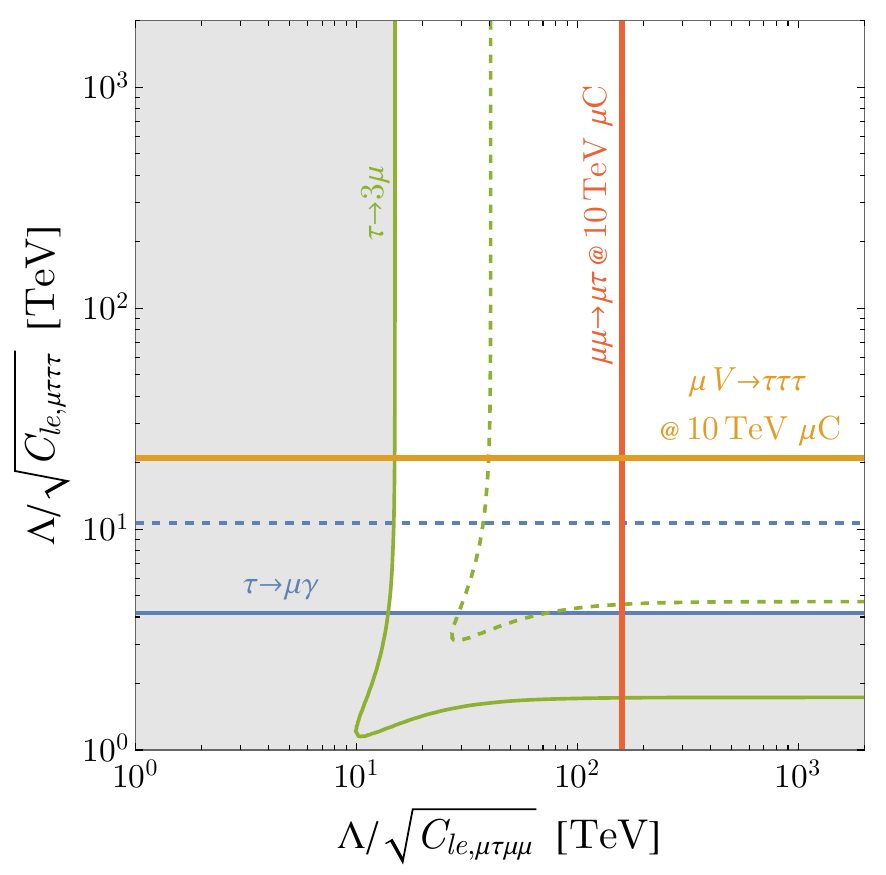} \\[0.5em]
\includegraphics[width=0.42\linewidth]{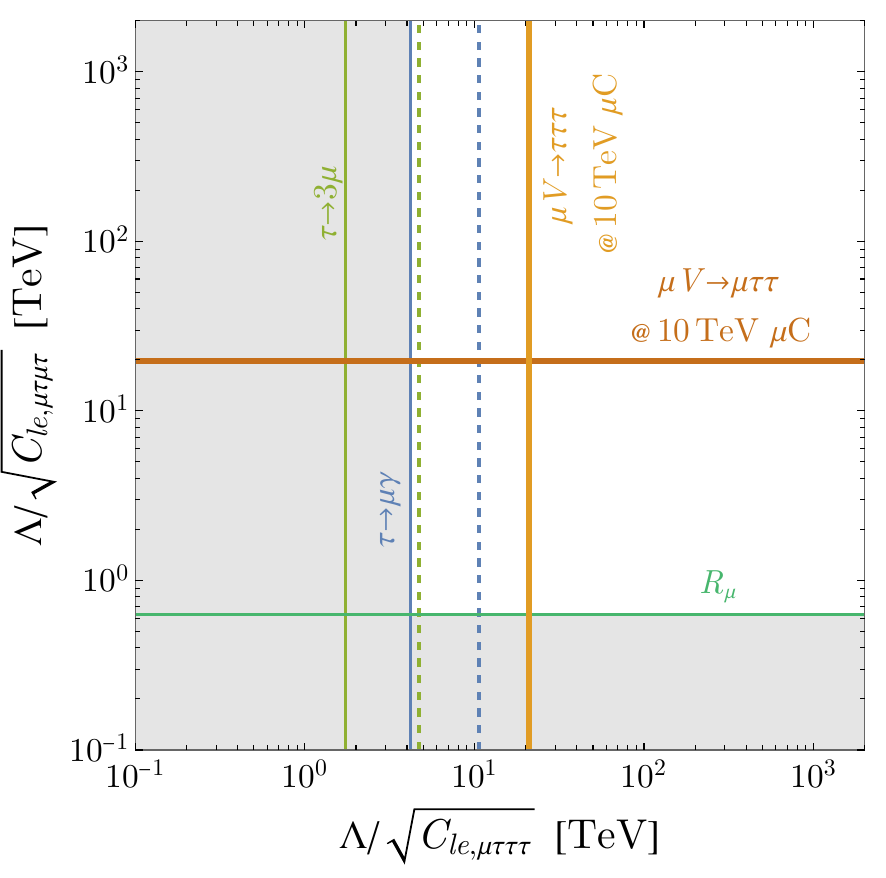}
\quad\quad
\includegraphics[width=0.42\linewidth]{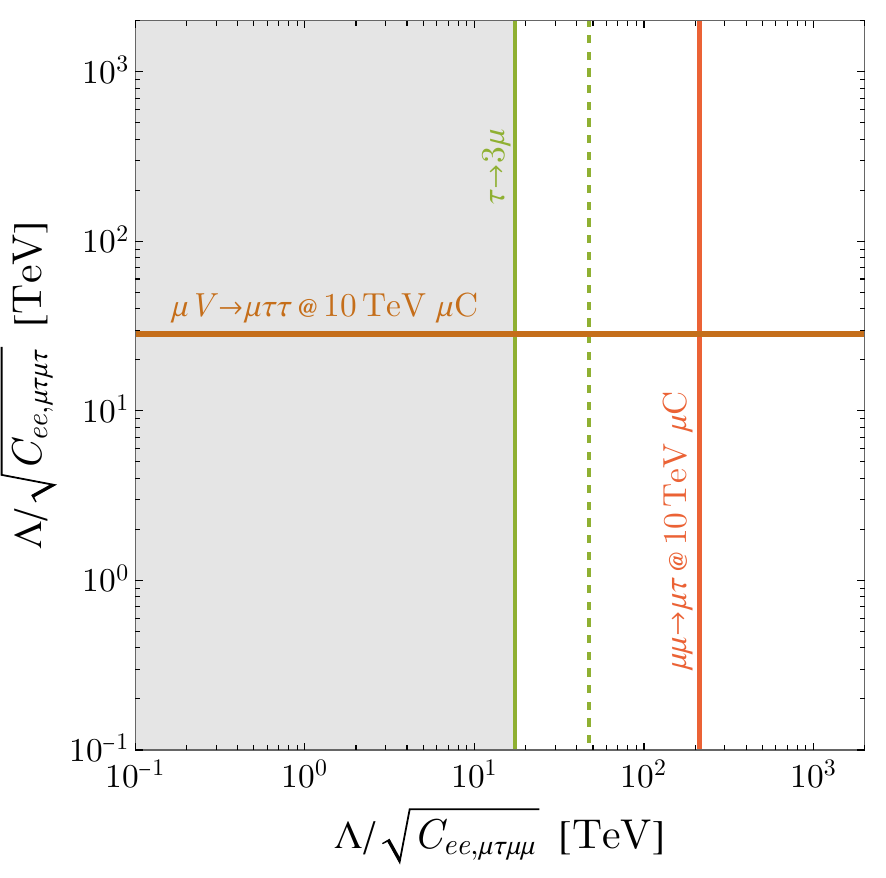}
\caption{Similar to Fig.~\ref{fig:2dplots_tamu}, but for different combinations of the four-fermion operators. The muon collider outperforms the low-energy probes for all the combinations of operators shown.}
\label{fig:2dplots_tamu_4ferm}
\end{figure}

The same exercise can be repeated for many different pairs of SMEFT operators. In Fig.~\ref{fig:2dplots_tamu} (Fig.~\ref{fig:2dplots_tamu_4ferm}) we show the reach of a high-energy muon collider on the planes of several representative pairs of operators involving leptons and electroweak bosons (four-fermion operators). 
Depending on the operators, different collider processes and low-energy probes dominate the sensitivity, underscoring the importance of a broad experimental program to capture different signatures. 
With the exception of the dipole operators (which, as discussed above, cannot be tested at high energies with the same sensitivity as $\tau \to \mu\gamma$) and the degeneracy between $C_{Hl}^{(1)}$ and $C_{Hl}^{(3)}$, the muon collider probes provide coverage of essentially all interesting directions in parameter space. 
The dips in sensitivity seen in some planes are the result of destructive interference between operators, which produces “flat directions” in the parameter space of certain observables.

As for the one-dimensional constraints considered in the previous section, the high-energy probes (even when considered with multiple operators at once) are unchanged by the interchange of flavor indices,\footnote{
This is strictly only the case when such an interchange is between two indices corresponding to fields with the same chirality. For $\mathcal{O}_{le,pqrs}$, by contrast, an interchange of the indices $q$ and $s$ exchanges the $\lab{SU}(2)_L$ doublet field with the singlet, which does have a mild impact on tree-level observables.} while the low-energy constraints change by factors of the lepton masses. 
In all the figures in this section, we are showing the flavor-index combination for which the low-energy constraints are most stringent.

%%%%%%%%%%%%%%%%%%%%%%%%%%%%%%%%%%%%%%%%%%%%%%%%%%%%%%%%%%%%%%%%%%%%%%%%%%
\subsection{Complementary Probes of Lepton Flavor Ans\"atze}
\label{subsec:summary_flavor}

We have seen that, depending on the effective operators responsible for flavor-violation, the scattering processes at a high-energy muon collider are particularly powerful at testing flavor-violating $\tau$\,--\,$\mu$ transitions, and in some of parameter space can even test beyond the reach of future measurements expected from Belle~II.
In contrast, the potential for high-energy muon colliders (or any more general high-energy collider experiment) for probing $\mu$\,--\,$e$ flavor transitions is much weaker than the extraordinary constraints attainable with precision muon decay or conversion experiments. 

\begin{figure}[t!]
\centering
\includegraphics[width=0.65\linewidth]{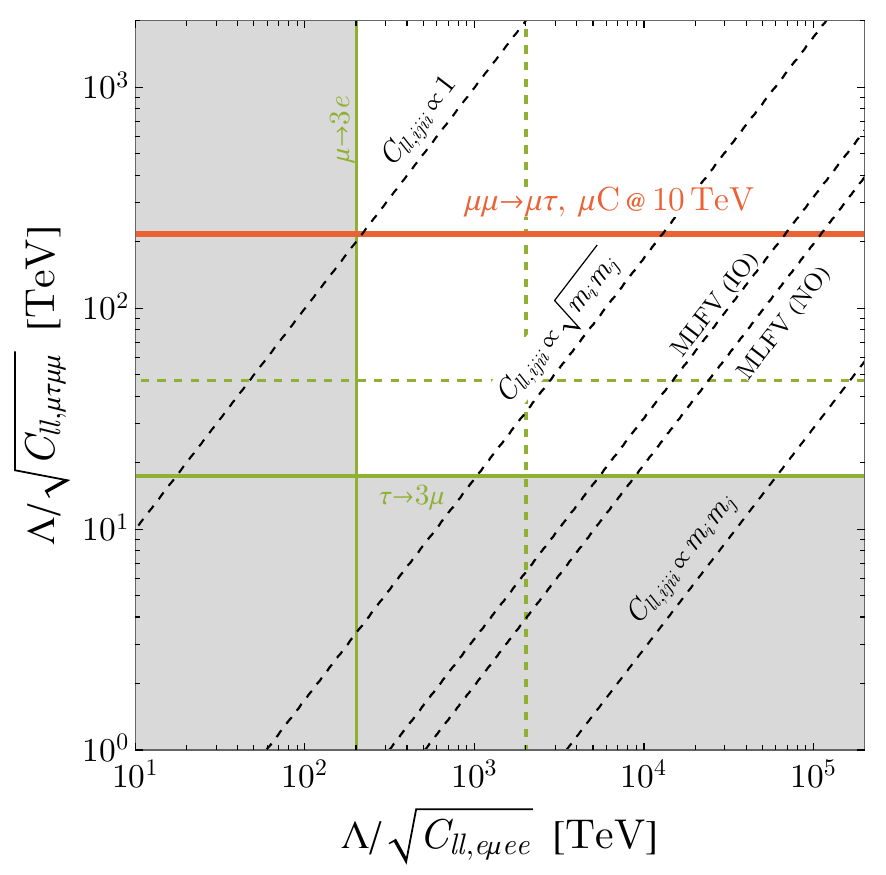}
\caption{Constraints on the left-handed four-fermion operator $C_{ll}$ with flavor indices $e\mu e e$ and $\mu\tau\mu\mu$. Muon collider constraint is shown in red, while precision measurements constraints from $\tau\to 3\mu$ and $\mu\to 3 e$ decay are shown in solid green (projected in dashed green). Different flavor ansatz are shown as dashed black lines. For anarchy ($C_{ll,ijii}\propto 1$), the projected $\mu\to 3e$ constraint on the scale of new physics is stronger than a 10 TeV muon collider, while for all other flavorful ansatz a 10 TeV muon collider would constrain higher scales.}
\label{fig:2dplots_flavor_highlight}
\end{figure}

\begin{figure}[hp]
\centering
\includegraphics[width=0.42\linewidth]{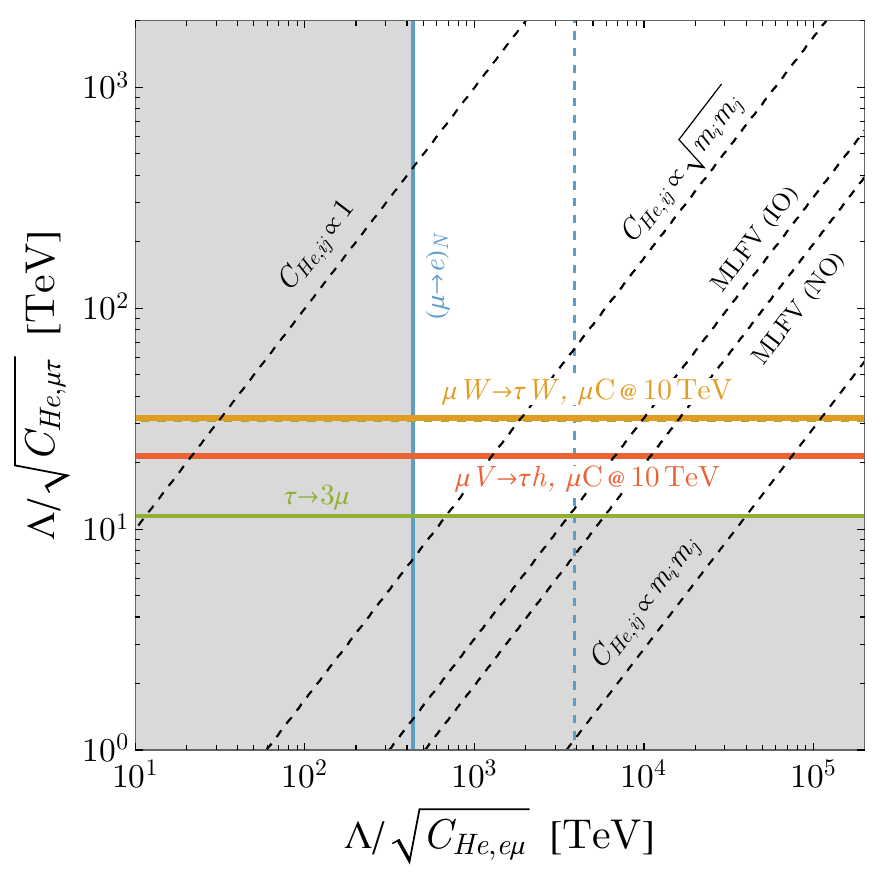}
\quad\quad
\includegraphics[width=0.42\linewidth]{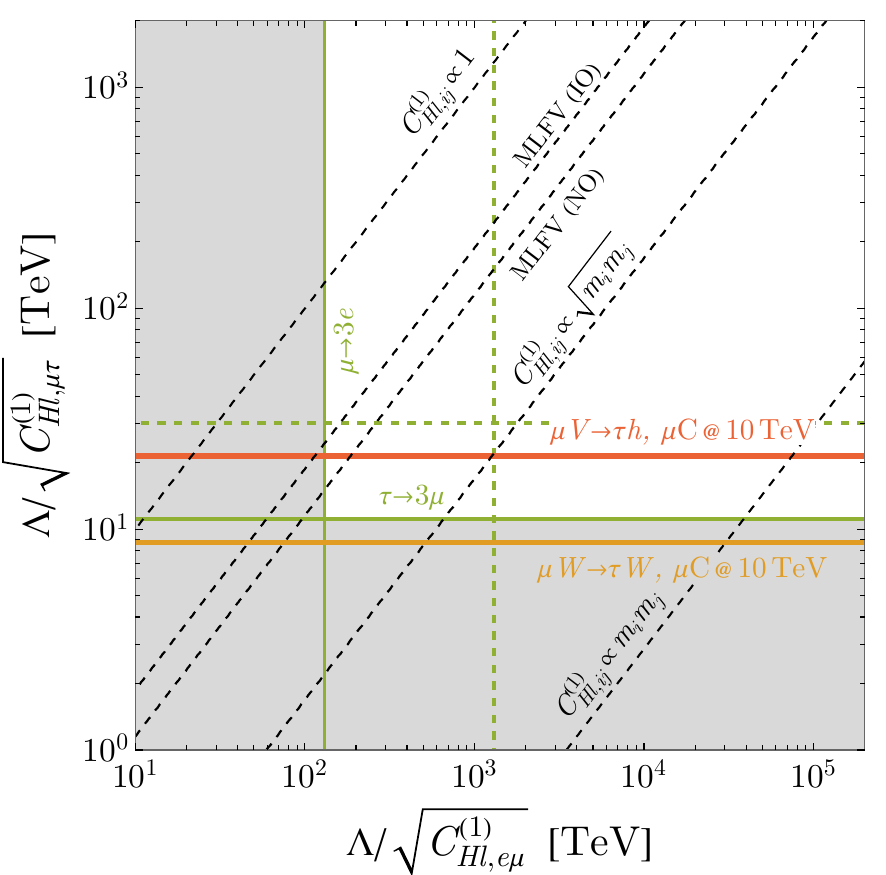} 
\\[0.5em]
\includegraphics[width=0.42\linewidth]{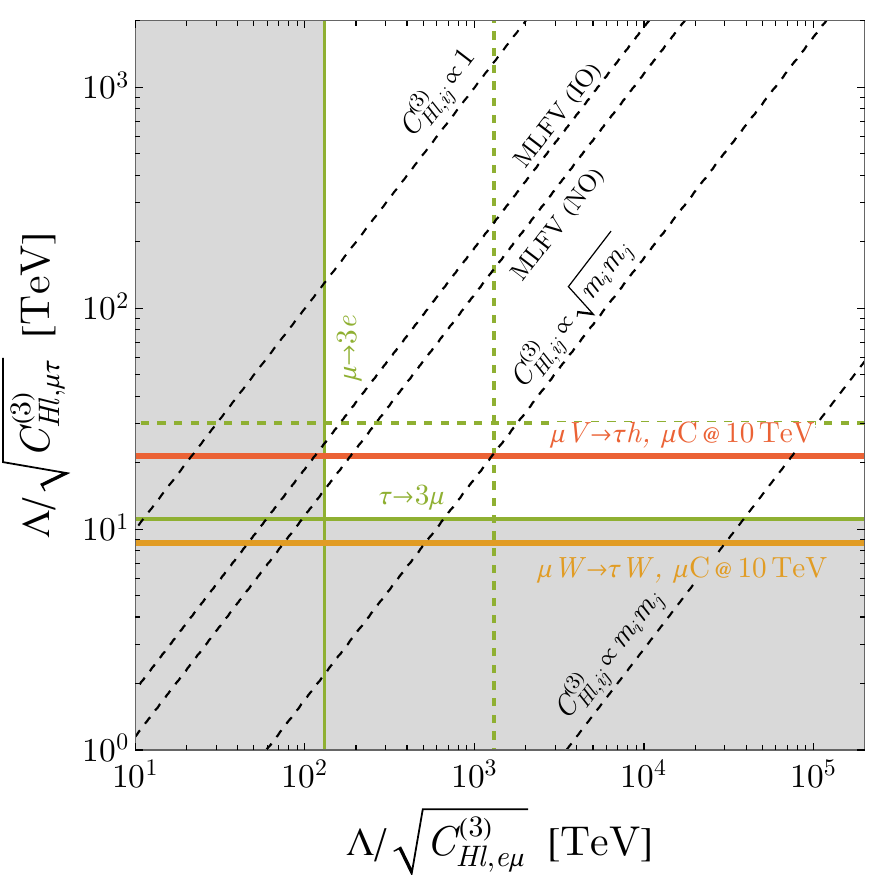}
\quad\quad
\includegraphics[width=0.42\linewidth]{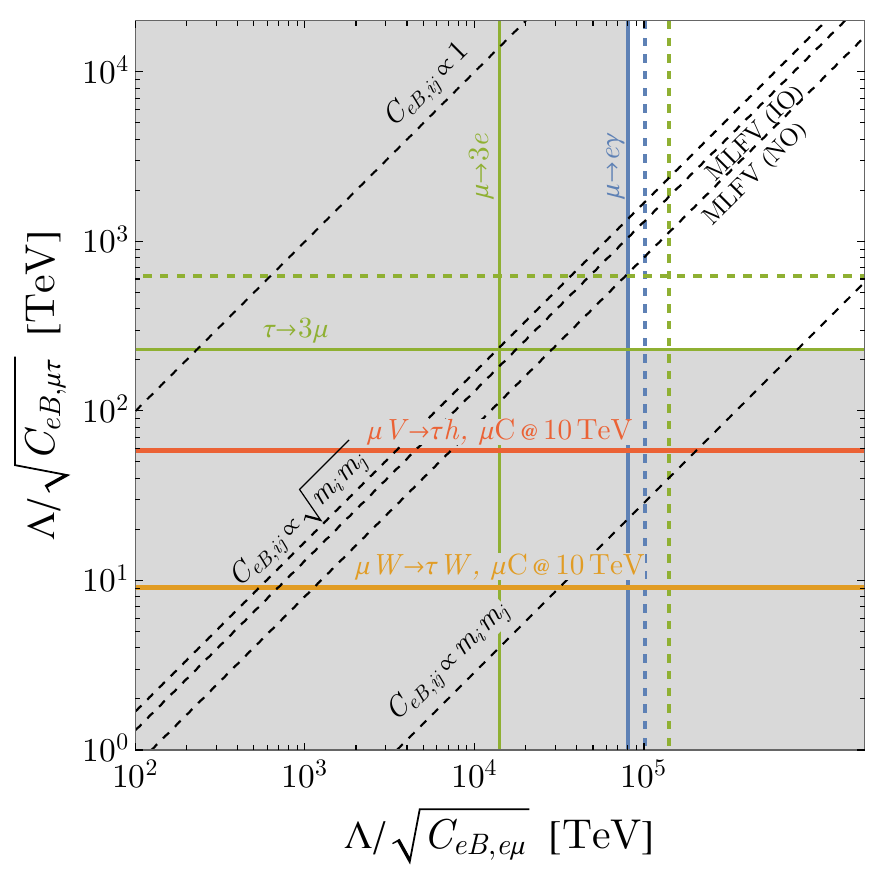} 
\\[0.5em]
\includegraphics[width=0.42\linewidth]{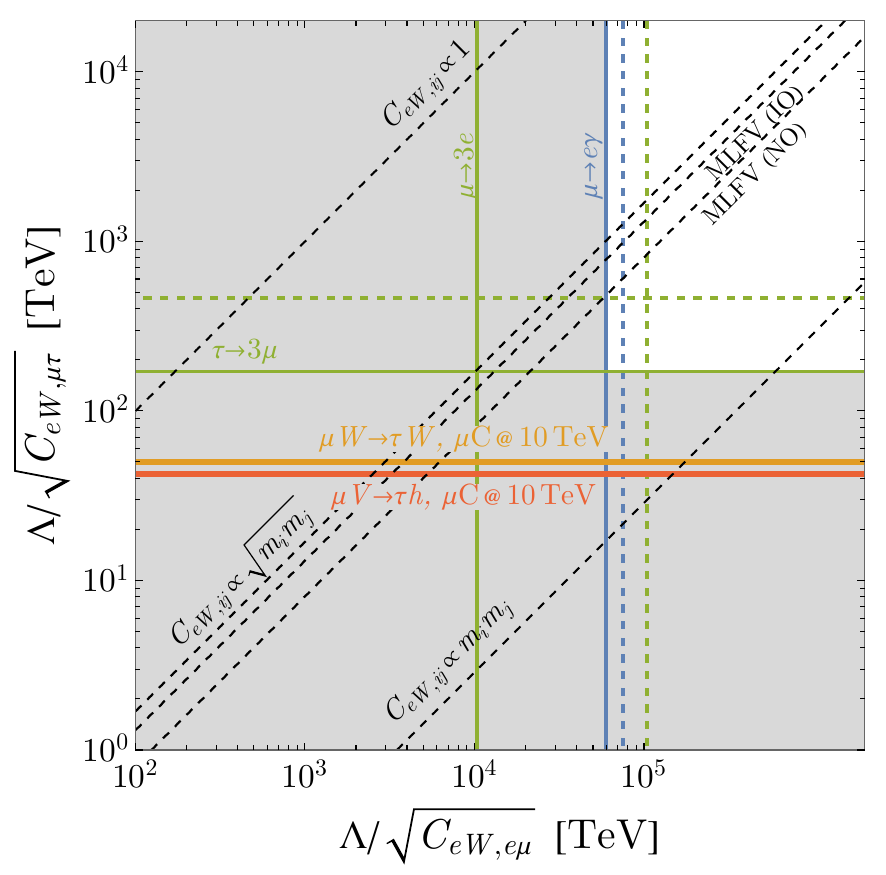}
\quad\quad
\includegraphics[width=0.42\linewidth]{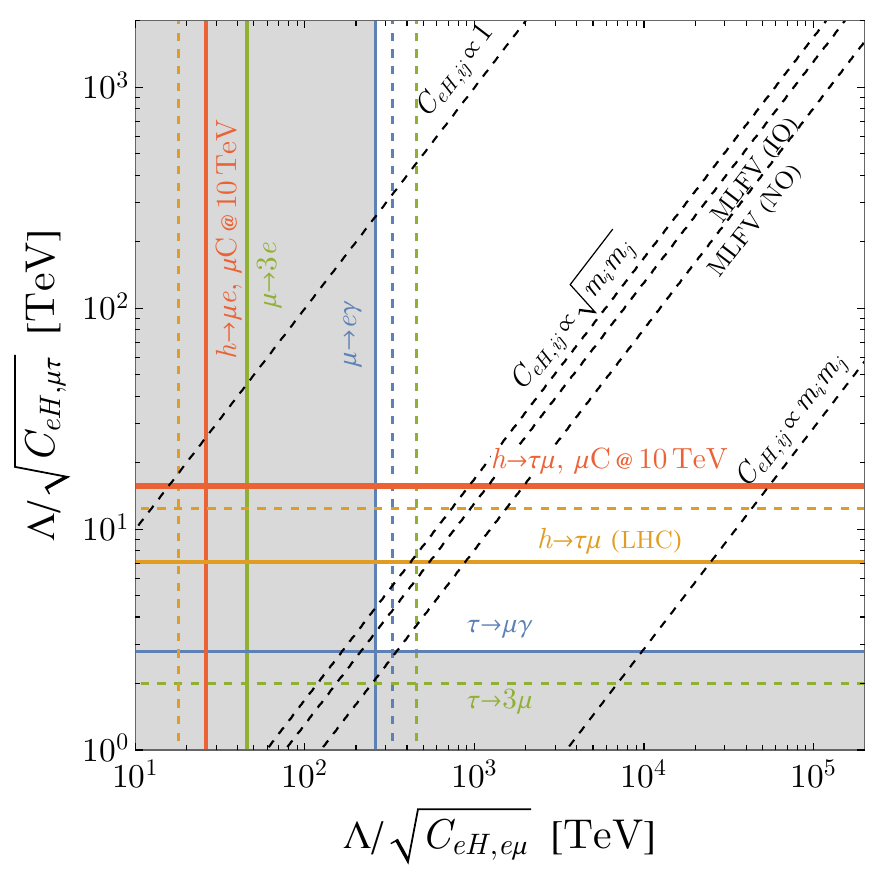}
\caption{Similar to Fig.~\ref{fig:2dplots_flavor_highlight}, but for a few different benchmark pair of operators involving the leptons and Higgs/electroweak bosons.}
\label{fig:2dplots_flavor}
\end{figure}

Nevertheless, it is a priori still unclear which set of constraints are the most important for constraining physics beyond the Standard Model: 
given that the charged lepton masses display enormous hierarchies, with $m_{\tau} \gtrsim 3000 \times m_e$, there is no reason to expect flavor-violating transitions to be of the same magnitude. Indeed, the SM quark sector exhibits flavor-changing transitions that vary by several orders of magnitude, and this hierarchical structure could extend to the lepton sector as well.

In the absence of an underlying model of flavor (which would dictate the pattern of flavor in both the SM and BSM interactions), we can only make different {\em ans\"atze} for the relative size of the flavor-violating transitions, and try to test them empirically. The most na\"ive ansatz is ``flavor anarchy'', which supposes that all the Wilson coefficients from new physics are $\mathcal{O}(1)$, independent of the flavor indices. In this case, $\mu$\,--\,$e$ transitions are the most powerful probe, given the incredible precision of the low-energy experiments. If there is hierarchical structure in the flavor-transitions that mimics the hierarchies in the charged lepton masses, however, $\tau$\,--\,$\mu$ transitions may be more constraining than $\mu$\,--\,$e$ transitions.

Another conservative ansatz is ``minimal lepton flavor violation'', or MLFV~\cite{Cirigliano:2005ck}. This imitates the philosophy of minimal flavor violation in the quark sector, where all BSM flavor transitions are related to the parameters of the mixing matrix in the Standard Model (in this case, the PMNS matrix). We work with the minimal field content scenario where there is no right-handed neutrino, and the flavor group of the lepton sector is $SU(3)_L\times SU(3)_E$. The flavor spurions are the charged lepton yukawa $\lambda_e$, which transforms as $\lambda_e\rightarrow V_R\lambda_eV_L^{\dagger}$, and the coefficient of the Weinberg operator $g_{\nu}$, which transforms as $g_{\nu}\rightarrow V_L^*g_{\nu}V_L^{\dagger}$. Since all the precision observables in our study have external charged lepton states, we work with the flavor basis where $\lambda_e$ is diagonal, leading to
\begin{equation}
    \lambda_e \propto {\rm{diag}}(m_e^2,m_{\mu}^2,m_{\tau}^2), \quad g_{\nu} \propto U^*{\rm{diag}}(m_{\nu_1}^2,m_{\nu_2}^2, m_{\nu_3}^2)U^{\dagger},
\end{equation}
where $U$ is the PMNS mixing matrix. From the transformation properties of the spurions, it is immediately clear that in MLFV, LFV operator coefficients are suppressed by more powers of $\lambda_e$ when there are more right-handed leptonic fields in the operator. In particular, operators with $LL$ lepton structure are suppressed by $\Delta \equiv g_{\nu}^{\dagger}g_{\nu}$, $LR$ ($RL$) by $\Delta\lambda_e^{\dagger}$ ($\lambda_e\Delta$), and $RR$ by $\lambda_e\Delta\lambda_e^{\dagger}$. The strong hierarchical structure in the numerical value of $\lambda_e$ means the flavor structure of the LFV operator coefficients in MLFV varies greatly between different operators. For $LL$ operators (such as $C_{Hl}^{(1)}$, $C_{Hl}^{(3)}$), $e\mu$ and $\mu\tau$ coefficients have a mild flavor hierarchy of about an order of magnitude, while for $RR$ operators (such as $C_{He}$, $C_{ee}$), the flavor hierarchy can be nearly five orders of magnitude.

\begin{figure}
\centering
\includegraphics[width=0.42\linewidth]{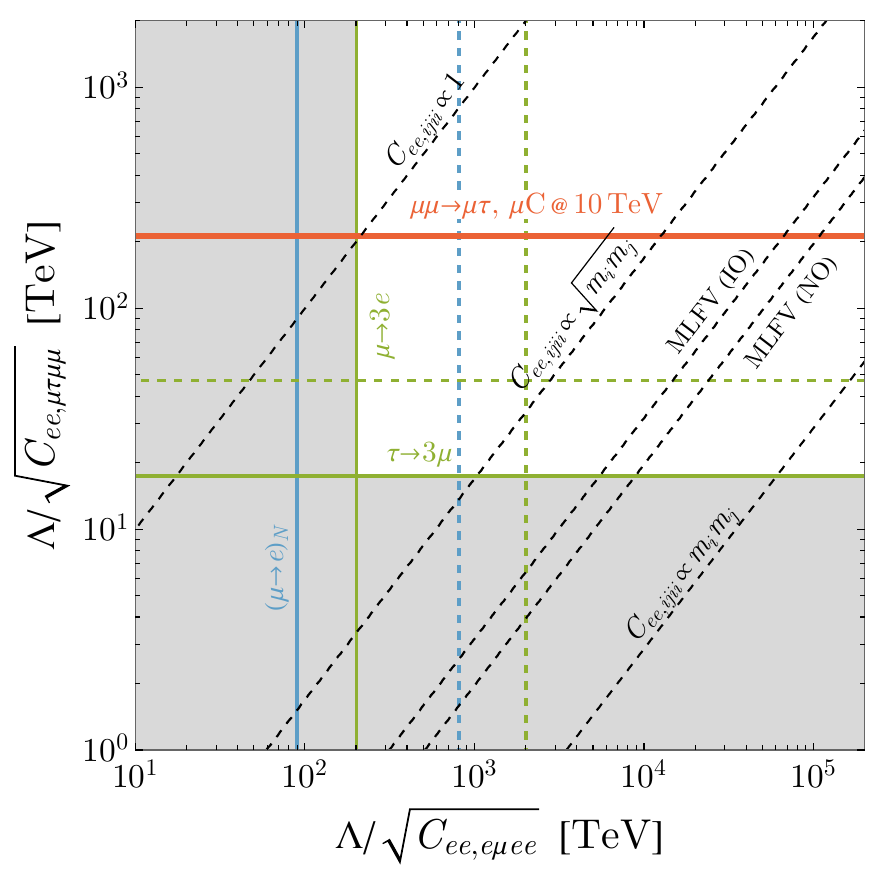}
\quad\quad
\includegraphics[width=0.42\linewidth]{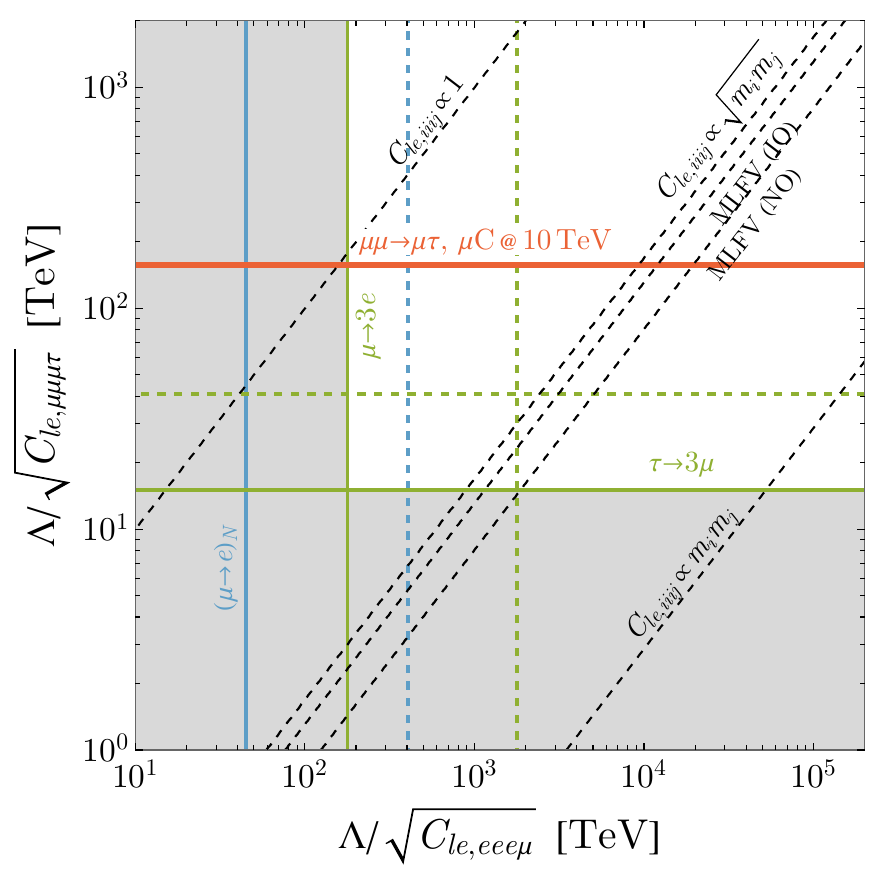} \\[0.5em]
\includegraphics[width=0.42\linewidth]{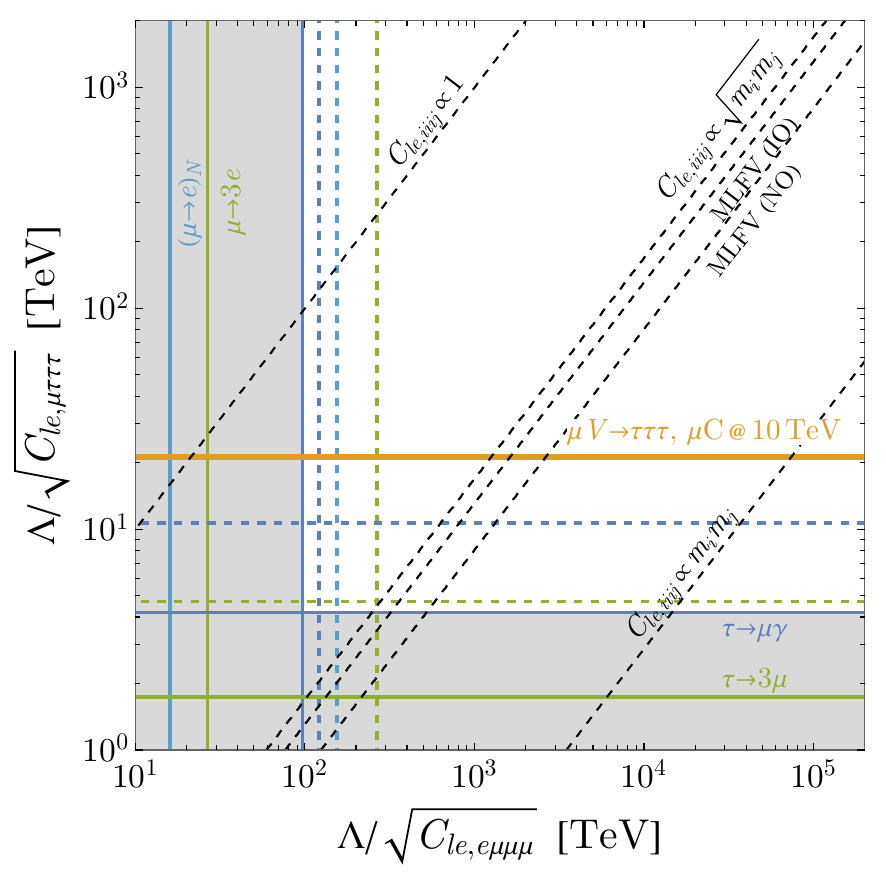}
\quad\quad
\includegraphics[width=0.42\linewidth]{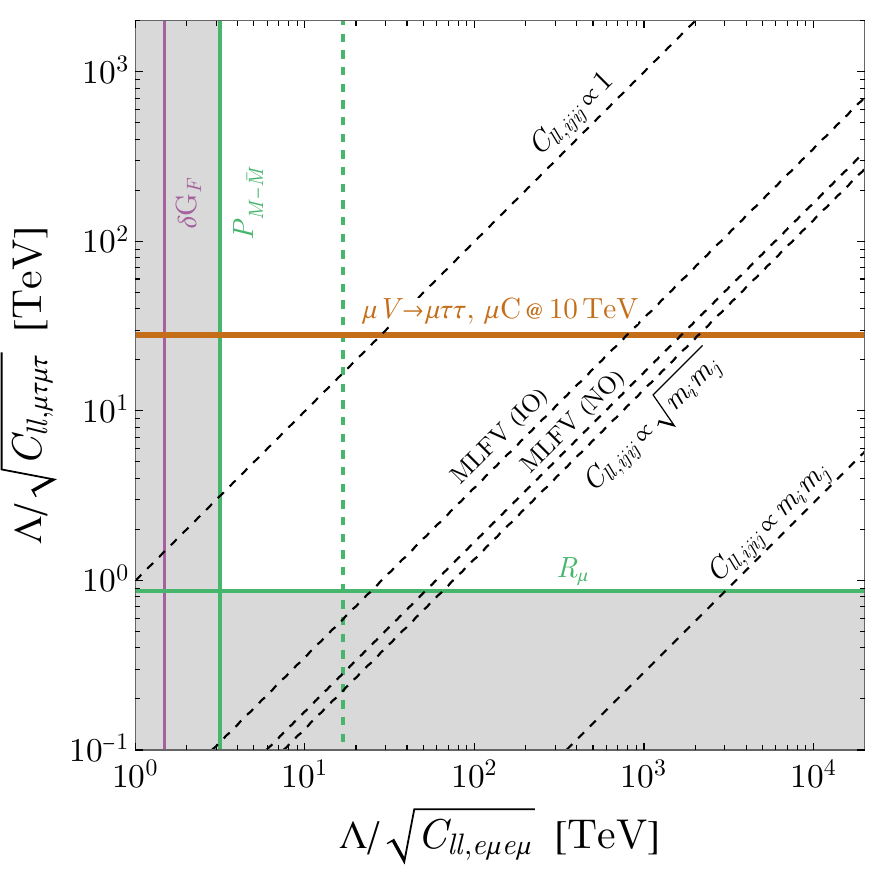}
\caption{Similar to Fig.~\ref{fig:2dplots_flavor}, but for combinations of the different four-fermion operators considered. A muon collider would outperform all projected constraints from low-energy probes under the assumption that $C_{ij} \propto \sqrt{m_i m_j}$, and for $C_{ll}$ it would surpass low-energy constraints even assuming flavor anarchy ($C_{ij} \propto 1$).}
\label{fig:2dplots_flavor_4ferm}
\end{figure}

To illustrate how different flavor {\em ans\"atze} could significantly change the comparison between different flavor transitions, we show the constraints on the scale associated with the operator $C_{ll}$ with flavor indices $e\mu ee$ and $\mu\tau\mu\mu$ in Fig.~\ref{fig:2dplots_flavor_highlight}. The precision measurements of $\mu\to 3e$ and $\tau \to 3\mu$ decays constrain $\Lambda/\sqrt{C_{ll,e\mu e e}}$ and $\Lambda/\sqrt{C_{ll,\mu\tau\mu\mu}}$ respectively, and a high-energy muon collider directly probes $\Lambda/\sqrt{C_{ll,\mu\tau\mu\mu}}$. These constraints are experimentally independent and therefore appear as vertical or horizontal lines on the plane. In  dashed black lines we show several flavor {\em ans\"atze} including those discussed earlier: $C_{ij} = 1$ (flavor anarchy), $C_{ij} \propto \sqrt{m_i m_j}$, $C_{ij} \propto m_i m_j$, MLFV with normal ordering in the neutrino masses (NO), and MLFV with inverted ordering (IO). NO and IO differ in the measured PMNS matrix elements and neutrino masses that go into $g_{\nu}$. All experimental values are taken from \cite{Esteban:2020cvm}. The flavor {\em ans\"atze} relate the dimensionless numbers $C_{ll,e\mu e e}$ and $C_{ll,\mu\tau\mu\mu}$, and they allow a comparison of the constraining power of these experiments on the underlying scale of new physics, $\Lambda$. For example, with flavor anarchy, we see that a 10 TeV muon collider has the same constraining power on $\Lambda$ as the current bound on $\mu\to 3 e$, while for all other flavorful ansatz, a 10 TeV muon collider constrains $\Lambda$ better than the projected bound from $\mu\to 3e$ (intersections of the muon collider line with the ansatz lines are further to the right than the projected $\mu\to 3e$ line). In Figs.~\ref{fig:2dplots_flavor} and \ref{fig:2dplots_flavor_4ferm}, we consider all other operators studied in this paper, and for each operator compare the constraints on the $ij = e\mu$ and $ij = \mu\tau$ flavor indices. The different constraints are shown in the same manner as in Section~\ref{subsec:summary_2d}, but for clarity, we omit some of the subleading constraints on each operator. As illustrated in Figs.~\ref{fig:2dplots_flavor} and \ref{fig:2dplots_flavor_4ferm}, the comparative strength of a high-energy muon collider varies greatly between different operators and flavor ansatz, but in general it is the most powerful constraint for four fermion operators, where there exist examples where a muon collider would be more constraining than precision experiments even in the flavor anarchy scenario (see e.g. the bottom right panel of Fig.~\ref{fig:2dplots_flavor_4ferm}).

While the exact comparison is different for each operator, generally, the $\tau$\,--\,$\mu$ probes at a high-energy muon collider are markedly the most important when $C_{ij} \propto m_i m_j$, while the constraints from muon colliders and future low-energy $\mu$\,--\,$e$ transition probes are competitive if $C_{ij} \propto \sqrt{m_i m_j}$. In the flavor anarchy scenario, the $\mu$\,--\,$e$ transitions are always the most important. The MLFV scenario exhibits all three possibilities depending on the chirality structure of the operators considered. This reinforces the preliminary results of Ref.~\cite{AlAli:2021let}, which considered only a single four-fermion operator and less detailed simulations of the signal and backgrounds. While more sophisticated {\em ans\"atze} or models of flavor might predict different patterns, it is clear that the powerful constraints on $\tau$\,--\,$\mu$ transitions can test a broad swath of interesting flavorful models of new physics.

%%%%%%%%%%%%%%%%%%%%%%%%%%%%%%%%%%%%%%%%%%%%%%%%%%%%%%%%%%%%%%%%%%%%%%%%%%
%%%%%%%%%%%%%%%%%%%%%%%%%%%%%%%%%%%%%%%%%%%%%%%%%%%%%%%%%%%%%%%%%%%%%%%%%%
\section{Conclusions \label{sec:conclusion}}

In this work, we compared the sensitivity of a future muon collider versus precision experiments in probing the scale of lepton flavor violation within the SMEFT framework. Since lepton flavor is already broken in the Standard Model, generic extensions of it are naturally expected to induce additional LFV effects, and it is important to quantify the discovery potential of such observables at future colliders. 

Our analysis focused on the nine dimension-six operators involving only leptons, electroweak gauge bosons, and the Higgs (see Eq.~\eqref{eq:smeft_operators}). Each of these operators can contribute to flavor violation between different generations and is sensitive to different constraints. In particular, while $\mu$\,--\,$e$ LFV is already stringently constrained by low-energy precision experiments, the dominant sensitivity to flavor-violating transitions involving the $\tau$ arises from colliders. For this reason, our study concentrated on collider probes of $\tau$\,--\,$e$ and $\tau$\,--\,$\mu$ LFV, where a muon collider offers unique opportunities.

We first calculated the contribution of these SMEFT operators to various low-energy lepton-flavor-violation observables. 
In doing so, we systematically matched the SMEFT operators onto the LEFT ones and incorporated the running effects for an accurate prediction. 
This systematic approach enables a clear comparison of the sensitivity of various low-energy observables to different SMEFT operators and LFV transitions, as shown in Figs.~\ref{fig:barchart} -- \ref{fig:barchart_4ferm}.
In particular, we find that while certain observables can probe very large scales of new physics in some LFV processes, the bounds from low-energy observables are considerably weaker for $\tau$\,--\,$e$ and $\tau$\,--\,$\mu$ LFV transitions. These can instead be probed in a high-energy muon collider.

To assess the sensitivity of a future muon collider for $\tau$\,--\,$\mu$ LFV operators, we simulated a variety of LFV processes, along with their SM backgrounds. We used \texttt{Delphes} fast simulation, and performed simple cut-and-count analyses involving kinematic variables, such as the transverse momentum of reconstructed objects, angular separations between visible objects and the missing four-momentum, and the $M_{T2}$ (or ``stransverse mass'') of the systems. 
Our main results are presented in Figs.~\ref{fig:barchart}--\ref{fig:2dplots_tamu_4ferm}, which illustrate the projected reach of different processes in probing the characteristic scales of new physics associated with each operator. 
For clarity, we provide below a brief summary of the processes and the operators to which they are sensitive: 
\begin{itemize}

    \item $h \to \tau l$: LFV Higgs decays provide a clean probe of the $\mathcal{O}_{eH}$ operator, which can introduce off-diagonal Yukawa couplings for the Higgs. 
    As shown in Fig.~\ref{fig:lfv_higgs_decay_summary}, $h \to \tau l$ at a muon collider can improve existing bounds on the LFV Yukawa coupling of the Higgs to $\tau$\,--\,$e$ and $\tau$\,--\,$\mu$ by roughly an order of magnitude compared to the LHC. 
    
    \item $\mu V \to \tau h$ ($V=Z,\gamma$): This process is sensitive to the operators $\mathcal{O}_{eB}$, $\mathcal{O}_{eW}$, $\mathcal{O}_{Hl}^{(1)}$, $\mathcal{O}_{Hl}^{(3)}$, and $\mathcal{O}_{He}$, and can probe scales up to $\Lambda \sim \mathcal{O}(10$–$100)$~TeV at a $10~\text{TeV}$ muon collider.

    \item $\mu W \to \tau W$: This process is also sensitive to the operators $\mathcal{O}_{eB}$, $\mathcal{O}_{eW}$, $\mathcal{O}_{Hl}^{(1)}$, $\mathcal{O}_{Hl}^{(3)}$, and $\mathcal{O}_{He}$, but to different linear combinations and thus provides complementary information. It also probes scales of $\mathcal{O}(10$–$100)$~TeV at a $10~\text{TeV}$ muon collider, depending on the operator.

    \item $\mu\mu \to \mu\tau$:
    This process is sensitive to four-fermion operators with a single $\mu$\,--\,$\tau$ transition. Since here the final state objects carry the full center-of-mass energy of the collision, this process leads to the strongest direct bounds, with the sensitivity of a $10~\textrm{TeV}$ collider reaching $\sim 200~\textrm{TeV}$.

    \item $\mu V \to \tau\tau\tau$:
    This process is sensitive to four-fermion operators involving a single muon and multiple $\tau$s. At $\sqrt{s} = 10~\textrm{TeV}$, the sensitivity reaches the $\sim 20 - 35~\textrm{TeV}$ scale.

    \item $\mu^{\mp} V \to \mu^{\pm}\tau^\mp\tau^\mp$:
    This process is uniquely sensitive to $\Delta L = 2$ four-fermion operators, with two separate $\mu$\,--\,$\tau$ transitions. The sensitivity  is at the $20-30~\textrm{TeV}$ level.
    
\end{itemize}

With the exception of the dipole operators $C_{eB}$ and $C_{eW}$, these sensitivities generically outperform the sensitivity from existing low-energy direct probes in $\tau$ decays. This sensitivity would allow for a robust follow-up to any potential discovery with the improved sensitivity at Belle~II. Additionally, for many operators involving the $\tau$, particularly four-fermion operators and combinations which are mass-suppressed at low energy, a muon collider would outperform even future limits from planned experiments.

To facilitate comparisons with the tests of $\mu$\,--\,$e$ flavor violation, we compared the reach of different searches under various flavor {\em ans\"atze} for the Wilson coefficients. 
The scales quoted above correspond to an anarchic ansatz, with all $C_{ij} = 1$. 
Beyond this, we considered several motivated alternatives: the assumption that $C_{ij} \propto \sqrt{m_i m_j}$, $C_{ij} \propto m_i m_j$, and the ansatz of minimal lepton flavor violation where the flavor structure of $C_{ij}$ depends greatly on the chirality structure of the operator.
For operators involving the leptons and Higgs/electroweak bosons, a muon collider is competitive with low-energy probes under the $C_{ij} \propto \sqrt{m_i m_j}$ ansatz in most cases, while for four-fermion operators, a muon collider outperforms all projected constraints from low-energy experiments under the $C_{ij} \propto \sqrt{m_i m_j}$ ansatz and even the $C_{ij} = 1$ ansatz in the case of $C_{ll}$ operator.

There are a number of clear directions for future study.
First, a crucial ingredient in obtaining a reliable estimate of the muon collider reach for rare processes such as LFV is a detailed understanding and accurate modeling of detector performance. 
While this is the subject of ongoing experimental efforts, dedicated studies of jet properties and their modeling are particularly well-motivated.
This includes both understanding the capabilities of detectors for studying jet substructure, and for systematic evaluations of tagging efficiencies. Our searches would benefit in particular from improved understanding of the capabilities in distinguishing boosted $h\to b\bar{b}$ jets, jets from hadronic $\tau$ decays, and hadronic jets originating from $W$ bosons or lighter QCD objects. Understanding the capability for tagging the charge of tau jets also has the potential to improve some of our studies.

Additionally, it is worthwhile to explore detector design strategies that could enhance the sensitivity of a high-energy muon collider to the LFV signals studied in this work. 
For example, several of the processes we consider ($\mu V \to \tau h$, $\mu V \to \tau \tau \tau$ and $\mu^{\mp} V \to \mu^{\pm}\tau\tau$) involve collinear radiation of a neutral electroweak boson off one of the muon beams, with the radiating muon traveling down the beam pipe.
It has recently been proposed \cite{Ruhdorfer:2023uea, Ruhdorfer:2024dgz} that a dedicated forward muon detector could significantly improve the tagging efficiency for such events, even with relatively light instrumentation. Given that any LFV signal would constitute an unambiguous indication of new physics beyond the Standard Model, it is highly motivated to assess the impact of such a forward detector on the sensitivity to LFV processes, as well as to investigate other potential detector augmentations that could further strengthen LFV searches.

While the SMEFT provides a useful framework for interpreting experimental bounds, we stress that it should ultimately be regarded as an intermediate step towards a long-term goal of constraining or discovering an explicit UV model.
In particular, since the bounds on the EFT cutoff are not parametrically beyond the center-of-mass energy of the collider, it is important to critically assess the applicability of the SMEFT in this intermediate regime. This is especially true in the context of models where the flavor-violating operators may be suppressed by small parameters.
It would be interesting to reassess the reach of the processes considered above in the context of explicit models.

Finally, we demonstrated that the assessment of the constraints on flavor-violating operators depends strongly on the choice of flavor ansatz. To go beyond MLFV, we introduced several ad-hoc {\em ans\"atze} for the relative size of different flavor-violating operators based on the charged lepton masses. It would be useful, however, to put these assumptions on firmer ground. One interesting direction would be to consider the predictions of different explicit models that {\em explain} the SM flavor hierarchies (see, for example, Refs.~\cite{Feldmann:2006jk, Bordone:2019uzc, Asadi:2023ucx, Cornella:2024jaw}). A more systematic investigation of motivated textures for BSM flavor violation would provide a more robust picture of the discovery potential of future colliders.

In summary, the combination of high center-of-mass energy and initial states dominated by second-generation leptons or electroweak gauge bosons makes a muon collider uniquely suited to probe LFV effects that are difficult to access at other high-energy colliders or even at the most sensitive small-scale experiments. As experimental efforts advance toward establishing the feasibility of such a machine, it becomes increasingly important to catalog additional signals and motivated BSM scenarios for which a high-energy muon collider offers truly distinctive discovery potential.

With experimental efforts toward a future muon collider gaining momentum, it is timely to explore the potential of such a machine to uncover new physics signals, including lepton-flavor violation. Beyond their intrinsic theoretical interest, these LFV signatures can provide valuable benchmarks to guide detector optimization and help shape the physics case for the collider from the outset. 
More broadly, studies such as ours illustrate how targeted analyses can steer the development of future searches for physics beyond the Standard Model, ensuring that the discovery potential of next-generation colliders is fully realized.

%%%%%%%%%%%%%%%%%%%%%%%%%%%%%%%%%%%%%%%%%%%%%%%%%%%%%%%%%%%%%%%%%%%%%%%%%%
%%%%%%%%%%%%%%%%%%%%%%%%%%%%%%%%%%%%%%%%%%%%%%%%%%%%%%%%%%%%%%%%%%%%%%%%%%
\section*{Acknowledgments}

We are grateful to Wolfgang Altmannshofer, Andr\'e de Gouv\^ea, Simon Knapen and Kevin Langhoff for useful discussions.
PA is partly supported by the U.S. Department of Energy grant number DE-SC0010107.
HB is supported in part by DOE grant DE-SC0013607 and by the National Science Foundation MPS-Ascend Postdoctoral Research Fellowship under Award No. 2503442. 
The work of KF was supported by: DOE grant DE-SC0013607, the Harvard GSAS Merit Fellowship, and the Miller Institute for Basic Research in Science, University of California Berkeley.
The work of SH was supported by the NSF grant PHY-2309456 and by the US-Israeli BSF grant 2016153. QL is supported in part by the NSF grant PHY-2210498 and PHY-2514611 and by the Simons Foundation. 
KF also thanks the The Munich Institute for Astro-, Particle and BioPhysics and the Aspen Center for Physics (which is supported by NSF grant PHY-2210452, Simons Foundation grant (1161654, Troyer), and Alfred P. Sloan Foundation grant G-2024-22395) for hospitality while working on this project. 
The computations in this paper were run on the FASRC Cannon cluster supported by the FAS Division of Science Research Computing Group at Harvard University and on the computing cluster at the Cornell Laboratory for Accelerator-based Sciences and Education (CLASSE).

%%%%%%%%%%%%%%%%%%%%%%%%%%%%%%%%%%%%%%%%%%%%%%%%%%%%%%%%%%%%%%%%%%%%%%%%%%
{\small
%\clearpage
\bibliography{lfv-muc-refs}
\bibliographystyle{utphys}
}

\end{document}

%% file: precision-summary-table.tex
% !TeX root = lfv-muc-draft-v0.tex

\begin{table}[!ht]
\centering
\renewcommand{\arraystretch}{1.2}
\begin{tabular}{c|ccc}
Observable & Current Limit (Reference) & Projected Limit (Reference)
\\[0.25em]
\hline \\[-1em]
$\BR(\mu \to e \gamma)$ & 
%$4.2 \times 10^{-13}$ (\textsc{meg}~\cite{TheMEG:2016wtm}) &
$1.5 \times 10^{-13}$ ({\footnotesize MEG-II~\cite{MEGII:2025gzr}}) & 
$6.0 \times 10^{-14}$ ({\footnotesize MEG-II~\cite{Baldini:2018nnn}}) \\
$\BR(\tau \to e \gamma)$ & 
$3.3 \times 10^{-8}$ ({\footnotesize BaBar~\cite{Aubert:2009ag}}) & 
$3.0 \times 10^{-9}$ ({\footnotesize Belle~II~\cite{Kou:2018nap}}) \\
$\BR(\tau \to \mu \gamma)$ & 
$4.2 \times 10^{-8}$ ({\footnotesize Belle~\cite{Belle:2021ysv}}) & 
$1.0 \times 10^{-9}$ ({\footnotesize Belle~II~\cite{Kou:2018nap}}) \\[0.75em]
%%%%%%%%%%%%%%%%%%%%
$\BR(\mu \to 3 e)$ & 
$1.0 \times 10^{-12}$ ({\footnotesize SINDRUM~\cite{Bellgardt:1987du}}) &
$\sim 1 \times 10^{-16}$ ({\footnotesize Mu3e~\cite{Mu3e:2020gyw}}) \\
$\BR(\tau \to 3 e)$ & 
$2.7 \times 10^{-8}$ ({\footnotesize Belle~\cite{Hayasaka:2010np}}) &
$4.6 \times 10^{-10}$ ({\footnotesize Belle~II~\cite{Kou:2018nap}}) \\
$\BR(\tau \to 3 \mu)$ & 
$1.9 \times 10^{-8}$ ({\footnotesize Belle~II~\cite{Belle-II:2024sce}}) &
$3.5 \times 10^{-10}$ ({\footnotesize Belle~II~\cite{Kou:2018nap}}) \\[0.75em]
%%%%%%%%%%%%%%%%%%%%
$\BR(\tau^- \to \mu^- e^+ e^-)$ & 
$1.6 \times 10^{-8}$ ({\footnotesize Belle~II~\cite{Belle-II:2025urb}}) & 
$3 \times 10^{-10}$ ({\footnotesize Belle~II~\cite{Kou:2018nap}}) \\
$\BR(\tau^- \to e^- \mu^+ \mu^-)$ & 
$2.4 \times 10^{-8}$ ({\footnotesize Belle~II~\cite{Belle-II:2025urb}}) & 
$5 \times 10^{-10}$ ({\footnotesize Belle~II~\cite{Kou:2018nap}}) \\[0.75em]
%%%%%%%%%%%%%%%%%%%%
$\CR(\mu \to e)_{\textrm{Ti}}$ & 
$4.3 \times 10^{-12}$ ({\footnotesize SINDRUM-II~\cite{SINDRUMII:1993gxf}}) & 
--- \\
$\CR(\mu \to e)_{\textrm{Au}}$ & 
$7.0 \times 10^{-13}$ ({\footnotesize SINDRUM-II~\cite{Bertl:2006up}}) & 
--- \\
$\CR(\mu \to e)_{\textrm{Al}}$ & 
--- & 
$\sim 1 \times 10^{-17}$ ({\footnotesize Mu2e, COMET~\cite{Bartoszek:2014mya, COMET:2018auw}}) \\[0.75em]
%%%%%%%%%%%%%%%%%%%%
$P(M_{\mu} \to \bar{M}_{\mu})$ & $8.3 \times 10^{-11}$~({\footnotesize MACS, PSI~\cite{Willmann:1998gd}}) & $\sim 1 \times 10^{-13}$~({\footnotesize MACE, J-PARC~\cite{Bai:2024skk}}) \\
$\delta G_F$ ($\mu \to e\nu\nu$) & $(1.2 \pm 17.0) \times 10^{-5}$ & --- \\
$R_{\mu}$ ($R_{\mu}^{\textrm{SM}} = 0.9726$) & $0.9735 \pm 0.0026$ ({\footnotesize Belle~II}~\cite{Belle-II:2024vvr}) & ---
\\[0.5em]
\hline\\[-1.25em]
%%%%%%%%%%%%%%%%%%%%
$\BR(Z \to \mu e)$ & 
$1.9 \times 10^{-7}$ ({\footnotesize CMS~\cite{CMS:2025wqy}}) &
$\sim 4 \times 10^{-8}$ ({\footnotesize HL-LHC*}) \\
$\BR(Z \to \tau e)$ & 
$5.0 \times 10^{-6}$ ({\footnotesize ATLAS~\cite{ATLAS:2021bdj}}) &
$\sim 1 \times 10^{-6}$ ({\footnotesize HL-LHC*}) \\
$\BR(Z \to \tau \mu)$ & 
$6.5 \times 10^{-6}$ ({\footnotesize ATLAS~\cite{ATLAS:2021bdj}}) &
$\sim 1 \times 10^{-6}$ ({\footnotesize HL-LHC*}) \\[0.75em]
%%%%%%%%%%%%%%%%%%%%
$\BR(h \to \mu e)$ & 
$4.4 \times 10^{-5}$ ({\footnotesize CMS~\cite{CMS:2023pte}}) &
$\sim 1 \times 10^{-5}$ ({\footnotesize HL-LHC*})\\
$\BR(h \to \tau e)$ & 
$2.0 \times 10^{-3}$ ({\footnotesize ATLAS~\cite{ATLAS:2023mvd}}) & 
$\sim 5 \times 10^{-4}$ ({\footnotesize HL-LHC*}) \\
$\BR(h \to \tau\mu)$ & 
$1.8 \times 10^{-3}$ ({\footnotesize ATLAS~\cite{ATLAS:2023mvd}}) & 
$\sim 5 \times 10^{-4}$ ({\footnotesize HL-LHC*}) \\
%
%\hline
\end{tabular}
\caption{Summary of current and future constraints on lepton-flavor-violating processes, all at $90\%~\textrm{C.L}$. The HL-LHC projections (marked with *) are obtained by scaling the LHC results to the projected luminosity of $3~\rm{ab}^{-1}$, assuming the searches remain statistics-limited. 
The constraint on $\delta G_F$ is determined from electroweak scale data; see text for details. 
The quoted value of $R_{\mu}$ combines the Belle~II determination with earlier values in the HFLAV combination~\cite{HFLAV:2022esi}.
}
\label{tab:precision_bounds}
\end{table}

%% file: lfv-higgs-summary-table.tex
\begin{table}[ht!]
\renewcommand{\arraystretch}{1.2}
\centering
\begin{tabular}{c c c | c c c}
\hline \hline 
\multicolumn{3}{c}{Process} & $\sigma \times \textrm{BR}~(\textrm{fb})$ & ~Efficiency~ & $N_{\textrm{events}}$ ($10\,\textrm{ab}^{-1}$) \\ 
\hline \hline
\multirow{8}{*}{$\mu e$} 
& \multicolumn{2}{c|}{Signal, $C_{eH,21} / \Lambda^2 = 1 / (10\,\textrm{TeV})^2$} 
& 0.58 & 0.47 & 2706 \\
\Cline{0.5pt}{2-6}
& \multirow{3}{*}{$\mu\nu e\nu$} 
  & (annihilation) & 14.6 & $2.8\times 10^{-5}$ & 4.1 \\
& & (charged VBF)  & 3.7 & $5.5\times 10^{-3}$ & 200 \\
& & (neutral VBF)  & 8.0 & $2.0\times 10^{-4}$ & 16 \\
\Cline{0.2pt}{2-6}
& \multirow{3}{*}{$\tau^{\pm}(\mu\nu\nu) \tau^{\mp}(e\nu\nu)$} 
  & (annihilation) & 0.042 & 0.0 & 0 \\
& & (charged VBF)  & 0.60 & $7.5\times 10^{-4}$ & 4.5 \\
& & (neutral VBF)  & 1.7 & 0.0 & 0 \\
\Cline{0.2pt}{2-6}
& \multicolumn{2}{c|}{Total Background} & & & 225 \\ 
\hline \hline
\multirow{8}{*}{$\tau e$} 
& \multicolumn{2}{c|}{Signal, $C_{eH,31} / \Lambda^2 = 1 / (10\,\textrm{TeV})^2$} 
& 0.25 & 0.53 & 1304 \\
\Cline{0.5pt}{2-6}
& \multirow{3}{*}{$\tau\nu e\nu$} 
  & (annihilation) & 0.14 & $5.9\times 10^{-5}$ & 0.1 \\
& & (charged VBF)  & 1.2 & $7.2\times 10^{-2}$ & 887 \\
& & (neutral VBF)  & 2.8 & $1.8\times 10^{-3}$ & 51 \\
\Cline{0.2pt}{2-6}
& \multirow{3}{*}{$\tau^{\pm}(\textrm{had.}) \tau^{\mp}(e\nu\nu)$} 
  & (annihilation) & 0.15 & 0.0 & 0 \\
& & (charged VBF)  & 2.9 & $9.4\times 10^{-2}$ & 2800 \\
& & (neutral VBF)  & 9.0 & $5.0 \times 10^{-5}$ & 4.5 \\
\Cline{0.2pt}{2-6}
& \multicolumn{2}{c|}{Total Background} & & & 3743 \\ 
\hline \hline
\multirow{8}{*}{$\tau\mu$} 
& \multicolumn{2}{c|}{Signal, $C_{eH,32} / \Lambda^2 = 1 / (10\,\textrm{TeV})^2$} 
& 0.29 & 0.53 & 1546 \\
\Cline{0.5pt}{2-6}
& \multirow{3}{*}{$\tau\nu\mu\nu$} 
  & (annihilation) & 7.8 & $2.0\times 10^{-3}$ & 157 \\
& & (charged VBF)  & 1.6 & $6.8\times 10^{-2}$ & 1050 \\
& & (neutral VBF)  & 3.4 & $1.7\times 10^{-3}$ & 59 \\
\Cline{0.2pt}{2-6}
& \multirow{3}{*}{$\tau^{\pm}(\textrm{had.}) \tau^{\mp}(\mu\nu\nu)$} 
  & (annihilation) & 0.16 & $1.5\times 10^{-5}$ & 0 \\
& & (charged VBF)  & 3.3  & $9.4\times 10^{-2}$ & 3141 \\
& & (neutral VBF)  & 10.4 & $9.6 \times 10^{-5}$ & 10 \\
\Cline{0.2pt}{2-6}
& \multicolumn{2}{c|}{Total Background} & & & 4417 \\ 
\hline \hline 
\end{tabular}
\caption{Summary of the expected number of events passing the selection cuts described in the text for the LFV Higgs decay signal and backgrounds. The reported cross sections account for the pre-selection requiring the correct number of leptons or jets in the central region ($|\eta|<2.5$) with $p_T > 30\,\textrm{GeV}$, for a fair comparison across the different backgrounds. For each channel, we include only one SMEFT operator $C_{eH,ij}$, noting that $C_{eH,ji}$ yields an identical contribution. }
\label{tab:lfv_higgs_summary}
\end{table}

%% file: high-energy-summary-table.tex
\begin{table}[p]
\renewcommand{\arraystretch}{1.2}
\centering
\scalebox{0.85}{
\begin{tabular}{c c c | c c c}
\hline \hline 
\multicolumn{3}{c}{Process} & $\sigma \times \textrm{BR}~(\textrm{fb})$ & ~Efficiency~ & $N_{\textrm{events}}$ ($10\,\textrm{ab}^{-1}$) \\ 
\hline \hline
\multirow{4}{*}{$\mu V \to \tau h$} 
& \multicolumn{2}{c|}{Signal, $C_{He,\mu\tau} / \Lambda^2 = 1 / (10\,\textrm{TeV})^2$} 
& $1.4\times 10^{-2}$ & 0.53 & 75.2 \\
& \multicolumn{2}{c|}{Signal, $C_{Hl,\mu\tau}^{(1)} / \Lambda^2 = 1 / (10\,\textrm{TeV})^2$} 
& $1.4 \times 10^{-2}$ & 0.54 & 74.8 \\
& \multicolumn{2}{c|}{Signal, $C_{Hl,\mu\tau}^{(3)} / \Lambda^2 = 1 / (10\,\textrm{TeV})^2$} 
& $1.4 \times 10^{-2}$ & 0.54 & 74.8 \\
& \multicolumn{2}{c|}{Signal, $C_{eB,\mu\tau} / \Lambda^2 = 1 / (10\,\textrm{TeV})^2$} 
& $0.79$ & 0.57 & $4.50 \times 10^{3}$ \\
& \multicolumn{2}{c|}{Signal, $C_{eW,\mu\tau} / \Lambda^2 = 1 / (10\,\textrm{TeV})^2$} 
& $0.21$ & 0.54 & $1.13 \times 10^{3}$ \\
\Cline{0.5pt}{2-6}
& \multicolumn{2}{c|}{$\mu\nu h(b\bar{b})$} 
& $1.1\times 10^{-2}$ & $1.2\times 10^{-2}$ & 1.4 \\
& \multicolumn{2}{c|}{$\mu\nu b\bar{b}$} 
& $6.4\times 10^{-2}$ & $3.0\times 10^{-3}$ & 1.8 \\
\Cline{0.2pt}{2-6}
& \multicolumn{2}{c|}{Total Background} & & & 3.4 \\ 
\hline \hline
\multirow{9}{*}{$\mu W \to \tau W$} 
& \multicolumn{2}{c|}{Signal, $C_{He,\mu\tau} / \Lambda^2 = 1 / (10\,\textrm{TeV})^2$} 
& $6.6\times 10^{-2}$ & 0.32 & 216 \\
& \multicolumn{2}{c|}{Signal, $C_{Hl,\mu\tau}^{(1)} / \Lambda^2 = 1 / (10\,\textrm{TeV})^2$} 
& $6.6 \times 10^{-2}$ & 0.233 & 155 \\
& \multicolumn{2}{c|}{Signal, $C_{Hl,\mu\tau}^{(3)} / \Lambda^2 = 1 / (10\,\textrm{TeV})^2$} 
& $6.6 \times 10^{-2}$ & 0.231 & 154 \\
& \multicolumn{2}{c|}{Signal, $C_{eB,\mu\tau} / \Lambda^2 = 1 / (10\,\textrm{TeV})^2$} 
& $8.9 \times 10^{-3}$ & 0.016 & 1.4 \\
& \multicolumn{2}{c|}{Signal, $C_{eW,\mu\tau} / \Lambda^2 = 1 / (10\,\textrm{TeV})^2$} 
& $0.6$ & 0.213 & $1.29\times 10^{3}$ \\%1290.4
\Cline{0.5pt}{2-6}
  & \multirow{1}{*}{$\tau\nu_{\tau}jj$} 
  & (annihilation) & $4.2 \times 10^{-2}$ & $2.5\times 10^{-3}$ & 0.98 \\
  \Cline{0.2pt}{2-6}
%& & (charged VBF) & 0.82 & $0.0$ & 0.0 \\
%& & (neutral VBF) & 1.75 & $0.0$ & 0.0 \\
  & \multirow{2}{*}{$\tau\tau$}
  & (annihilation) & $6.4\times 10^{-1}$ & $1.7\times 10^{-4}$ & 0.24 \\
& & (charged VBF) & 22.6 & $2.9\times 10^{-6}$ & 0.64 \\
%& & (neutral VBF) & 8.1 & $0.0$ & 0.0 \\ 
\Cline{0.2pt}{2-6}
& \multicolumn{2}{c|}{Total Background} & & & 1.87 \\ 
\hline\hline
\multirow{5}{*}{$\mu \mu \to \mu \tau$} 
& \multicolumn{2}{c|}{Signal, $C_{ee,\mu\mu\mu\tau} / \Lambda^2 = 1 / (10\,\textrm{TeV})^2$} &
$4.1 \times 10^{2}$ & 0.19 & $8.00 \times 10^{5}$ \\
& \multicolumn{2}{c|}{Signal, $C_{ll,\mu\mu\mu\tau} / \Lambda^2 = 1 / (10\,\textrm{TeV})^2$} 
& $4.1 \times 10^{2}$ & 0.21 & $8.76\times 10^{5}$ \\
& \multicolumn{2}{c|}{Signal, $C_{le,\mu\mu\mu\tau} / \Lambda^2 = 1 / (10\,\textrm{TeV})^2$} 
& $1.0 \times 10^{2}$ & 0.23 & $2.38\times 10^{5}$ \\
& \multicolumn{2}{c|}{Signal, $C_{le,\mu\tau\mu\mu} / \Lambda^2 = 1 / (10\,\textrm{TeV})^2$} 
& $1.0 \times 10^{2}$ & 0.25 & $2.57\times 10^{5}$ \\
\Cline{0.5pt}{2-6}
& \multicolumn{2}{c|}{$\mu\overline{\nu}_\mu \nu_{\tau}\tau$} 
& $8.1$ & $6.2 \times 10^{-5}$ & $5.0$ \\
\hline \hline
\multirow{4}{*}{$\mu V \to \tau \tau \tau$} 
& \multicolumn{2}{c|}{Signal, $C_{ee,\tau\tau\mu\tau} / \Lambda^2 = 1 / (10\,\textrm{TeV})^2$} &
$8.9\times 10^{-2}$ & 0.73 & $650$ \\
& \multicolumn{2}{c|}{Signal, $C_{ll,\tau\tau\mu\tau} / \Lambda^2 = 1 / (10\,\textrm{TeV})^2$} 
& $9.0\times 10^{-2}$ & 0.69 & $624$ \\
& \multicolumn{2}{c|}{Signal, $C_{le,\mu\tau\tau\tau} / \Lambda^2 = 1 / (10\,\textrm{TeV})^2$} 
& $1.1\times 10^{-2}$ & 0.71 & 78.2 \\
& \multicolumn{2}{c|}{Signal, $C_{le,\tau\tau\mu\tau} / \Lambda^2 = 1 / (10\,\textrm{TeV})^2$} 
& $1.1\times 10^{-2}$ & 0.72 & 80.8 \\
\hline \hline
\multirow{5}{*}{$\mu^\mp V \to \mu^\pm \tau^\mp \tau^\mp$} 
& \multicolumn{2}{c|}{Signal, $C_{ee,\mu\tau\mu\tau} / \Lambda^2 = 1 / (10\,\textrm{TeV})^2$} &
$4.6 \times 10^{-2}$ & 0.90 & 414 \\
& \multicolumn{2}{c|}{Signal, $C_{ll,\mu\tau\mu\tau} / \Lambda^2 = 1 / (10\,\textrm{TeV})^2$} 
& $4.7 \times 10^{-2}$ & 0.86 & 399 \\
& \multicolumn{2}{c|}{Signal, $C_{le,\mu\tau\mu\tau} / \Lambda^2 = 1 / (10\,\textrm{TeV})^2$} 
& $1.2 \times 10^{-2}$ & 0.83 & 95.8 \\
\Cline{0.5pt}{2-6}
& \multicolumn{2}{c|}{$4 \tau$ (annihilation; central $\mu$ and 2 $\tau$)} 
& $2.7\times 10^{-4}$ & 0.66 & 1.8\\
& \multicolumn{2}{c|}{$\nu_{\mu} \mu \nu_{\tau} \tau \tau \tau$ (central $\mu$ and 2 $\tau$)} 
& $7.2\times 10^{-3}$ & 0.04 & 3.2\\
& \multicolumn{2}{c|}{$\mu W \nu_{\mu} \tau \tau$} 
 & $1.3 \times 10^{-2}$ & 0.03 & 3.9 \\
& \multicolumn{2}{c|}{$\mu \nu_{\mu} \tau \nu_{\tau} j j$ (jet mistag)} 
& $1.8 \times 10^{-2}$ & 0.05 & 8.3 \\
\hline
& \multicolumn{2}{c|}{Total Background} 
& & & 17.2 \\
\hline \hline
\end{tabular}
}
\caption{
Table summarizing the signal and background efficiencies and expected event rates at a $\sqrt{s} = 10\,\textrm{TeV}$ muon collider for the various processes we have considered. For each process (first column), we show the cross section times branching ratio ($\sigma \times \textrm{BR}$), efficiency and number of expected events (assuming $10\,\textrm{ab}^{-1}$ integrated luminosity) for the signal (with a representative choice of operators) and the non-negligible background processes. The cross section times branching ratio is reported for the process {\em after} the pre-selection cuts requiring the correct number of reconstructed particle objects relevant for the process (including tagging efficiencies), while the efficiency is only the ratio of these events that pass cuts on higher-level variables. Both the process and its complex conjugate are included in calculating the cross section and the total event count. We elaborate on the cuts used for each process in the upcoming sections. See text for details.}
\label{tab:lfv_muc_summary}
\end{table}